%
%
%
%
%
%




\catcode`\@=13                                                         
\def@{\errmessage{AmS-TeX error: \string@ has no current use
     (use \string\@\space for printed \string@ symbol)}}
\catcode`\@=11                                                         
\def\@{\char'100 }
\catcode`\~=13                                                         


\def\err@AmS#1{\errmessage{AmS-TeX error: #1}}                         


\def\eat@AmS#1{}

\long\def\comp@AmS#1#2{\def\@AmS{#1}\def\@@AmS{#2}\ifx
   \@AmS\@@AmS\def\cresult@AmS{T}\else\def\cresult@AmS{F}\fi}          

\def\in@AmS#1#2{\def\intest@AmS##1#1##2{\comp@AmS##2\end@AmS\if T\cresult@AmS
   \def\cresult@AmS{F}\def\in@@AmS{}\else
   \def\cresult@AmS{T}\def\in@@AmS####1\end@AmS{}\fi\in@@AmS}%
   \def\cresult@AmS{F}\intest@AmS#2#1\end@AmS}                         


\let\relax@AmS=\relax                                                  


\def\magstep#1{\ifcase#1 \@m\or 1200\or 1440\or 1728\or 2074\or 2488\fi
     \relax@AmS}

\def\iterate{\body\let\next\iterate \else\let\next\relax@AmS\fi \next}

\def\enskip{\hskip.5em\relax@AmS}

\def\strut{\relax@AmS\ifmmode\copy\strutbox\else\unhcopy\strutbox\fi}

\let\+=\relax@AmS
\def\sett@b{\ifx\next\+\let\next\relax@AmS
    \def\next{\afterassignment\s@tt@b\let\next}%
  \else\let\next\s@tcols\fi\next}
\def\s@tt@b{\let\next\relax@AmS\us@false\m@ketabbox}

\def\smash{\relax@AmS 
  \ifmmode\def\next{\mathpalette\mathsm@sh}\else\let\next\makesm@sh
  \fi\next}


\def\define#1{\expandafter\ifx\csname\expandafter\eat@AmS\string#1\endcsname
   \relax@AmS\def\dresult@AmS{\def#1}\else
   \err@AmS{\string#1\space is already defined}\def
      \dresult@AmS{\def\garbage@AmS}\fi\dresult@AmS}                   

\def\predefine#1#2{\let#1=#2}


\chardef\plus=`+
\chardef\equal=`=
\chardef\less=`<
\chardef\more=`>


\let\ic@AmS=\/
\def\/{\unskip\ic@AmS}

\def\Space@AmS.{\futurelet\Space@AmS\relax@AmS}
\Space@AmS. 

\def~{\unskip\futurelet\tok@AmS\s@AmS}
\def\s@AmS{\ifx\tok@AmS\Space@AmS\def\next@AmS{}\else
        \def\next@AmS{\ }\fi\penalty 9999 \next@AmS}                  

\def\period{\unskip.\spacefactor3000 { }}

\def\srdr@AmS{\thinspace}                                             
\def\drsr@AmS{\kern .02778em }
\def\sldl@AmS{\kern .02778em}
\def\dlsl@AmS{\thinspace}

\def\lqtest@AmS#1{\comp@AmS{#1}`\if T\cresult@AmS\else\comp@AmS{#1}\lq\fi}


\def\qspace#1{\unskip
  \lqtest@AmS{#1}\let\fresult@AmS=\cresult@AmS\if T\cresult@AmS
     \def\qspace@AmS{\ifx\tok@AmS\Space@AmS\def\next@AmS{\dlsl@AmS`}\else
       \def\next@AmS{\qspace@@AmS}\fi\next@AmS}\else
     \def\qspace@AmS{\ifx\tok@AmS\Space@AmS\def\next@AmS{\drsr@AmS'}\else
       \def\next@AmS{\qspace@@AmS}\fi\next@AmS}\fi
    \futurelet\tok@AmS\qspace@AmS}                                    

\def\qspace@@AmS{\futurelet\tok@AmS\qspace@@@AmS}

\def\qspace@@@AmS{\if T\fresult@AmS  \ifx\tok@AmS`\sldl@AmS`\else
       \ifx\tok@AmS\lq\sldl@AmS`\else \dlsl@AmS`\fi \fi
                         \else  \ifx\tok@AmS'\srdr@AmS'\else
        \ifx\tok@AmS\rq\srdr@AmS'\else \drsr@AmS'\fi \fi
        \fi}

\def\{{\relax@AmS\ifmmode\delimiter"4266308 \else
    $\delimiter"4266308 $\fi}                            

\def\}{\relax@AmS\ifmmode\delimiter"5267309 \else$\delimiter"5267309 $\fi}

\def\AmSTeX{$\cal A$\kern-.1667em\lower.5ex\hbox{$\cal M$}\kern-.125em
     $\cal S$-\TeX}


\def\linebreak{\unskip\penalty-10000 }                                
\def\pagebreak{\vadjust{\penalty-10000 }}

\def\newline{\ifvmode \err@AmS{There's no line here to break}\else
     \hfil\penalty-10000 \fi}

\def\topspace#1{\insert\topins{\penalty100 \splittopskip=0pt
     \vbox to #1{}}}
\def\midspace#1{\setbox0=\vbox to #1{}\advance\dimen0 by \pagetotal
  \ifdim\dimen0>\pagegoal\topspace{#1}\else\vadjust{\box0}\fi}

\long\def\comment{\begingroup
 \catcode`\{=12 \catcode`\}=12 \catcode`\#= 12 \catcode`\^^M=12
   \catcode`\%=12 \catcode`^^A=14
    \comment@AmS}
\begingroup\catcode`^^A=14
\catcode`\^^M=12  ^^A
\long\gdef\comment@AmS#1^^M#2{\comp@AmS\endcomment{#2}\if T\cresult@AmS^^A
\def\comment@@AmS{\endgroup}\else^^A
 \long\def\comment@@AmS{\comment@AmS#2}\fi\comment@@AmS}\endgroup     


\def\text#1{\hbox{\rm#1}}

\def\quad{\relax@AmS\ifmmode
    \hbox{\hskip1em}\else\hskip1em\relax@AmS\fi}                      
\def\qquad{\quad\quad}
\def\,{\relax@AmS\ifmmode\mskip\thinmuskip\else$\mskip\thinmuskip$\fi}
\def\;{\relax@AmS
  \ifmmode\mskip\thickmuskip\else$\mskip\thickmuskip$\fi}

\def\frac#1#2{{#1\over#2}}

\mathchardef\:="603A                                                  


\def\big@AmS#1{{\hbox{$\left#1\vbox to\big@@AmS{}\right.\offspace@AmS$}}}
\def\Big@AmS#1{{\hbox{$\left#1\vbox to\Big@@AmS{}\right.\offspace@AmS$}}}
\def\bigg@AmS#1{{\hbox{$\left#1\vbox to\bigg@@AmS{}\right.\offspace@AmS$}}}
\def\Bigg@AmS#1{{\hbox{$\left#1\vbox to\Bigg@@AmS{}\right.\offspace@AmS$}}}
\def\offspace@AmS{\nulldelimiterspace0pt \mathsurround0pt }

\def\big@@AmS{8.5pt}                                
\def\Big@@AmS{11.5pt}
\def\bigg@@AmS{14.5pt}
\def\Bigg@@AmS{17.5pt}

\def\bigl{\mathopen\big@AmS}
\def\bigm{\mathrel\big@AmS}
\def\bigr{\mathclose\big@AmS}
\def\Bigl{\mathopen\Big@AmS}
\def\Bigm{\mathrel\Big@AmS}
\def\Bigr{\mathclose\Big@AmS}
\def\biggl{\mathopen\bigg@AmS}
\def\biggm{\mathrel\bigg@AMS}
\def\biggr{\mathclose\bigg@AmS}
\def\Biggl{\mathopen\Bigg@AmS}
\def\Biggm{\mathrel\Bigg@AmS}
\def\Biggr{\mathclose\Bigg@AmS}


{\catcode`'=13 \gdef'{^\bgroup\prime\prime@AmS}}
\def\prime@AmS{\futurelet\tok@AmS\prime@@AmS}
\def\prime@@@AmS#1{\futurelet\tok@AmS\prime@@AmS}
\def\prime@@AmS{\ifx\tok@AmS'\def\next@AmS{\prime\prime@@@AmS}\else
   \def\next@AmS{\egroup}\fi\next@AmS}


\def\topsmash{\relax@AmS\ifmmode\def\topsmash@AmS
   {\mathpalette\mathtopsmash@AmS}\else
    \let\topsmash@AmS=\maketopsmash@AmS\fi\topsmash@AmS}

\def\maketopsmash@AmS#1{\setbox0=\hbox{#1}\topsmash@@AmS}

\def\mathtopsmash@AmS#1#2{\setbox0=\hbox{$#1{#2}$}\topsmash@@AmS}

\def\topsmash@@AmS{\vbox to 0pt{\kern-\ht0\box0}}

\def\botsmash{\relax@AmS\ifmmode\def\botsmash@AmS
   {\mathpalette\mathbotsmash@AmS}\else
     \let\botsmash@AmS=\makebotsmash@AmS\fi\botsmash@AmS}

\def\makebotsmash@AmS#1{\setbox0=\hbox{#1}\botsmash@@AmS}

\def\mathbotsmash@AmS#1#2{\setbox0=\hbox{$#1{#2}$}\botsmash@@AmS}

\def\botsmash@@AmS{\vbox to \ht0{\box0\vss}}


\def\LimitsOnSums{\let\slimits@AmS=\displaylimits}                    
\def\NoLimitsOnSums{\let\slimits@AmS=\nolimits}

\LimitsOnSums

\mathchardef\coprod@AmS"1360       \def\coprod{\coprod@AmS\slimits@AmS}
\mathchardef\bigvee@AmS"1357       \def\bigvee{\bigvee@AmS\slimits@AmS}
\mathchardef\bigwedge@AmS"1356     \def\bigwedge{\bigwedge@AmS\slimits@AmS}
\mathchardef\biguplus@AmS"1355     \def\biguplus{\biguplus@AmS\slimits@AmS}
\mathchardef\bigcap@AmS"1354       \def\bigcap{\bigcap@AmS\slimits@AmS}
\mathchardef\bigcup@AmS"1353       \def\bigcup{\bigcup@AmS\slimits@AmS}
\mathchardef\prod@AmS"1351         \def\prod{\prod@AmS\slimits@AmS}
\mathchardef\sum@AmS"1350          \def\sum{\sum@AmS\slimits@AmS}
\mathchardef\bigotimes@AmS"134E    \def\bigotimes{\bigotimes@AmS\slimits@AmS}
\mathchardef\bigoplus@AmS"134C     \def\bigoplus{\bigoplus@AmS\slimits@AmS}
\mathchardef\bigodot@AmS"134A      \def\bigodot{\bigodot@AmS\slimits@AmS}
\mathchardef\bigsqcup@AmS"1346     \def\bigsqcup{\bigsqcup@AmS\slimits@AmS}

\def\LimitsOnInts{\let\ilimits@AmS=\displaylimits}
\def\NoLimitsOnInts{\let\ilimits@AmS=\nolimits}

\NoLimitsOnInts

\mathchardef\int@AmS"1352
\def\int{\gdef\intflag@AmS{T}\int@AmS\ilimits@AmS}                    

\mathchardef\oint@AmS"1348 \def\oint{\gdef\intflag@AmS{T}\oint@AmS\ilimits@AmS}

\def\inttest@AmS#1{\def\intflag@AmS{F}\setbox0=\hbox{$#1$}}

\def\intic@AmS{\mathchoice{\hbox{\hskip5pt}}{\hbox
          {\hskip4pt}}{\hbox{\hskip4pt}}{\hbox{\hskip4pt}}}           
\def\negintic@AmS{\mathchoice
  {\hbox{\hskip-5pt}}{\hbox{\hskip-4pt}}{\hbox{\hskip-4pt}}{\hbox{\hskip-4pt}}}
\def\intkern@AmS{\mathchoice{\!\!\!}{\!\!}{\!\!}{\!\!}}
\def\intdots@AmS{\mathchoice{\cdots}{{\cdotp}\mkern 1.5mu
    {\cdotp}\mkern 1.5mu{\cdotp}}{{\cdotp}\mkern 1mu{\cdotp}\mkern 1mu
      {\cdotp}}{{\cdotp}\mkern 1mu{\cdotp}\mkern 1mu{\cdotp}}}

\newcount\intno@AmS                                                   

\def\intii{\gdef\intflag@AmS{T}\intno@AmS=2\futurelet                 
              \tok@AmS\ints@AmS}
\def\intiii{\gdef\intflag@AmS{T}\intno@AmS=3\futurelet\tok@AmS\ints@AmS}
\def\intiv{\gdef\intflag@AmS{T}\intno@AmS=4\futurelet\tok@AmS\ints@AmS}
\def\intdotsint{\gdef\intflag@AmS{T}\intno@AmS=0\futurelet
    \tok@AmS\ints@AmS}

\def\ints@AmS{\findlimits@AmS\ints@@AmS}

\def\findlimits@AmS{\def\ignoretoken@AmS{T}\ifx\tok@AmS\limits
   \def\limits@AmS{T}\else\ifx\tok@AmS\nolimits\def\limits@AmS{F}\else
     \def\ignoretoken@AmS{F}\ifx\ilimits@AmS\nolimits\def\limits@AmS{F}\else
       \def\limits@AmS{T}\fi\fi\fi}

\def\multintlimits@AmS{\int@AmS\ifnum \intno@AmS=0\intdots@AmS
  \else \intkern@AmS\fi
    \ifnum\intno@AmS>2\int@AmS\intkern@AmS\fi
     \ifnum\intno@AmS>3 \int@AmS\intkern@AmS\fi \int@AmS}

\def\multint@AmS{\int\ifnum \intno@AmS=0\intdots@AmS\else\intkern@AmS\fi
   \ifnum\intno@AmS>2\int\intkern@AmS\fi
    \ifnum\intno@AmS>3 \int\intkern@AmS\fi \int}

\def\ints@@AmS{\if F\ignoretoken@AmS\def\ints@@@AmS{\if
    T\limits@AmS\negintic@AmS
 \mathop{\intic@AmS\multintlimits@AmS}\limits\else
    \multint@AmS\nolimits\fi}\else\def\ints@@@AmS{\if T\limits@AmS
   \negintic@AmS\mathop{\intic@AmS\multintlimits@AmS}\limits\else
    \multint@AmS\nolimits\fi\eat@AmS}\fi\ints@@@AmS}

\def\LimitsOnNames{\let\nlimits@AmS=\displaylimits}
\def\NoLimitsOnNames{\let\nlimits@AmS=\nolimits}

\LimitsOnNames


\def\operatornamewithlimits#1{\mathop{\mathcode`'="7027 \mathcode`-="702D
   \rm #1}\nlimits@AmS}

\def\liminj{\setbox0=\hbox{\rm lim}\mathop{\rm lim}
		\limits_{\topsmash{\hbox to \wd0{\leftarrowfill}}}}
\def\limproj{\setbox0=\hbox{\rm lim}\mathop{\rm lim}
		\limits_{\topsmash{\hbox to \wd0{\rightarrowfill}}}}


\newdimen\buffer@AmS
\buffer@AmS=\fontdimen13\tenex                                        
\newdimen\buffer
\buffer=\buffer@AmS

\def\resetbuffer{\fontdimen13 \tenex=\buffer@AmS \buffer=\buffer@AmS}


\def\Let@AmS{\relax@AmS\iffalse{\fi\let\\=\cr\iffalse}\fi}            

\def\align{\def\vspace##1{\noalign{\vskip ##1}}                       
  \,\vcenter\bgroup\Let@AmS\tabskip=0pt\openup3pt\mathsurround=0pt  
  \halign\bgroup\strut
  \hfil$\displaystyle{##}$&$\displaystyle{{}##}$\hfil\cr}        

\def\endalign{\strut\crcr\egroup\egroup}

\def\bunch{\def\vspace##1{\noalign{\vskip ##1}}
  \,\vcenter\bgroup\Let@AmS\tabskip=0pt\openup3pt\mathsurround=0pt
     \halign\bgroup\strut\hfil$\displaystyle{##}$\hfil\cr}

\def\endbunch{\strut\crcr\egroup\egroup}

\def\matrix{\catcode`\^^I=4 \futurelet\tok@AmS\matrix@AmS}            

\def\matrix@AmS{\relax@AmS\ifnum`}=0\fi\ifx\tok@AmS\format
   \def\next@AmS{\expandafter\matrix@@AmS\eat@AmS}\else
   \def\next@AmS{\matrix@@@AmS}\fi\next@AmS}

\def\matrix@@@AmS{
 \ifnum`{=0\fi\iffalse}\fi\,\vcenter\bgroup\Let@AmS\tabskip=0pt
    \normalbaselines\halign\bgroup $\strut\hfil##\hfil$&&\quad$\strut
  \hfil##\hfil$\cr\strut\cr\noalign{\kern-\baselineskip}}             

\def\matrix@@AmS#1\\{
   \def\premable@AmS{#1}\toks@{##}
 \def\c{$\copy\strutbox\hfil\the\toks@\hfil$}\def\r
   {$\copy\strutbox\hfil\the\toks@$}%
   \def\l{$\copy\strutbox\the\toks@\hfil$}%
\setbox0=
\hbox{\xdef\Preamble@AmS{\premable@AmS}}
 \def\vspace##1{\noalign{\vskip ##1}}\ifnum`{=0\fi\iffalse}\fi
\,\vcenter\bgroup\Let@AmS
  \tabskip=0pt\normalbaselines\halign\bgroup\span\Preamble@AmS\cr
    \mathstrut\cr\noalign{\kern-\baselineskip}}

\def\endmatrix{\crcr\mathstrut\cr\noalign{\kern-\baselineskip
   }\egroup\egroup\,\catcode`\^^I=10 }

\def\spacedots#1for#2{\multispan#2\leaders\hbox{$\mkern#1mu.\mkern
    #1mu$}\hfill}

\def\enabletabs{\catcode`\^^I=4 \enabletabs@AmS}
\def\enabletabs@AmS#1\disabletabs{#1\catcode`\^^I=10 }                

\def\smallmatrix{\futurelet\tok@AmS\smallmatrix@AmS}                  

\def\smallmatrix@AmS{\relax@AmS\ifnum`}=0\fi\ifx\tok@AmS\format
   \def\next@AmS{\expandafter\smallmatrix@@AmS\eat@AmS}\else
   \def\next@AmS{\smallmatrix@@@AmS}\fi\next@AmS}

\def\smallmatrix@@@AmS{
 \ifnum`{=0\fi\iffalse}\fi\,\vcenter\bgroup\Let@AmS\tabskip=0pt
    \baselineskip8pt\lineskip1pt\lineskiplimit1pt
  \halign\bgroup $\strut\hfil##\hfil$&&\;$\strut
  \hfil##\hfil$\cr\strut\cr\noalign{\kern-\baselineskip}}

\def\smallmatrix@@AmS#1\\{
   \def\premable@AmS{#1}\toks@{##}
 \def\c{$\copy\strutbox\hfil\the\toks@\hfil$}\def\r
   {$\copy\strutbox\hfil\the\toks@$}%
   \def\l{$\copy\strutbox\the\toks@\hfil$}%
\hbox{\xdef\Preamble@AmS{\premable@AmS}}
 \def\vspace##1{\noalign{\vskip ##1}}\ifnum`{=0\fi\iffalse}\fi
\,\vcenter\bgroup\Let@AmS
     \tabskip=0pt\baselineskip8pt\lineskip1pt\lineskiplimit1pt
\halign\bgroup\span\Preamble@AmS\cr
    \mathstrut\cr\noalign{\kern-\baselineskip}}

\def\endsmallmatrix{\crcr\mathstrut\cr\noalign{\kern-\baselineskip}
   \egroup\egroup\,}

\def\cases{\left\{ \,\vcenter\bgroup\Let@AmS\normalbaselines\tabskip=0pt
   \halign\bgroup$##\hfil$&\qquad$##\hfil$\cr}                        

\def\endcases{\crcr\egroup\egroup\right.}


\def\TagsOnLeft{\def\tagposition@AmS{L}}
\def\TagsOnRight{\def\tagposition@AmS{R}}
\def\TagsAsMath{\def\tagstyle@AmS{M}}
\def\TagsAsText{\def\tagstyle@AmS{T}}

\TagsOnLeft
\TagsAsText

\def\tag#1$${\if L\tagposition@AmS
    \leqno\else\eqno\fi\def\atag@AmS{T}\maketag@AmS#1\tagend@AmS$$}   

\def\maketag@AmS{\futurelet\tok@AmS\maketag@@AmS}                     
\def\maketag@@AmS{\ifx\tok@AmS[\def\next@AmS{\maketag@@@AmS}\else
      \def\next@AmS{\maketag@@@@AmS}\fi\next@AmS}
\def\maketag@@@AmS[#1]#2\tagend@AmS{\if F\atag@AmS\else             
   \if M\tagstyle@AmS\hbox{$#1$}\else\hbox{#1}\fi\fi
       \gdef\atag@AmS{F}}
\def\maketag@@@@AmS#1\tagend@AmS{\if F\atag@AmS \else
        \if T\autotag@AmS \setbox0=\hbox
    {\if M\tagstyle@AmS\tagform@AmS{$#1$}\else\tagform@AmS{#1}\fi}
                        \ifdim\wd0=0pt \tagform@AmS{*}\else
            \if M\tagstyle@AmS\tagform@AmS{$#1$}\else\tagform@AmS{#1}\fi
                     \fi\else
               \if M\tagstyle@AmS\tagform@AmS{$#1$}\else\tagform@AmS{#1}\fi
                     \fi
                  \fi\gdef\atag@AmS{F}}

\def\tagform@AmS#1{\hbox{\rm(#1\unskip)}}

\def\AutoTag{\def\autotag@AmS{T}}
\def\NoAutoTag{\def\autotag@AmS{F}}

\NoAutoTag

\def\inaligntag@AmS{F} \def\inbunchtag@AmS{F}                         

\def\CenteredTagsOnBrokens{\def\centerbroken@AmS{T}}                  
\def\TopOrBottomTagsOnBrokens{\def\centerbroken@AmS{F}}
\TopOrBottomTagsOnBrokens

\def\broken{\global\setbox0=\vbox\bgroup\Let@AmS\tabskip=0pt
 \if T\inaligntag@AmS\else
   \if T\inbunchtag@AmS\else\openup3pt\fi\fi\mathsurround=0pt
     \halign\bgroup\strut\hfil$\displaystyle{##}$&$\displaystyle{{}##}$\hfill
      \cr}
\def\endbroken{\strut\crcr\egroup\egroup
      \global\setbox7=\vbox{\unvbox0\setbox1=\lastbox
      \hbox{\unhbox1\unskip\setbox2=\lastbox
       \unskip\setbox3=\lastbox
         \global\setbox4=\copy3
          \box3\box2}}
  \if L\tagposition@AmS
     \if T\inaligntag@AmS
           \if T\centerbroken@AmS\gdef\broken@AmS
                {&\vcenter{\vbox{\moveleft\wd4\box7}}}
           \else
            \gdef\broken@AmS{&\vbox{\moveleft\wd4\vtop{\unvbox7}}}
           \fi
     \else                                                            
           \if T\centerbroken@AmS\gdef\broken@AmS
                {\vcenter{\box7}}%
           \else
              \gdef\broken@AmS{\vtop{\unvbox7}}%
           \fi
     \fi
  \else                                                  
      \if T\inaligntag@AmS
           \if T\centerbroken@AmS
              \gdef\broken@AmS{&\vcenter{\vbox{\moveleft\wd4\box7}}}%
          \else
             \gdef\broken@AmS{&\vbox{\moveleft\wd4\box7}}%
          \fi
      \else
          \if T\centerbroken@AmS
            \gdef\broken@AmS{\vcenter{\box7}}%
          \else
             \gdef\broken@AmS{\box7}%
          \fi
      \fi
  \fi\broken@AmS}

\def\cbroken{\xdef\centerbroken@@AmS{\centerbroken@AmS}%
                       \def\centerbroken@AmS{T}\broken}               
\def\endcbroken{\endbroken\def\centerbroken@AmS{\centerbroken@@AmS}}

\def\multline#1${\in@AmS\tag{#1}\if T\cresult@AmS
 \def\multline@AmS{\def\atag@AmS{T}\getmltag@AmS#1$}\else
   \def\multline@AmS{\def\atag@AmS{F}\setbox9=\hbox{}\multline@@AmS
    \multline@@@AmS#1$}\fi\multline@AmS}                              

\def\getmltag@AmS#1\tag#2${\setbox9=\hbox{\maketag@AmS#2\tagend@AmS}%
           \multline@@AmS\multline@@@AmS#1$}

\def\multline@@AmS{\if L\tagposition@AmS
     \def\lwidth@AmS{\hskip\wd9}\def\rwidth@AmS{\hskip0pt}\else
      \def\lwidth@AmS{\hskip0pt}\def\rwidth@AmS{\hskip\wd9}\fi}      

\def\multline@@@AmS{\def\vspace##1{\noalign{\vskip ##1}}%
 \def\shoveright##1{##1\hfilneg\rwidth@AmS\quad}                      
  \def\shoveleft##1{\setbox                                           
      0=\hbox{$\displaystyle{}##1$}%
     \setbox1=\hbox{$\displaystyle##1$}%
     \ifdim\wd0=\wd1
    \hfilneg\lwidth@AmS\quad##1\else
      \setbox2=\hbox{\hskip\wd0\hskip-\wd1}%
       \hfilneg\lwidth@AmS\quad\hskip-.5\wd2 ##1\fi}
     \vbox\bgroup\Let@AmS\openup3pt\halign\bgroup\hbox to \the\displaywidth
      {$\displaystyle\hfil{}##\hfil$}\cr\hfilneg\quad
      \if L\tagposition@AmS\hskip-1em\copy9\quad\else\fi}             

\def\endmultline{\if R\tagposition@AmS\quad\box9                 
   \hskip-1em\else\fi\quad\hfilneg\crcr\egroup\egroup}

\def\aligntag#1$${\def\inaligntag@AmS{T}\openup3pt\mathsurround=0pt   
 \Let@AmS
   \def\tag{\gdef\atag@AmS{T}&}                                       
   \def\vspace##1{\noalign{\vskip##1}}                                
    \def\xtext##1{\noalign{\hbox{##1}}}                               
   \def\break{\noalign{\penalty-10000 }}                              
   \def\nobreak{\noalign{\penalty 10000 }}
   \def\allowbreak{\noalign{\penalty 0 }}
   \def\goodbreak{\noalign{\penalty -500 }}
    \gdef\atag@AmS{F}%
\if L\tagposition@AmS\laligntag@AmS#1$$\else
   \raligntag@AmS#1$$\fi}

\def\raligntag@AmS#1$${\tabskip\centering
   \halign to \the\displaywidth
{\hfil$\displaystyle{##}$\tabskip 0pt
    &$\displaystyle{{}##}$\hfil\tabskip\centering
   &\llap{\maketag@AmS##\tagend@AmS}\tabskip 0pt\cr\noalign{\vskip-
     \lineskiplimit}#1\crcr}$$}

\def\laligntag@AmS#1$${\tabskip\centering                             
   \halign to \the\displaywidth
{\hfil$\displaystyle{##}$\tabskip0pt
   &$\displaystyle{{}##}$\hfil\tabskip\centering
    &\kern-\displaywidth\rlap{\maketag@AmS##\tagend@AmS}\tabskip
    \the\displaywidth\cr\noalign{\vskip-\lineskiplimit}#1\crcr}$$}

\def\bunchtag#1$${\def\inbunchtag@AmS{T}\openup3pt\mathsurround=0pt   
    \Let@AmS
   \def\tag{\gdef\atag@AmS{T}&}
   \def\vspace##1{\noalign{\vskip##1}}
   \def\xtext##1{\noalign{\hbox{##1}}}
   \def\break{\noalign{\penalty-10000 }}
   \def\nobreak{\noalign{\penalty 10000 }}
   \def\allowbreak{\noalign{\penalty 0 }}
    \def\goodbreak{\noalign{\penalty -500 }}
  \if L\tagposition@AmS\lbunchtag@AmS#1$$\else
    \rbunchtag@AmS#1$$\fi}

\def\rbunchtag@AmS#1$${\tabskip\centering
    \halign to \displaywidth {$\hfil\displaystyle{##}\hfil$&
      \llap{\maketag@AmS##\tagend@AmS}\tabskip 0pt\cr\noalign{\vskip-
       \lineskiplimit}#1\crcr}$$}

\def\lbunchtag@AmS#1$${\tabskip\centering
   \halign to \displaywidth
{$\hfil\displaystyle{##}\hfil$&\kern-
    \displaywidth\rlap{\maketag@AmS##\tagend@AmS}\tabskip\the\displaywidth\cr
    \noalign{\vskip-\lineskiplimit}#1\crcr}$$}




\def\numeratorleft#1{#1\hskip 0pt plus 1filll\relax@AmS}
\def\numeratorright#1{\hskip 0pt plus 1filll\relax@AmS#1}
\def\numeratorcenter#1{\hskip 0pt plus 1filll\relax@AmS
      #1\hskip 0pt plus 1filll\relax@AmS}

\def\cfrac@AmS#1,{\def\numerator@AmS{#1}\cfrac@@AmS*}                 

\def\cfrac@@AmS#1;#2#3\cfend@AmS{\comp@AmS\cfmark@AmS{#2}\if T\cresult@AmS
 \gdef\cfrac@@@AmS
  {\expandafter\eat@AmS\numerator@AmS\strut\over\eat@AmS#1}\else
  \comp@AmS;{#2}\if T\cresult@AmS\gdef\cfrac@@@AmS
  {\expandafter\eat@AmS\numerator@AmS\strut\over\eat@AmS#1}\else
\gdef\cfrac@@@AmS{\if R\cftype@AmS\hfill\else\fi
    \expandafter\eat@AmS\numerator@AmS\strut
    \if L\cftype@AmS\hfill\else\fi\over
       \eat@AmS#1\displaystyle {\cfrac@AmS*#2#3\cfend@AmS}}
     \fi\fi\cfrac@@@AmS}

\def\cfrac#1{\def\cftype@AmS{C}\cfrac@AmS*#1;\cfmark@AmS\cfend@AmS}

\def\cfracl#1{\def\cftype@AmS{L}\cfrac@AmS*#1;\cfmark@AmS\cfend@AmS}

\def\cfracr#1{\def\cftype@AmS{R}\cfrac@AmS*#1;\cfmark@AmS\cfend@AmS}


\def\overrightarrow{\mathpalette\overrightarrow@AmS}

\def\overrightarrow@AmS#1#2{\vbox{\halign{$##$\cr
    #1{-}\mkern-6mu\cleaders\hbox{$#1\mkern-2mu{-}\mkern-2mu$}\hfill
     \mkern-6mu{\to}\cr
     \noalign{\kern -1pt\nointerlineskip}
     \hfil#1#2\hfil\cr}}}

\def\overleftarrow{\mathpalette\overleftarrow@Ams}

\def\overleftarrow@Ams#1#2{\vbox{\halign{$##$\cr
     #1{\leftarrow}\mkern-6mu\cleaders\hbox{$#1\mkern-2mu{-}\mkern-2mu$}\hfill
      \mkern-6mu{-}\cr
     \noalign{\kern -1pt\nointerlineskip}
     \hfil#1#2\hfil\cr}}}

\def\overleftrightarrow{\mathpalette\overleftrightarrow@AmS}

\def\overleftrightarrow@AmS#1#2{\vbox{\halign{$##$\cr
     #1{\leftarrow}\mkern-6mu\cleaders\hbox{$#1\mkern-2mu{-}\mkern-2mu$}\hfill
       \mkern-6mu{\to}\cr
    \noalign{\kern -1pt\nointerlineskip}
      \hfil#1#2\hfil\cr}}}

\def\underrightarrow{\mathpalette\underrightarrow@AmS}

\def\underrightarrow@AmS#1#2{\vtop{\halign{$##$\cr
    \hfil#1#2\hfil\cr
     \noalign{\kern -1pt\nointerlineskip}
    #1{-}\mkern-6mu\cleaders\hbox{$#1\mkern-2mu{-}\mkern-2mu$}\hfill
     \mkern-6mu{\to}\cr}}}

\def\underleftarrow{\mathpalette\underleftarrow@AmS}

\def\underleftarrow@AmS#1#2{\vtop{\halign{$##$\cr
     \hfil#1#2\hfil\cr
     \noalign{\kern -1pt\nointerlineskip}
     #1{\leftarrow}\mkern-6mu\cleaders\hbox{$#1\mkern-2mu{-}\mkern-2mu$}\hfill
      \mkern-6mu{-}\cr}}}

\def\underleftrightarrow{\mathpalette\underleftrightarrow@AmS}

\def\underleftrightarrow@AmS#1#2{\vtop{\halign{$##$\cr
      \hfil#1#2\hfil\cr
    \noalign{\kern -1pt\nointerlineskip}
     #1{\leftarrow}\mkern-6mu\cleaders\hbox{$#1\mkern-2mu{-}\mkern-2mu$}\hfill
       \mkern-6mu{\to}\cr}}}


\def\dotsc{\mathinner{\ldotp\ldotp\ldotp}}
\def\dotsi{\mathinner{\cdotp\cdotp\cdotp}}
\def\dotsj{\mathinner{\ldotp\ldotp\ldotp}}
\def\dotsb{\mathinner{\cdotp\cdotp\cdotp}}

\def\binary@AmS#1{{\thinmuskip 0mu \medmuskip 1mu \thickmuskip 1mu    
      \setbox0=\hbox{$#1{}{}{}{}{}{}{}{}{}$}\setbox1=\hbox
       {${}#1{}{}{}{}{}{}{}{}{}$}\ifdim\wd1>\wd0\gdef\binary@@AmS{T}\else
       \gdef\binary@@AmS{F}\fi}}

\def\dots{\relax@AmS\ifmmode\def\dots@AmS{\mdots@AmS}\else
    \def\dots@AmS{\tdots@AmS}\fi\dots@AmS}

\def\mdots@AmS{\futurelet\tok@AmS\mdots@@AmS}

\def\mdots@@AmS{\def\thedots@AmS{\dotsj}%
  \ifx\tok@AmS\bgroup\else
  \ifx\tok@AmS\egroup\else
  \ifx\tok@AmS$\else
  \ifx\tok@AmS\\ \iffalse}\fi\else                      
  \ifx\tok@AmS&  \iffalse}\fi\else
  \ifx\tok@AmS\left\else
  \ifx\tok@AmS\right\else
  \ifx\tok@AmS,\def\thedots@AmS{\dotsc}\else
  \inttest@AmS\tok@AmS\if T\intflag@AmS\def\thedots@AmS{\dotsi}\else
  \binary@AmS\tok@AmS\if T\binary@@AmS\def\thedots@AmS{\dotsb}\else
   \def\thedots@AmS{\dotsj}\fi\fi\fi\fi\fi\fi\fi\fi\fi\fi\thedots@AmS}

\def\tdots@AmS{\unskip\ \tdots@@AmS}

\def\tdots@@AmS{\futurelet\tok@AmS\tdots@@@AmS}

\def\tdots@@@AmS{$\ldots\,
   \ifx\tok@AmS,$\else
   \ifx\tok@AmS.\,$\else
   \ifx\tok@AmS;\,$\else
   \ifx\tok@AmS:\,$\else
   \ifx\tok@AmS?\,$\else
   \ifx\tok@AmS!\,$\else
   $\ \fi\fi\fi\fi\fi\fi}


\def\leftset#1\mid#2\rightset{\hbox{$\displaystyle
\left\{\,#1\vphantom{#1#2}\;\right|\;\left.
    #2\vphantom{#1#2}\,\right\}\offspace@AmS$}}


\def\dotii#1{{\mathop{#1}\limits^{\vbox to -1.4pt{\kern-2pt
   \hbox{\tenrm..}\vss}}}}
\def\dotiii#1{{\mathop{#1}\limits^{\vbox to -1.4pt{\kern-2pt
   \hbox{\tenrm...}\vss}}}}
\def\dotiv#1{{\mathop{#1}\limits^{\vbox to -1.4pt{\kern-2pt
   \hbox{\tenrm....}\vss}}}}

\def\hatsymbol{{\mathchoice{\null}{\null}{\,\,\hbox{\lower 10pt\hbox
    {$\widehat{\null}$}}}{\,\hbox{\lower 20pt\hbox
       {$\hat{\null}$}}}}}


\def\overset#1\to#2{{\mathop{#2}^{#1}}}

\def\underset#1\to#2{{\mathop{#2}_{#1}}}

\def\oversetbrace#1\to#2{{\overbrace{#2}^{#1}}}
\def\undersetbrace#1\to#2{{\underbrace{#2}_{#1}}}


\def\theuproot{0 pt}

\def\therightroot{0mu}

\def\r@@t#1#2{\setbox\z@\hbox{$\m@th#1\sqrt{#2}$}%
  \dimen@\ht\z@ \advance\dimen@-\dp\z@ \advance\dimen@\theuproot
  \mskip5mu\raise.6\dimen@\copy\rootbox \mskip-10mu \mskip\therightroot
    \box\z@\gdef\theuproot{0 pt}\gdef\therightroot{0mu}}              


\def\boxed#1{\setbox0=\hbox{$\displaystyle{#1}$}\hbox{\lower.4pt\hbox{\lower
   3pt\hbox{\lower 1\dp0\hbox{\vbox{\hrule height .4pt \hbox{\vrule width
   .4pt \hskip 3pt\vbox{\vskip 3pt\box0\vskip3pt}\hskip 3pt \vrule width
      .4pt}\hrule height .4pt}}}}}}


\def\documentstyle#1{\input #1.sty}

\newif\ifretry@AmS
\def\y@AmS{y } \def\y@@AmS{Y } \def\n@AmS{n } \def\n@@AmS{N }
\def\ask@AmS{\message
  {Do you want output? (y or n, follow answer by return) }\loop
   \read-1 to\answer@AmS
  \ifx\answer@AmS\y@AmS\retry@AmSfalse\outputon
   \else\ifx\answer@AmS\y@@AmS\retry@AmSfalse\outputon
    \else\ifx\answer@AmS\n@AmS\retry@AmSfalse\outputoff
     \else\ifx\answer@AmS\n@@AmS\retry@AmSfalse\outputoff
      \else \retry@AmStrue\fi\fi\fi\fi
  \ifretry@AmS\message{Type y or n, follow answer by return: }\repeat}

\def\outputoff{\global\output{\setbox0=\box255 \deadcycles=0}}

\def\outputon{\global\output{\output@AmS}}

\catcode`\@=13


\catcode`\@=11



\normallineskiplimit=1pt
\parindent 10pt
\hsize 26pc
\vsize 42pc


\font\eightrm=cmr8
\font\sixrm=cmr6
\font\eighti=cmmi8 \skewchar\eighti='177
\font\sixi=cmmi6 \skewchar\sixi='177
\font\eightsy=cmsy8 \skewchar\eightsy='60
\font\sixsy=cmsy6 \skewchar\sixsy='60
\font\eightbf=cmbx8
\font\sixbf=cmbx6
\font\eightsl=cmsl8
\font\eightit=cmti8
\font\tensmc=cmcsc10


\font\ninerm=cmr9
\font\ninei=cmmi9 \skewchar\ninei='177
\font\ninesy=cmsy9 \skewchar\ninesy='60
\font\ninebf=cmbx9
\font\ninesl=cmsl9
\font\nineit=cmti9


\def\tenpoint{\def\pointsize@AmS{t}\normalbaselineskip=12pt            
 \abovedisplayskip 12pt plus 3pt minus 9pt
 \belowdisplayskip 12pt plus 3pt minus 9pt
 \abovedisplayshortskip 0pt plus 3pt
 \belowdisplayshortskip 7pt plus 3pt minus 4pt
 \def\rm{\fam0\tenrm}%
 \def\it{\fam\itfam\tenit}%
 \def\sl{\fam\slfam\tensl}%
 \def\bf{\fam\bffam\tenbf}%
 \def\smc{\tensmc}%
 \def\mit{\fam 1}%
 \def\cal{\fam 2}%
 \textfont0=\tenrm   \scriptfont0=\sevenrm   \scriptscriptfont0=\fiverm
 \textfont1=\teni    \scriptfont1=\seveni    \scriptscriptfont1=\fivei
 \textfont2=\tensy   \scriptfont2=\sevensy   \scriptscriptfont2=\fivesy
 \textfont3=\tenex   \scriptfont3=\tenex     \scriptscriptfont3=\tenex
 \textfont\itfam=\tenit
 \textfont\slfam=\tensl
 \textfont\bffam=\tenbf \scriptfont\bffam=\sevenbf
   \scriptscriptfont\bffam=\fivebf
\normalbaselines\rm}

\def\eightpoint{\def\pointsize@AmS{8}\normalbaselineskip=10pt
 \abovedisplayskip 10pt plus 2.4pt minus 7.2pt
 \belowdisplayskip 10pt plus 2.4pt minus 7.2pt
 \abovedisplayshortskip 0pt plus 2.4pt
 \belowdisplayshortskip 5.6pt plus 2.4pt minus 3.2pt
 \def\rm{\fam0\eightrm}%
 \def\it{\fam\itfam\eightit}%
 \def\sl{\fam\slfam\eightsl}%
 \def\bf{\fam\bffam\eightbf}%
 \def\mit{\fam 1}%
 \def\cal{\fam 2}%
 \textfont0=\eightrm   \scriptfont0=\sixrm   \scriptscriptfont0=\fiverm
 \textfont1=\eighti    \scriptfont1=\sixi    \scriptscriptfont1=\fivei
 \textfont2=\eightsy   \scriptfont2=\sixsy   \scriptscriptfont2=\fivesy
 \textfont3=\tenex   \scriptfont3=\tenex     \scriptscriptfont3=\tenex
 \textfont\itfam=\eightit
 \textfont\slfam=\eightsl
 \textfont\bffam=\eightbf \scriptfont\bffam=\sixbf
   \scriptscriptfont\bffam=\fivebf
\normalbaselines\rm}


\def\ninepoint{\def\pointsize@AmS{9}\normalbaselineskip=11pt
 \abovedisplayskip 11pt plus 2.7pt minus 8.1pt
 \belowdisplayskip 11pt plus 2.7pt minus 8.1pt
 \abovedisplayshortskip 0pt plus 2.7pt
 \belowdisplayshortskip 6.3pt plus 2.7pt minus 3.6pt
 \def\rm{\fam0\ninerm}%
 \def\it{\fam\itfam\nineit}%
 \def\sl{\fam\slfam\ninesl}%
 \def\bf{\fam\bffam\ninebf}%
 \def\mit{\fam 1}%
 \def\cal{\fam 2}%
 \textfont0=\ninerm   \scriptfont0=\sevenrm   \scriptscriptfont0=\fiverm
 \textfont1=\ninei    \scriptfont1=\seveni    \scriptscriptfont1=\fivei
 \textfont2=\ninesy   \scriptfont2=\sevensy   \scriptscriptfont2=\fivesy
 \textfont3=\tenex   \scriptfont3=\tenex     \scriptscriptfont3=\tenex
 \textfont\itfam=\nineit
 \textfont\slfam=\ninesl
 \textfont\bffam=\ninebf \scriptfont\bffam=\sevenbf
   \scriptscriptfont\bffam=\fivebf
\normalbaselines\rm}


\newcount\footmarkcount@AmS
\footmarkcount@AmS=0
\newcount\foottextcount@AmS
\foottextcount@AmS=0

\def\footnotemark{\unskip\futurelet\tok@AmS\footnotemark@AmS}
\def\footnotemark@AmS{\ifx [\tok@AmS \def\next@AmS{\footnotemark@@AmS}\else
   \def\next@AmS{\footnotemark@@@AmS}\fi\next@AmS}
\def\footnotemark@@AmS[#1]{{#1}}
\def\footnotemark@@@AmS{\global\advance\footmarkcount@AmS by 1
 \xdef\thefootmarkcount@AmS{\the\footmarkcount@AmS}$^{\thefootmarkcount@AmS}$}

\def\makefootnote@AmS#1#2{\insert\footins{\interlinepenalty100
   \eightpoint
  \splittopskip=6.8pt
  \splitmaxdepth=2.8pt
   \floatingpenalty=20000
   \leftskip = 0pt  \rightskip = 0pt
    \noindent {#1}\footstrut{\ignorespaces#2\unskip}\topsmash{\strut}}}

\def\footnotetext{\futurelet\tok@AmS\footnotetext@}
\def\footnotetext@{\ifx [\tok@AmS \def\next@AmS{\footnotetext@@AmS}\else
  \def\next@AmS{\footnotetext@@@AmS}\fi\next@AmS}
\def\footnotetext@@AmS[#1]#2{\makefootnote@AmS{#1}{#2}}
\def\footnotetext@@@AmS#1{\global\advance\foottextcount@AmS by 1
  \xdef\thefoottextcount@AmS{\the\foottextcount@AmS}%
\makefootnote@AmS{$^{\thefoottextcount@AmS}$}{#1}}

\def\footnote{\unskip\futurelet\tok@AmS\footnote@AmS}
\def\footnote@AmS{\ifx [\tok@AmS \def\next@AmS{\footnote@@AmS}\else
   \def\next@AmS{\footnote@@@AmS}\fi\next@AmS}
\def\footnote@@AmS[#1]#2{{\edef\sf{\the\spacefactor}%
  {#1}\makefootnote@AmS{#1}{#2}\spacefactor=\sf}}
\def\footnote@@@AmS#1{\ifnum\footmarkcount@AmS=\foottextcount@AmS\else
 \errmessage{AmS-TeX warning: last footnote marker was \the\footmarkcount@AmS,
   last footnote was
   \the\foottextcount@AmS}\footmarkcount@AmS=\foottextcount@AmS\fi
   {\edef\sf{\the\spacefactor}\footnotemark@@@AmS\footnotetext@@@AmS{#1}%
    \spacefactor=\sf}}

\def\adjustfootnotemark#1{\advance\footmarkcount@AmS by #1}           
\def\adjustfootnote#1{\advance\foottextcount@AmS by #1}


\def\topmatter@AmS{F}                                                 
\def\topmatter{\def\topmatter@AmS{T}}

\def\filhss@AmS{plus 1000pt}                                          
\def\overlong{\def\filhss@AmS{plus 1000pt minus1000pt}}

\newbox\titlebox@AmS

\setbox\titlebox@AmS=\vbox{}                                          

\def\title#1\endtitle{{\let\\=\cr                                     
  \global\setbox\titlebox@AmS=\vbox{\tabskip0pt\filhss@AmS
  \halign to \hsize
    {\tenpoint\bf\hfil\ignorespaces##\unskip\hfil\cr#1\cr}}}\def     
     \filhss@AmSs{plus 1000pt}}

\def\isauthor@AmS{F}                                                
\newbox\authorbox@AmS

\def\author#1\endauthor{\gdef\isauthor@AmS{T}{\let\\=\cr
 \global\setbox\authorbox@AmS=\vbox{\tabskip0pt
 \filhss@AmS\halign to \hsize
   {\tenpoint\smc\hfil\ignorespaces##\unskip\hfil\cr#1\cr}}}\def
      \filhss@AmS{plus 1000pt}}


\def\uctext@AmS#1{\uppercase@AmS#1\gdef                           
       \uppercase@@AmS{}${\hskip-2\mathsurround}$}
\def\uppercase@AmS#1$#2${\gdef\uppercase@@AmS{\uppercase@AmS}\uppercase
    {#1}${#2}$\uppercase@@AmS}

\newcount\Notes@AmS                                             

\def\sfootnote@AmS{\unskip\futurelet\tok@AmS\sfootnote@@AmS}
\def\sfootnote@@AmS{\ifx [\tok@AmS \def\next@AmS{\sfootnote@@@AmS}\else
    \def\next@AmS{\sfootnote@@@@AmS}\fi\next@AmS}
\def\sfootnote@@@AmS[#1]#2{\global\toks@{#2}\advance\Notes@AmS by 1
  \expandafter\xdef\csname Note\romannumeral\Notes@AmS @AmS\endcsname
   {\the\toks@}}
\def\sfootnote@@@@AmS#1{\global\toks@{#1}\global\advance\Notes@AmS by 1
  \expandafter\xdef\csname Note\romannumeral\Notes@AmS @AmS\endcsname
  {\the\toks@}}

\def\Sfootnote@AmS{\unskip\futurelet\tok@AmS\Sfootnote@@AmS}
\def\Sfootnote@@AmS{\ifx [\tok@AmS \def\next@AmS{\Sfootnote@@@AmS}\else
    \def\next@AmS{\Sfootnote@@@@AmS}\fi\next@AmS}
\def\Sfootnote@@@AmS[#1]#2{{#1}\advance\Notes@AmS by 1
  {\edef\sf{\the\spacefactor}\makefootnote@AmS{#1}{\csname
     Note\romannumeral\Notes@AmS @AmS\endcsname}\spacefactor=\sf}}
\def\Sfootnote@@@@AmS#1{\ifnum\footmarkcount@AmS=\foottextcount@AmS\else
 \errmessage{AmS-TeX warning: last footnote marker was \the\footmarkcount@AmS,
  last footnote was
   \the\foottextcount@AmS}\footmarkcount@AmS=\foottextcount@AmS\fi
 {\edef\sf{\the\spacefactor}\footnotemark@@@AmS \global\advance\Notes@AmS by 1
    \footnotetext@@@AmS{\csname
      Note\romannumeral\Notes@AmS @AmS\endcsname}\spacefactor=\sf}}

\def\TITLE#1\endTITLE                                           
{{\Notes@AmS=0 \let\\=\cr\let\footnote=\sfootnote@AmS
   \setbox0=\vbox{\tabskip\centering
  \halign to \hsize{\tenpoint\bf\ignorespaces##\unskip\cr#1\cr}}
 \Notes@AmS=0   \let\footnote=\Sfootnote@AmS
   \global\setbox\titlebox@AmS=\vbox{\tabskip0pt\filhss@AmS
\halign to \hsize{\tenpoint\bf\hfil
 \uctext@AmS{\ignorespaces##\unskip}\hfil\cr
          #1\cr}}}\def\filhss@AmS{plus 1000pt}}

\def\AUTHOR#1\endAUTHOR{\gdef\isauthor@AmS{T}{\Notes@AmS=0 \let\\=\cr
   \let\footnote=\sfootnote@AmS
 \setbox0 =\vbox{\tabskip\centering\halign to \hsize{\tenpoint\smc
   \ignorespaces##\unskip\cr#1\cr}}\Notes@AmS=0
   \let\footnote=\Sfootnote@AmS
  \global\setbox\authorbox@AmS=\vbox{\tabskip0pt\filhss@AmS\halign
  to \hsize{\tenpoint\smc\hfil\uppercase{\ignorespaces
     ##\unskip}\hfil\cr#1\cr}}}\def\filhss@AmS{plus 1000pt}}


\newcount\language@AmS                                            
\language@AmS=0
\def\german{\language@AmS=1}

\def\abstractword@AmS{\ifcase \language@AmS ABSTRACT\or ZUSAMMENFASSUNG\fi}
\def\logoword@AmS{\ifcase \language@AmS Typeset by \fi}
\def\subjclassword@AmS{\ifcase \language@AmS
     1980 Mathematics subject classifications \fi}
\def\keywordsword@AmS{\ifcase \language@AmS  Keywords and phrases\fi}
\def\Referenceword@AmS{\ifcase \language@AmS References\fi}

\def\isaffil@AmS{F}
\newbox\affilbox@AmS
\def\affil{\gdef\isaffil@AmS{T}\bgroup\let\\=\cr
   \global\setbox\affilbox@AmS
     =\vbox\bgroup\tabskip0pt\filhss@AmS
 \halign to \hsize\bgroup\tenpoint\hfil\ignorespaces##\unskip\hfil\cr}

\def\endaffil{\cr\egroup\egroup\egroup\def\filhss@AmS{plus 1000pt}}

\newcount\addresscount@AmS                                         
\addresscount@AmS=0

\def\address#1{\global\advance\addresscount@AmS by 1
  \expandafter\gdef\csname address\romannumeral\addresscount@AmS\endcsname
   {\noindent\eightpoint\ignorespaces#1\par}}

\def\isdate@AmS{F}                                                 
\def\date#1{\gdef\isdate@AmS{T}\gdef\date@AmS{\tenpoint\ignorespaces#1\unskip}}

\def\isthanks@AmS{F}
\def\thanks#1{\gdef\isthanks@AmS{T}\gdef\thanks@AmS{\eightpoint\ignorespaces
       #1\unskip}}

\def\keywords@AmS{}                                                
\def\keywords#1{\def\keywords@AmS{\noindent \eightpoint \it
\keywordsword@AmS .\enspace \rm\ignorespaces#1\par}}

\def\subjclass@AmS{}
\def\subjclass#1{\def\subjclass@AmS{\noindent \eightpoint\it
\subjclassword@AmS
(Amer.\ Math.\ Soc.)\/\rm: \ignorespaces#1\par}}

\def\isabstract@AmS{F}
\long\def\abstract#1{\gdef\isabstract@AmS{T}\long\gdef\abstract@AmS
   {\eightpoint \abstractword@AmS\period\ignorespaces #1\par}}        


\def\pretitle{}
\def\preauthor{}
\def\preaffil{}
\def\predate{}
\def\preabstract{}
\def\prepaper{}


\def\endtopmatter{\if F\topmatter@AmS \errmessage{AmS-TeX warning: You
    forgot the \string\topmatter, but I forgive you.}\fi
\hrule height 0pt \vskip -\topskip                                   
   \pretitle
   \vskip 24pt plus 12pt minus 12pt
   \unvbox\titlebox@AmS                                              
   \preauthor
   \if T\isauthor@AmS \vskip 12pt plus 6pt minus 3pt
       \unvbox\authorbox@AmS \else\fi
    \preaffil
   \if T\isaffil@AmS \vskip 10pt plus 5pt minus 2pt
       \unvbox\affilbox@AmS\else\fi
  \predate
   \if T\isdate@AmS \vskip 6pt plus 2pt minus 1pt
  \hbox to \hsize{\hfil\date@AmS\hfil}\else\fi
    \preabstract
\if T\isthanks@AmS
  \makefootnote@AmS{}{\thanks@AmS}\else\fi
   \if T\isabstract@AmS \vskip 15pt plus 12pt minus 12pt
 {\leftskip=16pt\rightskip=16pt
  \noindent \abstract@AmS}\else\fi
   \prepaper
     \vskip 18pt plus 12pt minus 6pt \tenpoint}


\newcount\addresnum@AmS                                               
\def\enddocument{\penalty10000 \sfcode`\.3000\vskip 12pt minus 6pt  
\keywords@AmS                                                         
\subjclass@AmS
\addresnum@AmS=0
  \loop\ifnum\addresnum@AmS<\addresscount@AmS\advance\addresnum@AmS by 1
  \csname address\romannumeral\addresnum@AmS\endcsname\repeat
\vfill\supereject\end}


\newbox\headingbox@AmS
\outer\def\heading{\medbreak\bgroup\let\\=\cr
\global\setbox\headingbox@AmS=\vbox\bgroup\tabskip0pt\filhss@AmS      
   \halign to \hsize\bgroup\tenpoint\smc\hfil\ignorespaces
            ##\unskip\hfil\cr}

\def\endheading{\cr\egroup\egroup\egroup\unvbox\headingbox@AmS
    \penalty10000 \def\filhss@AmS{plus 1000pt}\medskip}


\outer\def\proclaim#1{\xdef\curfont@AmS{\the\font}\medbreak        
  \noindent\smc\ignorespaces#1\unskip.\enspace\sl\ignorespaces}

\outer\def\proclaimnp#1{\xdef\curfont@AmS{\the\font}\medbreak      
  \noindent\smc\ignorespaces#1\enspace\sl\ignorespaces}

\def\finishproclaim{\par\curfont@AmS\ifdim\lastskip<\medskipamount 
 \removelastskip \penalty 55\medskip\fi}

\outer\def\demo#1{\par\ifdim\lastskip<\smallskipamount
  \removelastskip\smallskip\fi\noindent{\smc\ignorespaces#1\unskip:}\enspace
     \ignorespaces}

\outer\def\demonp#1{\ifdim\lastskip<\smallskipamount
  \removelastskip\smallskip\fi\noindent{\smc#1}\enspace\ignorespaces}

\newif\ifrunin@AmS                                                    
\runin@AmSfalse
\def\runin{\runin@AmStrue}
\def\conditions{\def\\##1:{\par\noindent                              
   \hbox to 1.5\parindent{\hss\rm\ignorespaces##1\unskip}%
      \hskip .5\parindent \hangafter1\hangindent2\parindent\ignorespaces}%
    \def\firstcon@AmS##1:{\ifrunin@AmS
     {\rm\ignorespaces##1\unskip}\ \ignorespaces
  \else\par\ifdim\lastskip<\smallskipamount\removelastskip\penalty55
     \smallskip\fi
     \\##1:\fi}\firstcon@AmS}
\def\endconditions{\par\smallbreak\runin@AmSfalse}                    


\def\refto#1{\in@AmS,{#1}\if T\cresult@AmS\refto@AmS#1\end@AmS\else   
    [{\bf#1}]\fi}
\def\refto@AmS#1,#2\end@AmS{[{\bf#1},#2]}

\def\Refs{\bigbreak\hbox to \hsize{\hfil\tenpoint
    \smc \Referenceword@AmS\hfil}\penalty 10000
      \bigskip\eightpoint\sfcode`.=1000 }                             

\newbox\nobox@AmS        \newbox\keybox@AmS        \newbox\bybox@AmS  
\newbox\bysamebox@AmS    \newbox\paperbox@AmS      \newbox\paperinfobox@AmS
\newbox\jourbox@AmS      \newbox\volbox@AmS        \newbox\issuebox@AmS
\newbox\yrbox@AmS        \newbox\pagesbox@AmS      \newbox\bookbox@AmS
\newbox\bookinfobox@AmS  \newbox\publbox@AmS       \newbox\publaddrbox@AmS
\newbox\finalinfobox@AmS

\def\refset@AmS#1{\expandafter\gdef\csname is\expandafter\eat@AmS     
  \string#1@AmS\endcsname{F}\expandafter
  \setbox\csname \expandafter\eat@AmS\string#1box@AmS\endcsname=\null}

\def\ref@AmS{\refset@AmS\no \refset@AmS\key \refset@AmS\by            
\gdef\isbysame@AmS{F}
 \refset@AmS\paper
  \refset@AmS\paperinfo \refset@AmS\jour \refset@AmS\vol
  \refset@AmS\issue \refset@AmS\yr
  \gdef\istoappear@AmS{F}
  \refset@AmS\pages
  \gdef\ispage@AmS{F}
  \refset@AmS\book
  \gdef\isinbook@AmS{F}
  \refset@AmS\bookinfo \refset@AmS\publ
  \refset@AmS\publaddr \refset@AmS\finalinfo \bgroup
     \ignorespaces}                                                   

\def\ref{\noindent\hangindent 20pt \hangafter 1 \def\refi@AmS{T}
  \def\refl@AmS{F}\def\\{\egroup\endref@AmS\gdef\refi@AmS{F}\ref@AmS}\ref@AmS}

\def\refdef@AmS#1#2{\def#1{\egroup\expandafter                        
  \gdef\csname is\expandafter\eat@AmS
  \string#1@AmS\endcsname{T}\expandafter\setbox
   \csname \expandafter\eat@AmS\string#1box@AmS\endcsname=\hbox\bgroup#2}}

\refdef@AmS\no{} \refdef@AmS\key{} \refdef@AmS\by{}
\def\bysame{\egroup\gdef\isbysame@AmS{T}\bgroup}                    
\refdef@AmS\paper\it
\refdef@AmS\paperinfo{} \refdef@AmS\jour{} \refdef@AmS\vol\bf
\refdef@AmS\issue{} \refdef@AmS\yr{}
\def\toappear{\egroup\gdef\istoappear@AmS{T}\bgroup}                
\refdef@AmS\pages{}
\def\page{\egroup\gdef\ispage@AmS{T}\setbox
                 \pagesbox@AmS=\hbox\bgroup}                        
\refdef@AmS\book{}
\def\inbook{\egroup\gdef\isinbook@AmS{T}\setbox
                               \bookbox@AmS=\hbox\bgroup}           
\refdef@AmS\bookinfo{} \refdef@AmS\publ{}
\refdef@AmS\publaddr{}
\refdef@AmS\finalinfo{}

\def\setpunct@AmS{\def\prepunct@AmS{, }}                              
\def\ppunbox@AmS#1{\prepunct@AmS\unhbox#1\unskip}                     

\def\endref@AmS{\def\prepunct@AmS{}
\if T\refi@AmS                                                      
  \if F\isno@AmS\hbox to 10pt{}\else                                
     \hbox to 20pt{\hss\unhbox\nobox@AmS\unskip. }\fi               
  \if T\iskey@AmS \unhbox\keybox@AmS\unskip\ \fi                    
  \if T\isby@AmS  \hbox{\unhcopy\bybox@AmS\unskip}\setpunct@AmS     
         \setbox\bysamebox@AmS=\hbox{\unhcopy\bybox@AmS\unskip}\fi  
  \if T\isbysame@AmS                                                
   \hbox to \wd\bysamebox@AmS{\leaders\hrule\hfill}\setpunct@AmS\fi
 \fi                                                                
  \if T\ispaper@AmS\ppunbox@AmS\paperbox@AmS\setpunct@AmS\fi          
  \if T\ispaperinfo@AmS\ppunbox@AmS\paperinfobox@AmS\setpunct@AmS\fi  
  \if T\isjour@AmS\ppunbox@AmS\jourbox@AmS\setpunct@AmS               
     \if T\isvol@AmS \ \unhbox\volbox@AmS\unskip\setpunct@AmS\fi    
     \if T\isissue@AmS \ \unhbox\issuebox@AmS\unskip\setpunct@AmS\fi
     \if T\isyr@AmS \ (\unhbox\yrbox@AmS\unskip)\setpunct@AmS\fi    
     \if T\istoappear@AmS \ (to appear)\setpunct@AmS\fi             
     \if T\ispages@AmS \ppunbox@AmS\pagesbox@AmS\setpunct@AmS\fi    
     \if T\ispage@AmS                                               
           \prepunct@AmS p.\ \unhbox\pagesbox@AmS\unskip\setpunct@AmS\fi
     \fi                                                            
  \if T\isbook@AmS \prepunct@AmS                                      
                     ``\unhbox\bookbox@AmS\unskip''\setpunct@AmS\fi
  \if T\isinbook@AmS \prepunct@AmS                                    
    \unskip\ in ``\unhbox\bookbox@AmS\unskip''\setpunct@AmS
       \gdef\isbook@AmS{T}\fi
  \if T\isbookinfo@AmS \ppunbox@AmS\bookinfobox@AmS\setpunct@AmS\fi   
  \if T\ispubl@AmS \ppunbox@AmS\publbox@AmS\setpunct@AmS\fi           
  \if T\ispubladdr@AmS \ppunbox@AmS\publaddrbox@AmS\setpunct@AmS\fi   
 \if T\isbook@AmS                                                     
  \if T\isyr@AmS \prepunct@AmS \unhbox\yrbox@AmS\unskip             
              \setpunct@AmS\fi
  \if T\istoappear@AmS \ (to appear)\setpunct@AmS\fi                
  \if T\ispages@AmS                                                 
    \prepunct@AmS pp.\ \unhbox\pagesbox@AmS\unskip\setpunct@AmS\fi
  \if T\ispage@AmS                                                  
    \prepunct@AmS p.\ \unhbox\pagesbox@AmS\unskip\setpunct@AmS\fi
 \fi
  \if T\isfinalinfo@AmS \period\unhbox\finalinfobox@AmS\else          
    \if T\refl@AmS .\else ; \fi\fi}

\def\endref{\egroup\gdef\refl@AmS{T}\endref@AmS\par}


\newif\ifguides@AmS
\guides@AmSfalse
\def\guidelines{\guides@AmStrue}
\def\noguidelines{\guides@AmSfalse}
\def\guidelinegap#1{\def\gwidth@AmS{#1}}
\def\gwidth@AmS{24pt}

\newif\iflogo@AmS
\def\nologo{\logo@AmSfalse}
\logo@AmStrue

\def\output@AmS{\ifnum\count0=1
 \shipout\vbox{\ifguides@AmS\hrule width \hsize \vskip\gwidth@AmS \fi
   \vbox to \vsize{\boxmaxdepth=\maxdepth\pagecontents}\baselineskip2pc
\iflogo@AmS \hbox to \hsize{\hfil\eightpoint \logoword@AmS\AmSTeX}\fi
     \ifguides@AmS \vskip\gwidth@AmS
\hrule width \hsize\fi}\vsize 44pc\else
   \shipout\vbox{\ifguides@AmS \hrule width \hsize \vskip\gwidth@AmS\fi
   \vbox to \vsize{\boxmaxdepth=\maxdepth\pagecontents}\baselineskip2pc\hbox to
  \hsize{\hfil \tenpoint\number\count0\hfil}\ifguides@AmS
    \vskip\gwidth@AmS\hrule width \hsize\fi}\fi\global\advance\count0 by 1
  \global\footmarkcount@AmS=0 \global\foottextcount@AmS=0
 \ifnum\outputpenalty>-20000 \else\dosupereject\fi}




\tenpoint

\catcode`\@=13

\output={\plainoutput}

\magnification=\magstep0
\baselineskip=10pt
\hoffset=-0.75truecm
\voffset=0.0truecm
\vsize=23.5truecm
\hsize=18.0truecm
\parskip=0.2cm
\parindent=1cm

\hfuzz=23pt

\def \bigbreak  {\goodbreak\bigskip}
\def \medbreak  {\goodbreak\medskip}
\def \smallbreak{\goodbreak\smallskip}
\def \header#1{\goodbreak\bigskip\centerline{\bf #1}\medskip\nobreak}
\def \subheader#1{\goodbreak\medskip\par\noindent{\bf #1}\smallskip\nobreak}

%
%
\def\pmb#1{\setbox0=\hbox{#1}%
  \kern-.025em\copy0\kern-\wd0
  \kern.05em\copy0\kern-\wd0
  \kern-.025em\raise.0433em\box0 }
%
%
%
\def\timedate{ {\tt
\count215=\time \divide\count215 by60  \number\count215
\multiply\count215 by-60 \advance \count215 by\time :\number\count215 \space
\number\day\space
\ifcase\month\or January\or February\or March\or April\or May\or June\or July
\or August\or September\or October\or November\or December\fi\space\number\year
}}
%
\def \etal {{\it et al.} }

%
%
\def\captpar{\dimen0=\hsize
             \advance\dimen0 by -1.0truecm
             \par\parshape 1 0.5truecm \dimen0 \noindent}
%
%
%
%
%
%

\def\s {\scriptscriptstyle}

\def\sqr#1#2{{\vcenter{\hrule height.#2pt
              \hbox{\vrule width.#2pt height#1pt \kern#1pt \vrule width.#2pt}
              \hrule height.#2pt}}}

\def\mathrelfun#1#2{\lower3.6pt\vbox{\baselineskip0pt\lineskip.9pt
  \ialign{$\mathsurround=0pt#1\hfil##\hfil$\crcr#2\crcr\sim\crcr}}}
\def\simlt{\mathrel{\mathpalette\mathrelfun <}}
\def\simgt{\mathrel{\mathpalette\mathrelfun >}}
\def\simpropto{\mathrel{\mathpalette\mathrelfun \propto}}
\def\ln {{\rm ln}}

\def\Si  {{\rm Si}}

\def\rmg {{\rm g}}

\def\rmp {{\rm p}}

\def\rms {{\rm s}}

\def\rmA {{\rm A}}

\def\rmN {{\rm N}}

\def\rmS {{\rm S}}

\def\bfk {{\bf k}}

\def\bfm {{\bf m}}
\def\bfn {{\bf n}}

\def\bfx {{\bf x}}

\def\calH {{\cal H}}

\def\calO {{\cal O}}

\def\hatbfm  {{\hat\bfm}}
\def\hatbfn  {{\hat\bfn}}

\def\Mpc {{\rm Mpc}}

\def\eV  {{\rm \hbox{e\kern-0.14em V}}}
\def\keV {{\rm \hbox{ke\kern-0.14em V}}}
\def\MeV {{\rm \hbox{Me\kern-0.14em V}}}
\def\GeV {{\rm \hbox{Ge\kern-0.14em V}}}

\def\etaobs{{  \eta_{\rm obs}}}

\def\bfxobs{{\bf x}_{\rm obs}}

\def\zmax  {{     z_{\rm max}}}
\def\rmax  {{     r_{\rm max}}}
\def\xmax  {{     x_{\rm max}}}

\def\rHZ{{r_{\s\rm HZ}}}


\font\FermiPPTfont=cmssbx10 scaled 1440
\font\FermiSmallfont=cmssq8 scaled 1200

\def\FNALpptheadnologo#1#2{
\null \vskip -1truein
\centerline{\hbox to 7.5truein {
\hskip 1.5cm
\vbox to 1in{\vfill
             \hbox{{\FermiPPTfont Fermi National Accelerator Laboratory}}
             \vfill}
\hfill
\vbox to 1in {\vfill \FermiSmallfont
              \hbox{#1}
              \hbox{#2}
              \vfill}
}}}

\FNALpptheadnologo{FERMILAB-Pub-96/328-A}{September 1996}

\topmatter
\title
Weak Lensing On the Celestial Sphere
\endtitle
\author
{Albert Stebbins}
\endauthor
\affil
NASA/Fermilab Astrophysics Center, 
FNAL, Box 500, Batavia, Illinois 60510, USA \\
\endaffil
\abstract{\ninepoint
This paper details a description of the pattern of galaxy image distortion over
the entire sky caused by the gravitational lensing which is the result of
large scale inhomogeneities in our universe.  We present a tensor spherical
harmonic formalism to describe this pattern, giving many useful formulae.  This
is applied to density inhomogeneities, where we compute the angular power
spectrum of the shear pattern, as well as the noise properties due to finite
galaxy sampling and cosmic variance.  We show that a detectable level of shear
is present for very nearby galaxies, $z\simlt0.2$.  For such a shallow sample
much of the largest signal-to-noise comes from very large angular scales,
$\theta\simgt10^\circ$, although it is in the form of very small shear at a
level $\simlt10^{-3}$.
}
\endtopmatter

\header{1. Introduction}

	In recent years gravitational lenses have provided a set of extremely
useful tools for understanding the universe around us.  One of these tools,
sometimes called the orientation correlation function or OCF, is a method
whereby one searches for alignments in the orientations of galaxies on the sky
(Tyson, Valdes, and Wenk 1990).  Such alignments will be caused by the
deflection of light by the gravitational field of mass concentrations in front
of the galaxies one is observing.  In the weak lensing approximation the OCF is
determined by the {\it shear} of the image deformation caused by this bending
of light.  Large format CCD's have made such observations of weak lensing shear
possible and there have been numerous studies of the mass concentration in
clusters of galaxies with this technique (Tyson, Valdes, and Wenk 1990,
Bonnett, Meiler \& Fort 1994, Fahlman \etal 1994, Smail, Ellis \& Fitchett
1994, Smail \& Dickinson 1995, Tyson and Fisher 1995, Squires \etal 1996a\&b,
Luppino \& Kaiser 1996).  It is in the direction of clusters of galaxies at
moderate redshift where the shear is liable to be greatest and the size of the
central parts of such clusters is a good fit with the typical field of view of
many CCD cameras, i.e. about $5'$.  Away from the direction of clusters of
galaxies the shear is liable to be much smaller.  In principle one should be
able to detect very small shears by looking for alignments among a greater
number of background galaxies. One might obtain these greater numbers either by
taking deeper images, which has the advantage that shear will increase with
depth, or by looking at larger areas on the sky.  With the advent of very large
CCD mosaic cameras with a large field of view one can expect much wider area
surveys looking for weak lensing shear.  A limiting factor in such wide surveys
is the small amplitude of the shear one is trying to measure, and further study
is required to see just how small one can reduce systematic errors when looking
for galaxy alignments.

	One of the most interesting applications of weak lensing which is to
map the mass distribution as close to the local neighborhood as possible. This
will allow a comparison of the galaxy distribution and the mass distribution
which should help us to understand galaxy formation and biasing.  Gould and
Villumsen (1994) have pointed out that the Sloan Digital Sky Survey (see Kent
1994 for a description of the ``SDSS'') which is imaging one quarter of the sky
should be able to measure the mass distribution around the Coma cluster.
Whether this goal is achievable depends on the level of non-correctable
systematic errors, however even if present imaging surveys are not successful
one can expect that very large area weak lensing surveys will be achieved
sometime in the future.

	Predictions of the angular distribution of shear and amplification on
the sky has been made in the context of the small areas on the sky where the
small angle approximation is utilized (Blandford \etal 1991, Miralda-Escud\'e
1991, Kaiser 1992, Kaiser and Squires 1993).  This is quite natural given that
most weak lensing observations would be limited to extremely small patches on
the sky.  However looking ahead to a time when the area on the sky surveyed
becomes large, such as in the SDSS, one will need to go beyond the small angle
approximation in order to characterize the weak lensing pattern one observes.
This is not to say that the lensing deflection angles will be large, they will
not, but rather that when the area on the sky becomes large one needs to take
into account the ``curvature'' of the celestial sphere when computing things
like shear-shear correlation functions.  Since the shear is not a scalar
quantity, but rather a rank-2 tensor, one can not simply use the the scalar
spherical harmonic expansion.  In this paper is presented a tensor spherical
harmonic expansion which the author feels is rather well suited to describing
the shear pattern.  Most published expositions of tensor spherical harmonics
were motivated by describing 3-d tensor gravitational fields and these
expositions are complicated by the additional formalism needed to describe the
3rd dimension (see Thorne (1980) for a review).  The formalism described here
is rather simpler as it is restricted to symmetric traceless tensor fields on
the 2-d sphere.  Although developed independently this formalism is very close
to that of Zerilli (1970).

	The paper is arranged as follows.  In \S2 we describe the decomposition
of the shear pattern into two geometrically distinct types in a formalism which
is easily generalizable from the small-angle approximation to the celestial
sphere.  In \S3 is considered the various approximation used when computing
the image distortion in lensing. In \S4 is presented the tensor spherical
harmonic decomposition of the shear, with a large number of useful identities.
In \S5 the tensor harmonic expansion is used to work out the formalism to
describe the shear pattern produced by density perturbations in an
Einstein-deSitter universe.  In \S7 we estimate the accuracy one might expect
to obtain in measurements of the shear due to the ``shot noise'' from the
finite number of galaxies that are available to probe the shear, and to
``cosmic variance'' (finite sampling).  In \S7 the formalism developed is
applied to a phenomenological model of the density perturbations in our
universe and illustrate the type of shear pattern one might expect to find and
how it compares to the shot noise and cosmic variance. In \S8 we summarize the
results.

\header{2. Scalar and Pseudo-Scalar Shear}

	The shear gives the degree to which images of objects we see on the
sky are elongated, after factoring out the intrinsic shape of the actual
objects.  The shear at a point has an amplitude and a direction and may be
described by a symmetric traceless matrix, i.e. 
$$\gamma_{ab}=\gamma\,\left(\matrix\cos2\varphi\ & \sin2\varphi\\
                                   \sin2\varphi\ &-\cos2\varphi\endmatrix
                            \right)
.\eqno(2.1)$$
Here $\gamma$ measures the amplitude and $\varphi$ the direction wrt to some
fiducial position angle on the sky.  Of course the direction is modulo $\pi$
not $2\pi$ since stretching an image in one direction is the same as stretching
it in the opposite direction. Note that rotating the direction of the shear by
$90^\circ$ changes the sign of the the shear, while rotating the shear by
$\pm45^\circ$ produces a shear which is orthogonal to the original, in the
sense that $\gamma^{ab}\gamma_{ab}'=0$.  (The Einstein summation convention is
used here and throughout this paper.) Whether one rotates $45^\circ$ to the
right or $45^\circ$ to the left changes the sign of the shear.  One can
construct any shear matrix by linear combinations of a given shear matrix and
one rotated by $45^\circ$.

	Here we are concerned with shear fields on the sky.  In this section we
consider the small angle approximation where the sky is approximated as a
Euclidean plane.  One may write any shear field on a plane as a Fourier
decomposition, e.g.
$$\gamma_{ab}(\vec{x})=-{1\over2\pi}\int d^2\vec{q}\,
\left[\phi^\oplus (\vec{q})\,\left(f_{,ab}-{1\over2}\nabla^2\delta_{ab}f\right)
     +\phi^\otimes(\vec{q})\,{1\over2}\left( f_{,ac}\epsilon^c{}_b
                                            +f_{,cb}\epsilon^c{}_a
                                                       \right)
      \right] \qquad f({\vec{q}},\vec{x})=e^{i\vec{q}\cdot\vec{x}}
\eqno(2.2)$$
where $\vec{q}$ is the 2-d wavenumber on the (planar) sky and
$$f_{,ab}={\partial^2\over\partial x^a\partial x^b}  \qquad
\epsilon^a{}_b=\left(\matrix 0 & 1 \\
                            -1 & 0 \endmatrix\right)
.\eqno(2.3)$$
The tensor $\epsilon^a{}_b$ is known as the Levi-Civita symbol in 2-d. The
$\gamma^\oplus$ modes have shear directed parallel to $\vec{q}$  while the
$\gamma^\otimes$ modes have shear rotated $45^\circ$ to the right from the
$\vec{q}$ direction.  This is all one needs to construct an arbitrary shear
field.


	Since the $\gamma^\oplus$ modes are just given by derivatives of the
scalar mode function $f({\vec{q}},\vec{x})$ we will refer to this part of the
shear as the {\it scalar shear}.  The $\gamma^\otimes$ modes are a
geometrically distinct component of the shear, given by the 2nd derivative
matrix of $f$ but multiplied by the $\epsilon^c{}_b$ which has the effect of
rotating the scalar shear by $45^\circ$ to the right in a right-handed
coordinate system or $45^\circ$ to the left in a left-handed coordinate system.
A change in the handedness would have the effect of multiplying the
pseudo-scalar shear tensor by $-1$.  This sign difference depending on the
handedness of the coordinate system means that the component of the shear
transforms as a pseudo-scalar rather than a scalar, and it is called the {\it
pseudo-scalar} component of the shear.  The two types of shear pattern are
illustrated in fig~1.  One can represent any pattern of shear by a sum of
scalar and pseudo-scalar shear

	The decomposition we have just described is similar to the
decomposition of a vector field into it's vortical and non-vortical parts.
Note however that in 3-d for each Fourier mode their are two linearly
independent vortical components and one must choose a azimuthal angle about the
wavenumber to specify one.  For this reason one cannot write these vortical
modes in terms of derivatives of the scalar modes $e^{i\bfk\cdot\bfx}$ plus a
handedness as this will not specify the azimuthal angle.  In 2-d one may
express arbitrary vector and tensor fields in terms of scalar functions
Levi-Civita symbols, as we have done above.  There is not the distinction
between scalar, vector, and tensor components of fields higher dimensions.

	In the usual treatment of gravitational (or non-gravitational) lensing,
one thinks of the lensing as being a displacement field on the sky, i.e. in a
given direction on the sky what one sees is at a true position in space which
is the apparent position displaced by some amount, $\vec{\Delta}$.  This is
what we will later call the {\sl displacement approximation}.  For arbitrary
gravitational perturbations it is problematic to define what the apparent
position is a hence what the displacement field is.  In the next section it is
argued that for weak fields it is an excellent approximation to treat the
lensing as a displacement field on an unperturbed space-time and this is the
approximation we will use here.  Let us denote the displacement from apparent
to true position on the planar sky by $\vec{\Delta}$. The $2\times2$
deformation matrix is given by
$$\psi_{ab}=\delta_{ab}+\Delta_{a,b}=
(1-\kappa)\,\delta_{ab}-\gamma_{ab}+\omega\,\epsilon_{ab}
\eqno(2.4)$$
which we have decomposed into term which gives the trace, a term which gives
the symmetric traceless part, and a term which gives the anti-symmetric part.
The trace gives the {\it expansion}, $\kappa$, the traceless symmetric part
give the shear, $\gamma_{ab}$, and the anti-symmetric part gives the {\it
rotation}, $\omega$.  The image amplification is
$1/((1-\kappa)^2-\gamma^2+\omega^2)$ or approximately $1+2\kappa$ in the weak
lensing limit.  An arbitrary displacement field can also be written as a
Fourier integral of scalar and pseudo-scalar components:
$$\Delta_a(\vec{x})=-{1\over2\pi}\int d^2\vec{q}\,
\left[\phi^\oplus (\vec{q})\,f_{,a}+\phi^\otimes(\vec{q})\,f_{,c}\epsilon^c{}_a
      \right] \qquad f({\vec{q}},\vec{x})=e^{i\vec{q}\cdot\vec{x}}
.\eqno(2.5)$$
Note that $f_{,c}\epsilon^c{}_a$ rotates the gradient of $f$ by $90^\circ$.
When the deformation is given by a displacement field on a Euclidean plane we
find the following relations
$$\nabla^2\kappa=+\gamma^{ab}{}_{,ab}                \qquad
  \nabla^2\omega=-\gamma^{ab}{}_{,ac}\epsilon^c{}_b
.\eqno(2.6)$$
Since $\nabla^2$ is invertible the scalar shear is given by the expansion,
$\kappa$, while the pseudo-scalar shear by the rotation, $\omega$. A scalar and
pseudo-scalar displacement field is illustrated in figure 1.

	It is unlikely that anything other than density inhomogeneities
contribute significantly to the shear in our universe.  These density
inhomogeneities produce scalar metric perturbations.  One shouldn't confuse the
``scalar''-ness of density perturbations which has to do with transformation
properties in 3 spatial dimensions, with the ``scalar''-ness of the shear which
has to do with transformation properties on the 2-dimensional sky. Nevertheless
it is true that 3-d scalar perturbations will only produce 2-d scalar shear, at
least in the weak lensing limit. In contrast tensor and vector metric
perturbations will contribute to both scalar and pseudo-scalar shear, however
we expect this to be very small. Thus we really do not expect there to be any
significant pseudo-scalar shear.  We have chosen to include pseudo-scalar shear
in our analysis for a variety of reasons.  Most importantly because the
pseudo-scalar shear is part of what one measures when one looks at galaxy
alignments, and one should not leave it out of ones analysis.  Also, one never
knows, maybe the pseudo-scalar shear field is not negligible, one should
measure it and see!
	
	Kaiser (1992) noted that the redundant information in the shear field
could provide a useful check of ones observation.  One can determine the
pseudo-scalar $\omega$ in the same way one determines the scalar $\kappa$ 
if one first rotates the galaxy position angles by $45^\circ$, and then one
should check that the derived $\omega$ is is consistent with zero.  (Kaiser
\etal 1994; A. Tyson private communications; Stebbins, McKay, and Frieman
1996). Just such a procedure has been implemented in Luppino and Kaiser (1996)
where they compare $\gamma^{ab}{}_{,ab}$ to $\gamma^{ab}{}_{,ac}\epsilon^c{}_b$
showing that the latter is indeed much smaller than the former in the field of
the cluster of galaxies ms1054-03, and consistent with zero.

\header{3. Approximations}

	In this paper we make predictions of extremely small image deformation.
These are calculated using a number of approximations and one must be careful
that the approximations are accurate enough so that these predictions are
accurate. In other words one must be sure that there are not other small but
significant contribution to the lensing which are not included.  Many
approximations become more accurate as the image deformation becomes smaller,
while others do not. In this section we identify most of the approximations
used here. None of these approximations are new to this work, and not all
studies of gravitational lensing make use of all of the approximations used
here.  Below we will use the geometric optics approximation in weak field
gravity for perturbations from Friedman-Robertson-Walker space-times.  These
are clearly excellent approximations for cosmology.  It is generally assumed
that we are in the ``weak lensing'' limit, namely that the deformation tensor
of eqs~(2.4) \& (4.1) are close to the identity matrix, although many of the
formulae are true even for strong lensing.  For isolated regions around galaxy
clusters, galaxies, and stars weak lensing will not be a valid approximation,
and generally as one goes to smaller angular scales the stronger the lensing
becomes.  To what extent effects from small scale strong lensing creep into
larger scales is a subject which needs further study.

	When computing image distortion we integrate the perturbation to a beam
of photons along the unperturbed path of the photons. To do otherwise would be
to include terms which are formally 2nd order in the metric perturbation so one
can argue that this is part of the weak lensing approximation.  Even when the
image distortion is large this may also be a good approximation if the
statistical properties of matter field which one traverses on the perturbed
trajectory are quite similar to that traversed on the unperturbed trajectory.
However the argument that density perturbations only produce scalar patterns of
shear rests largely on the assumption that it is valid to use the unperturbed
trajectory.  In general for strong lensing one can expect density perturbations
to produce both scalar and pseudo-scalar shear.  An exception occurs when the
mass which produce the strong lensing is localized to a narrow range of
distances.  In this case one can still use the unperturbed trajectory to
compute the total deflection accurately.  Thus in studies of lensing around
individual galaxies or galaxy clusters one can assume that the shear pattern is
purely scalar since it is unlikely that two large mass concentrations will
line-up sufficiently to contributed significantly to the lensing.  However on
small enough angular scales the strong lensing will be due to a variety of
objects along the line-of-sight and the shear pattern will contain both scalar
and pseudo-scalar components.  Again further study should be done on the
contamination by small scale strong lensing of larger scale weak lensing.

	One approximation which is important for this paper is what will be
called the {\it deflection approximation}.  Namely that one can think of the
effect of gravitational lensing as a deflection, where the light has been
deflected from a ``straight path'' in an underlying flat geometry. While this
is a perfectly valid description for lenses made of refractive materials in
flat space, it is only an approximation for gravitational lensing.  This is
because there really is no underlying flat geometry, after all the light is
deflected because the space is not flat.  Below we will argue that in the cases
of interest the space is flat enough so that this deflection approximation is a
good one.  Even though in this paper we consider very weak lensing, the image
distortions we consider are still much weaker than the intrinsic distortion
pattern caused by the curvature of spacetime. Note that this is not an
intrinsic property of the weak lensing approximation, but rather has more to do
with how the amplitude of metric perturbations in our universe vary with
scale. If we had very large perturbations entering the horizon today much of
the image distortion would be a result of the global geometry rather than the
deflection.

	Let us try to make these ideas more precise.  In \S5 of this paper we
consider the lensing due to density perturbations.  The deflection as a
function of distance from the observer is computed using the metric evaluated
on the unperturbed path, i.e. by twice integrating the ``gravitational
acceleration'' of the photon perpendicular to the unperturbed path.  With this
deflection we compute the intersection of both the unperturbed trajectory and
the ``deflected trajectory'' with a surface defined by a fixed coordinate
distance from the observer, i.e.  $r=\sqrt{x^2+y^2+z^2}$, which we will call
the ``source sphere''.  We use the the Jacobian of the mapping of the Newtonian
coordinates from the deflected position on the source sphere to the undeflected
position as the deformation tensor in eqs~(2.4) or (4.1).  This procedure has
some obvious failings.  Firstly if the photons do not intersect the source
sphere at right angles then even for photons which are not focused or sheared
the Jacobian matrix will contain an eigenvalue of $1/\cos\theta$ where $\theta$
is angle between the normal to the surface and the direction of the perturbed
photon trajectory.  Thus even if there is no true image distortion one would
infer a de-amplification and shear.  This error is not large, in the sense that
the leading error is $\propto\theta^2$, so is an insignificant error in the
sense of weak lensing.  In cosmology the deflections are rarely greater than
$10^{-4}$ radians even when the lensing is strong.  A more worrisome error
comes from the fact that the source sphere defined above is not truly spherical
in geometry. Scalar perturbations in Newtonian gauge have an isotropic metric
so the intrinsic geometry of the source sphere does not lead to any spurious
shear.  However a small coordinate patch spanning a coordinate area $A$ on the
source will actually spans a physical area $(1-2\Phi/c^2)A$ where $\Phi$ is the
Newtonian gravitational potential.  Unless one corrects for this one will add a
spurious contribution to the expansion, $\kappa$, of $2\Phi/c^2$.  Since this
contribution to $\kappa$ is linear in the metric perturbation it does not
become arbitrarily smaller than the true signal in the weak lensing limit.
However, for our own universe we know that $\Phi\simlt10^{-5}c^2$ while, as we
will see below, $\kappa\simgt10^{-4}$.  Thus by ignoring the intrinsic geometry
of the source sphere we only make a small error.  As mentioned above, the
reason for this fortuitous numerology has to do with the shape of the power
spectrum.  No such fortuitous numerology applies to the case of gravitational
lensing from gravitational radiation, so one cannot use the deflection
approximation in this case.  We do not consider lensing from gravitational
radiation in this paper.  We will ignore the intrinsic geometry of the source
sphere in the rest of the paper.  While it is not terribly cumbersome to
include this term, it would invalidate the simple and accurate relationship
between the different components of the image distortion, $\kappa$, $\omega$,
and $\gamma_{ab}$ as expressed in eq~(2.6), or implicitly in eq~(4.10).

	A more general treatment of image distortion, can be gotten from the
optical scalar equation (e.g. Hawking \& Ellis 1973; Schneider, Ehlers, \&
Falco 1992) whereby one measure the distortion of a bundle of light rays with
respect to an orthonormal basis parallel transported along with the light rays
from the observer.  This approach makes no approximation other than assuming
the validity of geometric optics.  It would be too cumbersome for our purposes
to use this approach on unperturbed trajectories, i.e. to use ray-tracing.  One
could approximate the optical scalar equation by implementing the same
procedure along an unperturbed trajectory. One property of the optical scalar
equation which is something of a drawback is that since the parallel
transported basis vectors will rotate with the photons, this method does not
yield any information on image rotation.  While in the case of strong
gravitational fields it is difficult to make a sensible definition of image
rotation, in the case of weak fields where the image distortion is generally
much larger than the metric perturbation there is a real sense in which images
may be rotated and it would be useful to include this effect in ones
computations.

	In \S5 and afterwards a set of assumptions and simplifications are
implemented.  We do only consider a flat matter-dominated Einstein-deSitter
cosmology, although one could clearly generalize all of formulae to open or
closed cosmologies with or without a cosmological constant.  The formulae used
in \S5\&\S7 assume that the perturbations grow according to linear theory.
This is a good approximation on large scales $\simgt10\,h^{-1}$Mpc but we when
we consider the shear at shallow depths we do venture into a regime where
linear theory is not appropriate.  In fact the model power spectrum in \S7 is
no-linear on scales of interest.  One could try to correct this by having the
power spectrum vary as one looks back in time, i.e. with radius.  Since the
results presented here are meant to be illustrative this approach is not
warranted.  In any case since the emphasis is of this paper is on shallow
surveys these sorts of corrections are liable to be small.

\header{4. Decomposition of Shear on the Celestial Sphere}

	Even though the coherence in the shear field falls off with separation
there will be some small coherence even over very large angles on the sky.  To
understand this large angle coherence one must take into account that the sky
is a sphere and not a plane.  Here we consider the weak lensing on the
celestial sphere, i.e. the sphere of the sky.  For the moment we consider the
image deformation at a fixed distance but later generalize the analysis to
include the fact that the galaxies whose images are distorted span a range of
distances.  Thus we have the a mapping from the direction in which you are
looking, $\hatbfn$, to the direction you are looking at, $\hatbfm$, at a given
distance, $r$.  The effect of the lensing is to displace the photon trajectory
at a given distance by the 3-d vector $\Delta=\hatbfm-\hatbfn$.  If the angle
by which the light rays are bent are sufficiently small then $\Delta$ and
$\hatbfn$ are nearly perpendicular so that we may approximate $\Delta$ as being
in the tangent space of the the direction sphere $\hatbfn$, i.e. $\Delta(r)$ is
a vector field on the sphere.  As in \S2 we may define the deformation tensor
and it's various components
$$g_{ab}+\Delta_{a:b}=\psi_{ab}=(1-\kappa)\,g_{ab}-\gamma_{ab}
+\omega\epsilon_{ab}
\eqno(4.1)$$
where and $\Delta_{a:b}$ refers to a covariant derivative of $\Delta$ wrt
the metric $g_{ab}$ on the sphere. \footnote{All of the indexed quantities are
components of tensors on the sphere. Below we will use the coordinate bases
defined by spherical polar coordinates, $(\theta,\phi)$ on the sphere.
Although the two basis vector are orthogonal they are not orthonormal, and in
particular, tensor components with a $\phi$ index will contain extra factors of
$\sin^2\theta$ when compared to tensors defined with respect to an orthonormal
basis in the same direction.}  Here $\kappa$, $\gamma_{ab}$, and $\omega$
give the expansion, shear, and rotation.  The shear is defined such that
$g{}^{ab}\gamma_{ab}=0$ which fully specifies $\kappa$. From the shear field
one may define two new  quantities
$$\gamma_\rms=\nabla^{-2}\gamma_{ab}{}^{:ab}  \qquad
  \gamma_\rmp=\nabla^{-2}\gamma_{ab}{}^{:bc}\epsilon^a{}_c
.\eqno(4.2)$$
where $\nabla^{-2}$ is the inverse Laplace operator on the sphere, i.e.
$$\nabla^2\phi{}\equiv\phi_{:a}{}^a \qquad
  \nabla^{-2}         \phi_{:a}{}^a=\phi+{\rm constant}
.\eqno(4.3)$$
We must add the arbitrary constant because the $\nabla^{-2}$ is not completely
defined since $\nabla^2$ has a zero eigenvalue which corresponds to the
constant, $l=0$, eigenmode.  Here $l$ gives the eigenvalue of the Laplace
operator which is $-l(l+1)$.  The quantity $\gamma_\rms$ is contributed to only
by the scalar part of the shear, as described in \S2, while the quantity
$\gamma_\rmp$ is contributed only to by the pseudo-scalar part of the shear.
From $\gamma_\rms$ and $\gamma_\rmp$ one can fully reconstruct $\gamma_{ab}$.
However from $\gamma_{ab}$ or from $\gamma_\rms$ and $\gamma_\rmp$, one cannot,
in general, fully reconstruct the displacement vector $\vec{\Delta}$.  As we
shall see, the $l=1$ eigenmodes do effect $\Delta$ but not $\gamma_{ab}$.  The
fact that the $l=0$ eigenmodes also do not contribute to $\gamma_{ab}$ does not
matter since it also does not contribute to $\vec{\Delta}$.  By definition
$\kappa$ and $\omega$ have zero mean and therefore no contribution from the
$l=0$ eigenmode.  It therefore makes since to define $\gamma_\rms$ and
$\gamma_\rmp$ to have zero mean and therefore no contribution for the $l=0$
eigenmode.  By construction they have not contribution from the $l=1$
eigenmodes.

	Above we have used the completely antisymmetric tensor defined by
$$\epsilon_{ab}=\sqrt{g}\left(\matrix 0 & 1 \\
                                     -1 & 0 \endmatrix\right) \qquad
\epsilon^a{}_b={1\over\sqrt{g}}\left(\matrix g_{12}& g_{22}\\
                                            -g_{11}&-g_{21}\endmatrix\right)
\qquad\epsilon^{ab}={1\over\sqrt{g}}\left(\matrix 0 & 1 \\
                                                 -1 & 0 \endmatrix\right)
\qquad g\equiv\Vert g_{ab}\Vert
\eqno(4.4)$$
which commutes with covariant differentiation, i.e. $\epsilon^{ab}{}_{:c}=0$.
When $\epsilon_{ab}$ is contracted with a 1-index vector it has the effect of
rotating the vector by $90^\circ$. Whether the rotation is to the left of
right can be chosen arbitrarily, but whatever choice one makes the sign of
$\epsilon_{ab}$ is multiplied by $-1$ under a parity transformation, and thus
a leftward rotation is transformed to a rightward and vice-versa.  Thus
quantities containing odd numbers of $\epsilon_{ab}$'s are odd under parity and
we refer to them a pseudo-scalar part of the shear.  Note also that the shear
$\epsilon_a{}^2b\gamma_{ab}$ yields another traceless symmetric tensor which if
interpreted as a shear tensor has it's shear rotated by $45^\circ$ wrt that of
the original shear tensors.  Rotating the displacement vectors by $90^\circ$
corresponds to rotating the shear by $45^\circ$.  Performing either of these
rotations will transform a purely scalar patter to a purely pseudo-scalar one,
and vice-versa.

\subheader{Spherical Harmonic Expansion}

	In analogy with the Fourier expansion of \S2 we may expand the
displacement vector on the sphere in terms of a spherical harmonic functions
$Y_{\s(l,m)}(\hatbfn)$.  These functions are defined such that 
$$\nabla^2Y_{\s(l,m)}(\hatbfn)=-l(l+1)Y_{\s(l,m)}(\hatbfn) \qquad
\int d^2\hatbfn\,Y_{\s(l,m)}\,Y^*_{\s(l',m')}=\delta_{ll'}\delta_{mm'}
.\eqno(4.5)$$
Using 2-d spherical geometry this implies the identities
$$\eqalign{
&Y_{\s(l,m)}{}^{:ab}{}_{:ab}=\nabla^2\nabla^2Y_{\s(l,m)}+\nabla^2Y_{\s(l,m)}
                            =l\,(l+1)\,(l(l+1)-1)\,Y_{\s(l,m)}              \cr
&\int d^2\hatbfn\,Y_{\s(l,m):ab}\,Y^*_{\s(l',m')}{}^{:ab}=
                                l\,(l+1)\,(l(l+1)-1)\,\delta_{ll'}\delta_{mm'}
           }.\eqno(4.6)$$
The spherical harmonic analog of eq~(2.5) is
$$\Delta_a=-\sum_{l=1}^\infty\sum_{m=-l}^{+l}
                      ( \phi_{\s(l,m)}^\oplus  Y_{\s(l,m)}{}_{:a}
                       +\phi_{\s(l,m)}^\otimes Y_{\s(l,m)}{}_{:b}
                                                   \epsilon^b{}_a)
\eqno(4.7)$$
including both scalar and pseudo-scalar displacement.  Note that we do not
include a $l=m=0$ term in the sum since $Y_{\s(0,0)}{}_{:a}=0$.  Furthermore
since $\Delta_a$ is real we require
$$\phi_{\s(l,m)}^\oplus =(-1)^m{\phi_{\s(l,m)}^\oplus }^* \qquad
  \phi_{\s(l,m)}^\otimes=(-1)^m{\phi_{\s(l,m)}^\otimes}^* \qquad {\rm since}
  \qquad Y_{\s(l,-m)}^*=(-1)^m\,Y_{\s(l,m)}
.\eqno(4.8)$$
In the above decomposition the $\phi_{\s(l,m)}^\oplus$ terms give the scalar
part of the displacement vector which is proportional to the gradient of the
spherical harmonic, $Y_{\s(l,m)}$.  The pseudo-scalar part, given by the
$\phi^\otimes$ terms, is proportional to the gradient of $Y_{\s(l,m)}$ rotated
by 90$^\circ$.  The deformation tensor is then given by
$$\psi_{ab}=g_{ab}
-\sum_{l=1}^\infty\sum_{m=-l}^{+l}( \phi_{\s(l,m)}^\oplus  Y_{\s(l,m)}{}_{:ab}
                                   +\phi_{\s(l,m)}^\otimes Y_{\s(l,m)}{}_{:ac}
                                                         \epsilon{}^c{}_b)
\eqno(4.9)$$
and it's component parts by
$$\eqalign{
\kappa=&-{1\over2}
  \sum_{l=1}^\infty\sum_{m=-l}^{+l}l(l+1)\phi_{\s(l,m)}^\oplus  Y_{\s(l,m)} \cr
\omega=&-{1\over2}
  \sum_{l=1}^\infty\sum_{m=-l}^{+l}l(l+1)\phi_{\s(l,m)}^\otimes Y_{\s(l,m)} \cr
\gamma_{ab}=&\sum_{l=2}^\infty\sum_{m=-l}^{+l}\left(
  \phi_{\s(l,m)}^\oplus (          Y_{\s(l,m)}{}_{:ab}
                   -{1\over2}g_{ab}Y_{\s(l,m)}{}_{:c}{}^{:c})
 +\phi_{\s(l,m)}^\otimes{1\over2}( Y_{\s(l,m)}{}_{:ac}\epsilon{}^c{}_b
                                  +Y_{\s(l,m)}{}_{:bc}\epsilon{}^c{}_a)
                                                     \right)
           }\eqno(4.10)$$
where we have used eq~(4.5).  In the shear we do not include the $l=1$ terms,
since the deformation tensor from these modes do not contribute to the shear as
we will see below.  Combining (4.2,6,\&10) one finds
$$\eqalign{\gamma_\rms=(-\nabla^{-2}\psi_{ab}{}^{:ab}+\kappa)
&={1\over2}\sum_{l=2}^\infty\sum_{m=-l}^{+l}\,(l+2)(l-1)\,
                                         \phi_{\s(l,m)}^\oplus\,Y_{\s(l,m)} \cr
\gamma_\rmp=-(\nabla^{-2}\psi_{ab}{}^{:bc}\epsilon^a{}_c
             -\nabla^{-2}(\omega \epsilon_{ab})^{:bc}\epsilon^a{}_c)
=&{1\over2}\sum_{l=1}^\infty\sum_{m=-l}^{+l}(l+2)(l-1)\,
                                        \phi_{\s(l,m)}^\otimes\,Y_{\s(l,m)} \cr
           }\eqno(4.11)$$
and that the $l=1$ terms do not contribute to either $\gamma_\rms$ or
$\gamma_\rmp$.   For large $l$ we recover the small angle result, i.e.
$\gamma_\rms=\kappa$ and $\gamma_\rmp=-\omega$.  These relations hold for any
deformation of our sky onto the 3-d space, but they are nevertheless they are
non-trivial consistency checks.  For example one can hope to measure both
$\gamma_\rms$ and $\kappa$, the former from galaxy alignments and the later
from galaxy brightness distributions in various bands.  These being independent
measures of the same thing would provide a consistency check. If they were
found to be inconsistent then would discover that one is not measuring
variations in alignments and brightnesses due to an image deformation.
Measurement error, Galactic extinction,  intrinsic alignments, intrinsic galaxy
clusterings and other effects will all contribute to inconsistencies.

\subheader{Decomposition of Distortion Pattern}

	Given combinations of $\Delta^a$, $\kappa$, $\omega$, $\gamma_\rms$,
$\gamma_\rmp$, or $\gamma^{ab}$ one can compute the mode coefficients via
$$\eqalign{
\phi_{\s(l,m)}^\oplus
          =-&{1\over l(l+1)}\int d^2\hatbfn\,Y_{\s(l,m)}^*{}_{:a}\,\Delta^a
          =+ {1\over l(l+1)}\int d^2\hatbfn\,Y_{\s(l,m)}^*\,\Delta^a{}_{:a}
          =- {2\over l(l+1)}\int d^2\hatbfn\,Y_{\s(l,m)}^*\,\kappa          \cr
\phi_{\s(l,m)}^\otimes
           =+&{1\over l(l+1)}\int d^2\hatbfn\,Y_{\s(l,m)}^*{}_{:a}\,\Delta^b 
                                                             \epsilon^a{}_b
           =- {1\over l(l+1)}\int d^2\hatbfn\,Y_{\s(l,m)}^*\,\Delta^b{}_{:a}
                                                              \epsilon^a{}_b
           =  {2\over l(l+1)}\int d^2\hatbfn\,Y_{\s(l,m)}^*\,\omega         \cr
\phi_{\s(l,m)}^\oplus
 =+&{1\over l(l+1)(l(l+1)-1)}\int d^2\hatbfn\,Y_{\s(l,m)}^*{}_{:ab}\,\psi^{ab}
 =  {1\over l(l+1)-1}\int d^2\hatbfn\,Y_{\s(l,m)}^*\,(\gamma_\rms+\kappa)   \cr
\phi_{\s(l,m)}^\otimes
 =+&{1\over l(l+1)(l(l+1)-1)}\int d^2\hatbfn\,Y_{\s(l,m)}^*{}_{:ac}\,
                                                     \epsilon^c{}_b\,\psi^{ab}
 =  {1\over l(l+1)-1}\int d^2\hatbfn\,Y_{\s(l,m)}^*\,(\gamma_\rmp-\omega)   \cr
\phi_{\s(l,m)}^\oplus
 =+&{2\over(l+2)(l+1)l(l-1)}\int d^2\hatbfn\,Y_{\s(l,m)}^*{}_{:ab}\,\gamma^{ab}
 =- {2\over(l+2)      (l-1)}\int d^2\hatbfn\,Y_{\s(l,m)}^*\,\gamma_\rms     \cr
\phi_{\s(l,m)}^\otimes
 =+&{2\over(l+2)(l+1)l(l-1)}\int d^2\hatbfn\,Y_{\s(l,m)}^*{}_{:ac}\,
                                                     \epsilon^{cd}\,\gamma^{ad}
 =- {2\over(l+2)      (l-1)}\int d^2\hatbfn\,Y_{\s(l,m)}^*\,\gamma_\rmp     \cr
           }\eqno(4.12)$$
using the orthonormality relation of eqs~(4.5\&6).  We see that one cannot
obtain the $l=1$ terms from the shear since substituting $l=1$ in the last four
expressions one finds that the prefactors are infinite while the integrals are
zero.

\subheader{Mean Square Expansion, Rotation, and Shear}

	For a given realization of the shear pattern one can construct the
quantities
$$\widehat{C_l^\oplus }
        ={l^2(l+1)^2\over4(2l+1)}\sum_{m=-l}^l|\phi^\oplus_{ \s(l,m)}|^2 \qquad
  \widehat{C_l^\otimes}
        ={l^2(l+1)^2\over4(2l+1)}\sum_{m=-l}^l|\phi^\otimes_{\s(l,m)}|^2
.\eqno(4.13)$$
The $\widehat{\phantom{C_l}}$ notation is used to indicate that these
quantities may be used as estimators of the power spectra coefficients,
$C_l^\oplus$ and $C_l^\otimes$, defined below.  Combining the decomposition in
eq~(4.10) with the orthonormality relations of eqs~(4.5\&6) we find that the
mean square $\kappa$, $\omega$, and $\gamma_{ab}$ averaged over the sky is 
$$\eqalign{
\overline{\kappa^2}=&\sum_{l=1}^\infty{2l+1\over4\pi}\widehat{C_l^\oplus }  \cr
\overline{\omega^2}=&\sum_{l=1}^\infty{2l+1\over4\pi}\widehat{C_l^\otimes}  \cr
\overline{\gamma^2}=\overline{-\Vert\gamma^a{}_b\Vert}
                   ={1\over2}\overline{\gamma^{ab}\gamma_{ab}}
                   =&\sum_{l=2}^\infty{2l+1\over4\pi}\,{(l+2)(l-1)\over l(l+1)}
                             \,(\widehat{C_l^\oplus}+\widehat{C_l^\otimes})
           }\eqno(4.14)$$
where $\Vert\gamma^a{}_b\Vert$ indicates the determinant of the matrix 
$\gamma^a{}_b$.  We use the notation $\overline{\phantom{\kappa}}$ to indicate
averages over the sky.  One expects the mean square shear to receive it's
largest contribution from small angular scale so that the large $l$ terms
dominate these sums.  If this is the case then
$\overline{\gamma^2}\approx\overline{\kappa^2}+\overline{\omega^2}$.  One could
construct equations analogous to eq~(4.14) for mean-square values after
convolution with a window function by one could include a $|W_l|^2$ factor to
each of the terms of the sum, where $W_l$ is the spherical harmonic
decomposition of the window function.

\subheader{Explicit Construction of the Shear Tensor}

	Here we give the explicit expression for the shear tensor in terms of
the spherical harmonic mode coefficients.  Using spherical polar coordinates,
$(\theta,\phi)$, the metric and anti-symmetric tensor are
$$g_{ab}=\left(\matrix 1 & 0            \\
                       0 & \sin^2\theta \endmatrix\right) \qquad
  g=\sin^2\theta \qquad
\epsilon^a{}_b=\left(\matrix         0           & \sin\theta \\
                             -{1\over\sin\theta} &     0      \endmatrix\right)
\eqno(4.15)$$
so (4.10) becomes
$$\eqalign{\gamma_{ab}=
&+\sum_{l=2}^\infty\sum_{m=-l}^{+l}\phi_{\s(l,m)}^\oplus
\left(\matrix  {1\over2}              Y_{\s(l,m)}{}_{:\theta\theta}
              -{1\over2\sin^2\theta}\,Y_{\s(l,m)}{}_{:\phi  \phi  } 
             & Y_{\s(l,m)}{}_{:\theta\phi  }  \\
               Y_{\s(l,m)}{}_{:\theta\phi  }
             & {1\over2}              Y_{\s(l,m)}{}_{:\phi  \phi  } 
              -{1\over2}\sin^2\theta\,Y_{\s(l,m)}{}_{:\theta\theta}
      \endmatrix\right)                                                     \cr
&+\sum_{l=2}^\infty\sum_{m=-l}^{+l}\phi_{\s(l,m)}^\otimes
\left(\matrix -{1\over  \sin\theta}\,Y_{\s(l,m)}{}_{:\theta\phi  }
             & {1\over2}\sin\theta \,Y_{\s(l,m)}{}_{:\theta\theta}
              -{1\over2 \sin\theta}\,Y_{\s(l,m)}{}_{:\phi  \phi  }    \\
               {1\over2}\sin\theta \,Y_{\s(l,m)}{}_{:\theta\theta}
              -{1\over2 \sin\theta}\,Y_{\s(l,m)}{}_{:\phi  \phi  }
             &          \sin\theta \,Y_{\s(l,m)}{}_{:\theta\phi  }
      \endmatrix\right)
           }.\eqno(4.16)$$
To continue we need the explicit form of the spherical harmonics, which we take
to be
$$Y_{\s(l,m)}(\theta,\phi)
      =\sqrt{{2l+1\over4\pi}{(l-m)!\over(l+m)!}}\,P_l^m(\cos\theta)\,e^{im\phi}
\eqno(4.17)$$
following the conventions of Jackson (1975).

	Using this explicit form of the spherical harmonics one can show that
the deflection vectors $Y_{\s(1,0)}{}_{:b} \epsilon^b{}_a$,
$ (Y_{\s(1,+1)}{}_{:b}+Y_{\s(1,-1)}{}_{:b})\epsilon^b{}_a$, and
$i(Y_{\s(1,+1)}{}_{:b}-Y_{\s(1,-1)}{}_{:b})\epsilon^b{}_a$, are generators of
solid body rotations of the sphere. Clearly solid body rotation generates no
amplification or shear, but does translate and/or rotate the images and thus
we see why the $l=1$ pseudo-scalar displacement contributes nothing to the
shear, $\gamma_{ab}$.   The $l=1$ scalar displacement is just the pseudo-scalar
displacement rotated by $90^\circ$ which would yield a shear rotated by
$45^\circ$.  However since the $l=1$ pseudo-scalar shear is zero, so must be
the $l=1$ scalar shear.

	Let us defining two auxiliary function via
$$\eqalign{\sqrt{{2l+1\over4\pi}{(l-m)!\over(l+m)!}}\,
  G^+_{\s(l,m)}(\cos\theta)\,e^{im\phi}&=
                    {1\over2}            Y_{\s(l,m)}{}_{:\theta\theta}
                   -{1\over2\sin^2\theta}Y_{\s(l,m)}{}_{:\phi  \phi  }      \cr
           \sqrt{{2l+1\over4\pi}{(l-m)!\over(l+m)!}}\,
  G^-_{\s(l,m)}(\cos\theta)\,e^{im\phi}&=Y_{\s(l,m)}{}_{:\theta\phi}
           }\eqno(4.18)$$
or more explicitly
$$\eqalign{
G^+_{\s(l,m)}(x)
=&-\left(  {l-m^2\over1-x^2}+{1\over2}l(l-1)\right)\,P_l^m    (x)
         +(l+m){x\over1-x^2}                       \,P_{l-1}^m(x)           \cr
G^-_{\s(l,m)}(x)=&m\,\left((l-1)\,{x\over1-x^2}\,P_l^m    (x)
                          -(l+m)\,{1\over1-x^2}\,P_{l-1}^m(x)\right)
           }\eqno(4.19)$$
we find from substituting (4.18) into (4.16) that
$$\eqalign{
\gamma_{\theta\theta}=-{\gamma_{\phi\phi}\over\sin^2\theta}
=&\sum_{l=2}^\infty\sqrt{2l+1\over4\pi}
  \sum_{m=-l}^{+l}\sqrt{(l-m)!\over(l+m)!}\,
\left(  \phi_{\s(l,m)}^\oplus \,G^+_{\s(l,m)}(\cos\theta)
      -i\phi_{\s(l,m)}^\otimes\,G^-_{\s(l,m)}(\cos\theta)\right)\,e^{im\phi}\cr
{\gamma_{\theta\phi}\over\sin\theta}={\gamma_{\phi\theta}\over\sin\theta}
=&\sum_{l=2}^\infty\sqrt{2l+1\over4\pi}
  \sum_{m=-l}^{+l}\sqrt{(l-m)!\over(l+m)!}\,
 \left(i\phi_{\s(l,m)}^\oplus \,G^-_{\s(l,m)}(\cos\theta)
       +\phi_{\s(l,m)}^\otimes\,G^+_{\s(l,m)}(\cos\theta)\right)\,e^{im\phi}\cr
           }.\eqno(4.20)$$
Note the symmetry relations
$$G^\pm_{\s(l,-m)}=\pm(-1)^m{(l+m)!\over(l-m)!}\,G^\pm_{\s(l,m)}
\eqno(4.21)$$
and the orthonormality relations
$$\eqalign{
{1\over2}\int_{-1}^1dx\,( G^+_{\s(l,m)}(x)\,G^+_{\s(l',m)}(x)
                         +G^-_{\s(l,m)}(x)\,G^-_{\s(l',m)}(x))
      &={1\over4(2l+1)}{(l+2)!\over(l-2)!}{(l+m)!\over(l-m)!}\,\delta_{ll'} \cr
{1\over2}\int_{-1}^1dx\,( G^+_{\s(l,m)}(x)\,G^-_{\s(l',m)}(x)
                         +G^-_{\s(l,m)}(x)\,G^+_{\s(l',m)}(x))&=0
           }.\eqno(4.22)$$
From these identities one can rederive the formula for $\overline{\gamma^2}$ in
eq~(4.14).

Below we will find that it is of particular interest to calculate the shear at
the pole of the coordinate system.  In spite of the fact that the coordinates
system is singular at the pole, mode expansion at the pole takes a particularly
simple form.  The coordinate singularity means that one should consider only
$\gamma_{\theta\theta}$ in the limit $\theta\rightarrow0$ for a given value of
$\phi$.  In this limit the value $\gamma_{\theta\theta}$ will retain an 
$\phi$-dependence and from this one has the angular dependence of the shear.
One finds that in the $\theta=0$ limit that only $m=\pm2$ contributes
$$G^\pm_{\s(l,m)}(1)=   {1\over4}{(l+2)!\over(l-2)!}\,\delta_{m,2}
                     \pm{1\over4}\,\delta_{m,-2}
\eqno(4.23)$$
which on can use to compute the 2-point correlation function of the shear.

\subheader{Correlation Functions and Power Spectrum}

	Henceforth we shall explicitly include the radial $r$-dependence of the
deformation since we may be interested in correlating the different components
of the correlation at different redshifts.  We shall also be considering a
statistical distribution of deformation fields.  Averaging over realizations,
one can characterize any set of 2-point functions in terms of the power spectra
which we choose to define by  
$$\left\langle\phi^\oplus_{\s(l,m)}     (r )\,
             {\phi^\oplus_{\s(l',m')}}^*(r')\right\rangle
            ={4C_l^\oplus(r,r')\over l^2(l+1)^2}\delta_{ll'}\delta_{mm'} \qquad
  \left\langle\phi^\otimes_{\s(l,m)}     (r )\,
             {\phi^\otimes_{\s(l',m')}}^*(r')\right\rangle
            ={4C_l^\otimes(r,r')\over l^2(l+1)^2}\delta_{ll'}\delta_{mm'}
.\eqno(4.24)$$
The Kronecker $\delta$-functions and the lack of $m$-dependence follow from the
assumed homogeneity and isotropy of the distributions we are considering.  We
include a dependence of the deformation with depth $r$.  We will also assume
that the distribution of deformations is even under parity as it would be for
gravitationally induced deformations.  Since $\phi^\oplus$ is even under parity
and $\phi^\otimes$ is odd under parity it follows that
$$\left\langle\phi^\oplus_{\s(l,m)}     (r )
             {\phi^\otimes_{\s(l',m')}}^*(r')\right\rangle=0
.\eqno(4.25)$$
Both $C_l^\oplus(r,r')$ and $C_l^\otimes(r,r')$ are real and are non-negative
when $r=r'$.


	The correlation functions of various quantities can be computed in
terms of the correlation function
$$\eqalign{
C_\kappa(r,r',\vartheta)
&=\left\langle\kappa(r,\hatbfn)\,\kappa(r',\hatbfn')\right\rangle
 ={1\over4\pi}\sum_{l=1}^\infty(2l+1)\,P_l(\cos\vartheta)\,C_l^\oplus (r,r')\cr
C_\omega(r,r',\vartheta)
&=\left\langle\omega(r,\hatbfn)\,\omega(r',\hatbfn')\right\rangle
 ={1\over4\pi}\sum_{l=1}^\infty(2l+1)\,P_l(\cos\vartheta)\,C_l^\otimes(r,r')\cr
  \left\langle\kappa(r,\hatbfn)\,\omega(r',\hatbfn')\right\rangle
&=0                                                                         \cr
C_\gamma(r,r',\vartheta,\varphi,\varphi')
 ={1\over2\pi}\sum_{l=2}^\infty{2l+1\over l^2(l+1)^2}\,\biggl[
  &(C_l^\oplus (r,r')\,G^+_{\s(l,2)}(\cos\vartheta)
   +C_l^\otimes(r,r')\,G^-_{\s(l,2)}(\cos\vartheta))\,\cos2\varphi
                                                    \,\cos2\varphi'         \cr
 +&(C_l^\oplus (r,r')\,G^-_{\s(l,2)}(\cos\vartheta)
   +C_l^\otimes(r,r')\,G^+_{\s(l,2)}(\cos\vartheta))\,\sin2\varphi
                                                    \,\sin2\varphi'\biggr]  \cr
C_{\gamma\kappa}(r,r',\vartheta,\varphi)
&=-{1\over4\pi}\sum_{l=2}^\infty{2l+1\over l(l+1)}\,C_l^\oplus(r,r')\,
                                         P_l^2(\cos\vartheta)\,\cos2\varphi \cr
C_{\gamma\omega}(r,r',\vartheta,\varphi)
&=+{1\over4\pi}\sum_{l=2}^\infty{2l+1\over l(l+1)}\,C_l^\otimes(r,r')\,
                                         P_l^2(\cos\vartheta)\,\sin2\varphi
           }.\eqno(4.26)$$
In each of the above expressions $\vartheta=\angle(\hatbfn,\hatbfn')$, i.e. the
angle between the two points.  $C_\gamma$ measures the correlation of
$\gamma_{ab}$ with itself, while $C_{\gamma\kappa}$ and $C_{\gamma\omega}$
measures the correlation of $\gamma_{ab}$ with $\kappa$ and $\omega$,
respectively.  The meaning of $\varphi$ and $\varphi'$ are described in fig~2.
Setting $C_l^\otimes=0$ in $C_\gamma$ we obtain a correlation function similar
to those in Kaiser (1993).  The $C_{\s++}(\alpha)$, $C_{\s+\times}(\alpha)$,
and $C_{\s\times\times}(\alpha)$ of that paper corresponds to setting
$\varphi=\varphi'=\alpha$, $\varphi=\varphi'+{\pi\over4}=\alpha$,
$\varphi=\varphi'=\alpha-{\pi\over4}$.  Kaiser (1992) paper did not include
pseudo-scalar shear since this is not contributed to by density perturbations.

	From the above correlation function one may calculate the the mean
square quantities at a given distance
$$\eqalign{
\langle\kappa^2\rangle&=C_\kappa(r,r,0)
 =\sum_{l=1}^\infty{2l+1\over4\pi}\,C_l^\oplus (r,r)                     \qquad
\langle\omega^2\rangle =C_\omega(r,r,0)
 =\sum_{l=1}^\infty{2l+1\over4\pi}\,C_l^\otimes(r,r)                        \cr
\langle\gamma^2\rangle=\langle{1\over2}\gamma^{ab}\gamma_{ab}\rangle
&=C_\gamma(r,r,\vartheta,0,0)+C_\gamma(r,r,\vartheta,{\pi\over4},{\pi\over4})
 =\sum_{l=2}^\infty{2l+1\over4\pi}\,{(l+2)(l-1)\over l(l+1)}\,
  \left[C_l^\oplus (r,r)+C_l^\otimes(r,r)\right]                            \cr
           }.\eqno(4.27)$$
Note that the factor $(l+2)(l-1)/(l(l+1))=1+\calO(l^{-2})$, and since most of
the shear is expected to be generated at very small scales we find
$$\langle\gamma^2\rangle\approx\langle\kappa^2\rangle
                              +\langle\omega^2\rangle
.\eqno(4.28)$$
Remember that although (4.14) is quite similar to (4.27) the former refers to
averages over the sphere for a given realization, which could be a observed
quantity, while the latter refers to an average over realization and is a
purely theoretical construct.  The reasons that the formulae are so similar is
because, by construction, 
$$\langle\widehat{C_l^\oplus }(r,r')\rangle=C_l^\oplus (r,r') \qquad
  \langle\widehat{C_l^\otimes}(r,r')\rangle=C_l^\otimes(r,r')
,\eqno(4.29)$$
i.e $\widehat{C_l^\oplus}$ and $\widehat{C_l^\otimes}$ are unbiased estimators
of $C_l^\oplus$ and $C_l^\otimes$, respectively.

	Comparing (4.26) with the orthonormality relations, (4.5\&6) as well as
$${1\over2}\int_{-1}^1 dx\,P_l(x)\,P_{l'}(x)={1\over2l+1}\delta_{ll'}
\eqno(4.30)$$
we see that if one knows the correlation functions one may calculate the power
spectra using
$$\eqalign{
C_l^\oplus (r,r')=&2\pi\int_{-1}^1 dx\,P_l(x)\,C_\kappa(r,r'\cos^{-1}x)
                 =-{2\over(l+2)(l-1)}
                       \int_{-1}^1 dx\,\int_0^{2\pi}d\varphi\,P_l^2(x)\,
                       C_{\gamma\kappa}(r,r',\cos^{-1}x,\varphi)\,\cos2\phi \cr
C_l^\otimes(r,r')=&2\pi\int_{-1}^1 dx\,P_l(x)\,C_\omega(r,r'\cos^{-1}x)
                 =-{2\over(l+2)(l-1)}
                       \int_{-1}^1 dx\,\int_0^{2\pi}d\varphi\,P_l^2(x)\,
                       C_{\gamma\omega}(r,r',\cos^{-1}x,\varphi)\,\sin2\phi \cr
C_l^\oplus (r,r')=&{4\over\pi}\,{1\over(l+2)^2(l-1)^2}\,
                   \int_{-1}^1 dx\,\int_0^{2\pi}d\varphi
                                 \,\int_0^{2\pi}d\varphi'\,
                           C_\gamma(r,r',\cos^{-1}x,\varphi,\varphi')\times \cr
&\hskip160pt\left( G^+_{\s(l,2)}(x)\,\cos2\varphi\,\cos2\varphi'
                  +G^-_{\s(l,2)}(x)\,\sin2\varphi\,\sin2\varphi'\right)     \cr
C_l^\otimes(r,r')=&{4\over\pi}\,{1\over(l+2)^2(l-1)^2}\,
                   \int_{-1}^1 dx\,\int_0^{2\pi}d\varphi
                                 \,\int_0^{2\pi}d\varphi'\,
                           C_\gamma(r,r',\cos^{-1}x,\varphi,\varphi')\times \cr
&\hskip160pt\left( G^+_{\s(l,2)}(x)\,\sin2\varphi\,\sin2\varphi'
                  +G^-_{\s(l,2)}(x)\,\cos2\varphi\,\cos2\varphi'\right)     \cr
        }.\eqno(4.31)$$

\subheader{Small-Angle Limit}

	Now let us consider the small angle limit of the formulae just derived.
Over small angles the surface of the sphere is approximately a plane, and the 
wavenumber, $l$, can be consider modulus of a 2-d vector wavenumber in this
plane in a Fourier decomposition of the displacement if $l\gg1$.  This Fourier
representation is context in which most analyses of image deformation has been
analyzed, and the formulae derived above should approach this Fourier limit for
$l\gg1$ over regions much smaller than one radian.  In particular the angular
correlation function should approach the Fourier limit for $l\gg1$ and
$\vartheta\ll1$.  In the small-angle large-$l$ limit one can show
$$\eqalign{G^\pm_{\s(l,m)}(\cos\theta)
&=(-1)^m {1\over4}l^{6-m}(J_{m-2}(l\theta)\pm J_{m+2}(l\theta))      
                                     +\calO({1\over l},{m\over l},\theta^2) \cr
P^m_l(\cos\theta)
&=(-1)^m{(l+m)!\over(l-m)!}J_m(l\theta)l^{-m}
                               +\calO({1\over l^{m+2}},{m\over l},\theta^2)
           }.\eqno(4.32)$$
Thus in the small-angle approximation the correlation functions become
$$\eqalign{C_\kappa(r,r',\vartheta)
&\approx{1\over2\pi}\int^\infty dl\,l\,J_0(l\vartheta)\,C_l^\oplus (r,r')\qquad
  C_{\gamma\kappa}(r,r',\vartheta,\varphi)
\approx-{1\over2\pi}\int^\infty dl\,l\,J_2(l\vartheta)\,C_l^\oplus(r,r')\,
                                                               \cos2\varphi \cr
  C_\omega(r,r',\vartheta)
&\approx{1\over2\pi}\int^\infty dl\,l\,J_0(l\vartheta)\,C_l^\otimes(r,r')\qquad
  C_{\gamma\omega}(r,r',\vartheta,\varphi)
\approx+{1\over2\pi}\int^\infty dl\,l\,J_2(l\vartheta)\,C_l^\otimes(r,r')\,
                                                               \sin2\varphi \cr
C_\gamma(r,r',\vartheta,\varphi,\varphi')
&\approx{1\over4\pi}\int^\infty dl\,l\,\biggl[
             ( C_l^\oplus (r,r')\,(J_0(l\vartheta)+J_4(l\vartheta))
              +C_l^\otimes(r,r')\,(J_0(l\vartheta)-J_4(l\vartheta)))
                                               \,\cos2\varphi\,\cos2\varphi'\cr
&\hskip58pt +( C_l^\oplus (r,r')\,(J_0(l\vartheta)-J_4(l\vartheta))
              +C_l^\otimes(r,r')\,(J_0(l\vartheta)+J_4(l\vartheta)))
                                               \,\sin2\varphi\,\sin2\varphi'
                                              \biggr]
           }\eqno(4.33)$$
while the small angle limit of (4.31) is
$$\eqalign{
C_l^\oplus (r,r')\approx&2\pi\int_0^\infty d\vartheta\,\vartheta\,
                                   J_0(l\vartheta)\,C_\kappa(r,r'\vartheta)
                 \approx-2   \int_0^\infty d\vartheta\,\vartheta\,
                             \int_0^{2\pi}d\varphi\,J_2(l\vartheta)\,
                        C_{\gamma\kappa}(r,r',\vartheta,\varphi)\,\cos2\phi \cr
C_l^\otimes(r,r')\approx&2\pi\int_0^\infty d\vartheta\,\vartheta\,
                                   J_0(l\vartheta)\,C_\omega(r,r'\vartheta)
                 \approx-2   \int_0^\infty d\vartheta\,\vartheta\,
                             \int_0^{2\pi}d\varphi\,J_2(l\vartheta)\,
                        C_{\gamma\omega}(r,r',\vartheta,\varphi)\,\sin2\phi \cr
C_l^\oplus (r,r')\approx&{1\over\pi}\,
        \int_0^\infty d\vartheta\,\vartheta\,\int_0^{2\pi}d\varphi
                                           \,\int_0^{2\pi}d\varphi'\,
                           C_\gamma(r,r',\vartheta,\varphi,\varphi')
                     \left( J_0(l\vartheta)\,\cos2(\varphi-\varphi')
                           +J_4(l\vartheta)\,\cos2(\varphi+\varphi')\right) \cr
C_l^\otimes(r,r')\approx&{1\over\pi}\,
        \int_0^\infty d\vartheta\,\vartheta\,\int_0^{2\pi}d\varphi
                                           \,\int_0^{2\pi}d\varphi'\,
                           C_\gamma(r,r',\vartheta,\varphi,\varphi')
                     \left( J_0(l\vartheta)\,\cos2(\varphi-\varphi')
                           -J_4(l\vartheta)\,\cos2(\varphi+\varphi')\right) \cr
        }.\eqno(4.34)$$
Using a 2-d definition of a power spectrum where
$$\langle f(\hatbfn)\,f(\hatbfn')\rangle
=2\pi \int_0^\infty dl\,l\,P_f(l)\,J_0(l\vartheta) \qquad 
                                             \vartheta=\angle(\hatbfn,\hatbfn')
\eqno(4.35)$$
we see that
$$P_\kappa(l)={1\over4\pi^2}\,C_l^\oplus  \qquad
  P_\omega(l)={1\over4\pi^2}\,C_l^\otimes
\eqno(4.36)$$
and that $\gamma_{ab}$ is described by these same two power spectra, e.g.
$$\langle\gamma^2\rangle\approx\langle\kappa^2\rangle+\langle\omega^2\rangle
\approx2\pi        \int^\infty dl\,l\,(P_\kappa(l)+P_\omega(l))
\approx{1\over2\pi}\int^\infty dl\,l\,(C_l^\oplus(l)+C_l^\otimes)
.\eqno(4.37)$$

\subheader{Visibility Functions}

	To measure the shear one typically looks for correlations in galaxy
ellipticity orientation, and one needs enough galaxies to find the correlation
in the noise generated by the random orientation of the galaxies.  Usually one
doesn't know the distance to the galaxies except in a statistical sense, and
even if one did, one would have to sum over galaxies at different distances to
obtain a significant signal.  Thus the shear at a given distance is not what
one measures, but rather some weighted average of the shear at different
distances. We may represent this distribution of distances by a visibility
function, $V(r)$, normalized so that
$$\int_0^\infty dr\,V(r)=1
,\eqno(4.38)$$
so that the average expansion, rotation, and shear is
$$\overline{\kappa}(\hatbfn)=\int_0^\infty dr\,V(r)\,\kappa(r,\hatbfn) \qquad
  \overline{\omega}(\hatbfn)=\int_0^\infty dr\,V(r)\,\omega(r,\hatbfn) \qquad
 \overline{\gamma}_{ab}(\hatbfn)=\int_0^\infty dr\,V(r)\,\gamma_{ab}(r,\hatbfn)
.\eqno(4.39)$$
One can apply all of the formulae derived above by replacing the quantities at
fixed distance with the same quantities, averaged over distance, e.g. the mode
coefficients
$$\overline{\phi}^\oplus_{\s(l,m)}
                      =\int_0^\infty dr\,V(r)\,\phi^\oplus_{\s(l,m)} (r) \qquad
\overline{\phi}^\otimes_{\s(l,m)}
                      =\int_0^\infty dr\,V(r)\,\phi^\otimes_{\s(l,m)}(r)
,\eqno(4.40)$$
or the correlation functions
$$C_{\overline{x}}(\vartheta,\cdots)=
\int_0^\infty dr\,V(r)\,\int_0^\infty dr'\,V(r')\,C_x(r,r',\vartheta,\cdots)
.\eqno(4.41)$$
where $x$ might be $\kappa$, $\omega$, or $\gamma$.

	It is not the purpose of this paper to be terribly realistic about how
one samples the shear with a given galaxy sample.  The only visibility function
which we will consider explicitly in this paper corresponds to uniform sampling
in space up to a given distance, $r_{\rm max}$.  The visibility function in
this case is
$$V(r)=3{r^2\over r_{\rm max}^3}
\eqno(4.42)$$
where $r_{\rm max}$ gives the maximum comoving distance which one includes in
ones sample.

\header{5. Application To Density Inhomogeneities}

	Here we consider the image distortion from a linear growing mode
density field in  a matter-dominated Einstein-deSitter cosmology.  For these
growing modes the Newtonian gravitational potential is constant with time at a
given comoving position.  It is convenient to use the Spherical Harmonic - 
Spherical Bessel function expansion so the gravitational potential may be
expanded
$$\eqalign{\Phi(r,\theta,\phi)
&=\sum_{l=0}^\infty\sum_{m=-l}^l \int_0^\infty dk\,k^2
\widetilde{\Phi}_{\s(l,m)}(k)\,j_l(kr)\,Y_{\s(l,m)}(\theta,\phi)\cr
\widetilde{\Phi}_{\s(l,m)}(k)
&={2\over \pi}\int_0^\infty dr\,r^2\,j_l(kr)
  \int_0^\pi d\theta\,\sin\theta 
  \int_{-\pi}^\pi d\phi\,Y_{\s(l,m)}^*(\theta,\phi)\,\Phi(r,\theta,\phi)
           }.\eqno(5.1)$$
where $r$ is the comoving distance from the observer.  For the spatially flat
geometry we are considering, these modes are orthogonal eigenfunctions of the
Laplace operator.  For an statistically homogeneous and isotropic distribution
the correlation function of mode amplitudes is of the form
$$\langle\widetilde{\Phi}_{  \s(l ,m )}(k )\,
         \widetilde{\Phi}^*_{\s(l',m')}(k')\rangle
=(4\pi)^2P_\Phi(k)\,{1\over k^2}\,\delta(k-k')\,\delta_{ll'}\,\delta_{mm'}
\eqno(5.2)$$
where $P_\Phi$ is the power spectrum of the potential fluctuations. Using the
relation between the potential and overdensity,
$$\nabla^2\Phi=4\pi a^2G\delta\rho={3\over2}{H_0^2\over a^2}\delta
,\eqno(5.3)$$
we may also relate the potential power spectrum to the density power spectrum: 
$$P_{\Phi}(k)={9\over4}\left({H_0\over ak}\right)^4\,P_\delta(k)
.\eqno(5.4)$$
Below we will use the notation $\hatbfn$ for $(\theta,\phi)$.  Clearly these
scalar modes will lead only to scalar components to the image deformation.
Thus we may describe the deformation by a scalar function
$$\Delta_a=-\phi^\oplus{}_{:a} \qquad 
\phi^\oplus{}(\hatbfn)
                  =\sum_{l=1}^\infty\phi^\oplus_{\s(l,m)}\,Y_{\s(l,m)}(\hatbfn)
\qquad \phi^\otimes_{\s(l,m)}=0
.\eqno(5.5)$$
Comparing eq~(5.5) with eq~(4.12) we find
$$\phi^\oplus(r,\hatbfn)
=2\int_{\etaobs-r}^\etaobs d\eta'\,{r-(\etaobs-\eta')\over r(\etaobs-\eta')}
                                 \,\Phi(\bfxobs+(\etaobs-\eta')\,\hatbfn,\eta')
=2\int_0^r dr'\,{r-r'\over rr'}\,\Phi(r',\hatbfn)
,\eqno(5.6)$$
and comparing this with (5.1) we see that
$$\phi^\oplus_{\s(l,m)}(r)
=2\int_0^\infty dk\,k^2 \widetilde{\Phi}_{\s(l,m)}(k)\,
  \int_0^r dr'\,{r-r'\over rr'}\,j_l(kr')
=2\int_0^\infty dk\,k^2 \widetilde{\Phi}_{\s(l,m)}(k)\,I_l(kr)
\eqno(5.7)$$
where
$$I_l(x)=\int_0^1 {dy\over y}\,(1-y)\,j_l(xy) \qquad l>0
.\eqno(5.8)$$
There is no need for an $l=0$ term since there is no $l=0$ contribution to
shear or expansion.  In the appendix we present analytic expressions for
$I_l(x)$ as well as series and asymptotic expansions. We will argue below that
in most practical applications $I_l$ is well approximated by eq~(A6), i.e.
$$I_l(x)\approx\left(a_l-{b_l\over x}\right)\,\calH(x-{b_l\over a_l})
   \qquad a_l={\sqrt{\pi}\over2l}{\Gamma({l+2\over2})\over\Gamma({l+3\over2})}
   \quad  b_l={\sqrt{\pi}\over l}{\Gamma({l+1\over2})\over\Gamma({l  \over2})}
\eqno(5.9)$$
which we call the {\it asymptotic approximation}.  Combining (4.24), (5.2),
(5.4), and (5.7) we find
$$\eqalign{C_l^\oplus(r,r')
&=16\pi^2 l^2(l+1)^2\int_0^\infty dk\,k^2 \,I_l(kr)\,I_l(kr')\,P_\Phi(k)    \cr
&=36\pi^2 l^2(l+1)^2\left({H_0\over a}\right)^4
                \int_0^\infty{dk\over k^2}\,I_l(kr)\,I_l(kr')\,P_\delta(k)  \cr
           }.\eqno(5.10)$$
Taking into account a visibility function one finds
$$\overline{C_l^\oplus}
=36\pi^2 l^2(l+1)^2\left({H_0\over a}\right)^4
  \int_0^\infty{dk\over k^2}\,|\overline{I_l}(k)|^2\,P_\delta(k)
\eqno(5.11)$$
where
$$\overline{I_l}(k)=\int_0^\infty dr\,V(r)\,I_l(kr)
.\eqno(5.12)$$
From the asymptotic series of eqs~(A4\&5) we see that
$$\lim_{k\rightarrow0}\overline{I_l}(k)={k^l\overline{r^l}\over(2l+1)!!\,(l+1)}
\qquad \overline{r^l}\equiv\int_0^\infty dr\,V(r)\,r^l \qquad
\lim_{k\rightarrow\infty}\overline{I_l}(k)=
\lim_{k\rightarrow\infty}          I_l (k)=
               {\sqrt{\pi}\over2l}{\Gamma({l+2\over2})\over\Gamma({l+3\over2})}
,\eqno(5.13)$$
i.e. $\overline{I_l}(k)$ goes grows like $k^l$ for small $k$ and goes to a
constant for large $k$.  Most visibility functions one might consider are
peaked around a typical distance which we denote by $r_*$.  The transition
region between these two asymptotic forms of eq~(5.13) occurs at $kr_*\sim l$.
In general $\overline{I_l}(k)$ is a monotonically increasing function of $k$.


\subheader{Deep Large-Angle Shear}

	Let us define the spectral index $n$ by
$$n\equiv k{d\over dk}P_\delta(k)
.\eqno(5.14)$$
Comparing the asymptotic form of eq~(5.11) with eq~(5.13) we see that the
$C_l^\oplus$ is well defined if $n<1$ for large $k$ and $n>1-2l$ for small $k$.
It seems likely in our own universe that $n\simlt-1$ at large $k$ while
$n\sim1$ for small $k$ so there is no convergence problem. 

	In most cases the dominant contribution to $\overline{C_l^\oplus}$ is
predominantly from wavenumbers $k\sim l/r_*$.  The only exception to this rule
that one is likely to encounter in practice is when one considers small $l$ and
deep samples.  This arises because we expect to that on large scales the power
spectrum will go to an $n=1$ Harrison-Zel'dovich, i.e.
$$P(k)\approx A k \qquad k\rHZ\simlt1
.\eqno(5.15)$$
The evidence for this comes only from measurements of large-angle CMBR
anisotropies (Smoot \etal 1992).  Studies of galaxy clustering have not yet
reached large enough scales to see the spectrum approach $n\approx1$.  With
weak lensing, if $r_*/l\simgt\rHZ$, we expect to probe larger scales than
$\rHZ$.  In this case the integral of eq~(5.11) will have it's a largest
contribution from the region where $n\approx1$ and $\overline{I_l}(k)$ is
approximately constant, i.e. the large $k$ limit of $\overline{I_l}(k)$ but the
small $k$ limit of $P_\delta(k)$, and the integral will be approximated by
$$\overline{C_l^\oplus}\approx
36\pi^3 \left({H_0\over a}\right)^4 A\,\left[\ln{r_*\over l\rHZ}\right]
\,\left({\Gamma({l+2\over2})\over\Gamma({l+1\over2})}\right)^2
{\longrightarrow \atop{\scriptstyle l\gg1}}
18\pi^3 \left({H_0\over a}\right)^4 A\,l\,\left[\ln{r_*\over l\rHZ}\right]
 \qquad r_*\gg l\rHZ
.\eqno(5.16)$$
Since the contribution to this part of the integral is only logarithmically
dominant we never expect this to be an extremely accurate approximation.  On
the other hand the this expression is not very sensitive to $\rHZ$ or the way 
the power spectrum turns over from $n=1$ at small scales.  The contribution to
$\overline{C_l^\oplus}$ comes from the wavenumbers in a broad range from
$k\sim l/r_*$ to $k\sim1/\rHZ$ rather than from a narrow range of wavenumbers
near $k\sim l/r_*$.  In our universe it is likely that the lo-$l$ shear
averaged over a depth of a gigaparsecs will receive significant contribution
from scales of a few hundred megaparsecs to gigaparsecs.

\subheader{Volume Limited Samples With Power Law Spectra}

	As an example we compute the angular power spectrum of the shear in a
volume limited sample of galaxies in the case where the density power spectrum
is a power law in wavenumber, i.e.
$$P(k)=A k^n
.\eqno(5.17)$$
For a pure power law both $C_l^\oplus$ and $\overline{C_l^\oplus}$ have an
ultra-violet (large $k$) divergence if $n\ge1$ and an infrared (small $k$)
divergence if $n+2l\le1$.  Combining the visibility function of (4.42) with the
asymptotic approximation of eqs~(A6) or (B9) one finds the approximate result
$$\eqalign{\overline{C_l^\oplus}
&\approx81\pi^3 2^n{l^{3-n}\over(l+1)^{1-n}}
                   {(11-3n)(6-n)\Gamma(1-n)\over\Gamma(8-n)}
                \left({\Gamma({l\over2})\over\Gamma({l+1\over2})}\right)^{4-2n}
                \left({H_0\over a}\right)^4 A r_{\rm max}^{1-n}             \cr
&{\longrightarrow \atop{\scriptstyle l\gg1}}\ 
 324\pi^3(11-3n)(6-n)\,{\Gamma(1-n)\over\Gamma(8-n)}\,l^n\,
                \left({H_0\over a}\right)^4 A r_{\rm max}^{1-n}
           } \qquad    1-2l<n<1
.\eqno(5.18)$$
We illustrate that the error in this approximate result is small in fig~2. We
see that for $n\simlt1$ eq~(5.18) is an excellent approximation for all $l$
while for smaller $n$ there are only significant corrections from small $l$.
In our universe the density power spectrum is not an exact power law but may be
approximately so at small and large scales. It is probable than on scales
$\ll100h^{-1}$Mpc that $n\approx-1$ while at large scales $n\approx1$.  We see
that eq~(5.18) diverges as $n\rightarrow1$.  This logarithmic divergence
is regulated by a cutoff at small scales as explained above.

	One can argue that the asymptotic approximation of eqs~(A6) and (B9) is
a very good approximation when applied to our own universe.  This follows from
the fact that for deep galaxy catalogs it is the large scales which are
relevant for the small $l$.  At these large scales $n\approx1$, which is where 
eq~(5.18) is extremely accurate for all $l$.  It is true that this asymptotic
approximation may be off by a $\sim30$\% at $l=2$ for extremely shallow
surveys, but as we shall see below one needs a fairly deep survey to measure
the shear.

\subheader{Small-Angle Limit}

	The small-angle shear has been studied extensively elsewhere and
provides a useful check on our results so far.	The large-$l$ limit of $I_l$ in
eq~(A7) is appropriate for small angle correlation functions.  Taking this
limit we find that (5.9) becomes
$$C_l^\oplus(r,r')\approx 18\pi^3l\,\left({H_0\over a}\right)^4
  \int_{\max({l\over r},{l\over r'})}^\infty{dk\over k^2}\,
  \left(1-{l\over kr }\right)\,
  \left(1-{l\over kr'}\right)\,P_\delta(k) \qquad l\gg1
.\eqno(5.16)$$
Thus eq~(4.37) becomes
$$\langle\gamma^2\rangle\approx\langle\kappa^2\rangle
\approx{1\over2\pi}\int_0^\infty dl\,l\,C_l^\oplus(r,r)      
={3\pi^2\over10}\left({H_0\over a}\right)^4r^3\int_0^\infty dk\,k\,P_\delta(k)
\eqno(5.17)$$
and using
$${H_0 r\over a}=2\left(1-{1\over\sqrt{1+z}}\right)
\eqno(5.18)$$
we finds that the rms shear at the redshift $z$ is
$$\sqrt{\langle\kappa^2\rangle}=\sqrt{\langle\gamma^2\rangle}
=2\pi\left(1-{1\over\sqrt{1+z}}\right)^{3\over2}
     \left[{3\over5}{H_0\over a}\int_0^\infty dk\,k\,P(k)\right]^{1\over2}
\eqno(5.19)$$
which agrees with eq~(83) of Blandford \etal (1991) and eq~(2.3.6) of
Kaiser~(1992).  Note that the polarization, $p$, used in these papers is twice
the shear, $\gamma$, used here.  Furthermore Blandford \etal (1991) uses units
where $H_0/a=1$ while Kaiser~(1992) uses units where ${1\over2}H_0/a=1$.
Finally note that $P(k)$ in Kaiser~(1992) is defined to be a factor $(2\pi)^3$
larger than that used here, while Blandford \etal (1991) use the same Fourier
convention as here.

\subheader{Distance-Limited Sample at Small Angles}

	Now let us consider the average shear in a spatially uniform sample
going to a maximum distance $\rmax$, i.e. use the visibility function of
eq~(4.42). In the small angle limit we obtain
$$\overline{C_l^\oplus}\approx 18\pi^3l\,\left({H_0\over a}\right)^4
  \int_{l\over \rmax}^\infty{dk\over k^2}\,
  \left(1+{l^3\over2(k\rmax)^3}-{3l\over2k\rmax}\right)^2\,P_\delta(k)
\eqno(5.20)$$
and thus
$$\eqalign{\langle\overline{\gamma}^2\rangle
\approx    \langle\overline{\kappa}^2\rangle
\approx{1\over2\pi}\,\int_0^\infty dl\,l\,\overline{C_l^\oplus}
=&{17\pi^2\over140}\,\left({H_0\over a}\right)^4\rmax^3
                                           \int_0^\infty dk\,k\,P_\delta(k) \cr
=&{34\pi^2\over35}\,\left(1-{1\over\sqrt{1+\zmax}}\right)^3{H_0\over a}
                                           \int_0^\infty dk\,k\,P_\delta(k) \cr
           }.\eqno(5.21)$$
Comparing (5.21) to (5.17) we see that the average shear in a distance limited
sample is roughly 0.64$\ldots$ times the shear at the edge of the sample.

\subheader{Power Law Power Spectra for a Distance-Limited Sample}

	Now let us also specify the power spectrum, choosing a power-law
overdensity correlation function 
$$\xi(r)\equiv\langle\delta(\bfx)\delta(\bfx')\rangle
             =\left({r_0\over r}\right)^{3+n} \qquad r=|\bfx-\bfx'|
\eqno(5.22)$$
which is equivalent to the power-law power spectrum
$$P_\delta(k)
   ={r_0^{3+n}\Gamma(-{n\over2})\over2^{3+n}\pi\sqrt{\pi}\Gamma({3+n\over2})}\,
   k^n \qquad -3<n<0
.\eqno(5.23)$$
Substituting this into (5.20) we obtain
$$l^2\overline{C_l^\oplus}\approx
324\pi\sqrt{\pi}\,(11-3n)\,(6-n)\,{1\over l}\,
   {\Gamma(-{n\over2})\Gamma(1-n)\over\Gamma({3+n\over2})\Gamma(8-n)}\,
                     \left({H_0\rmax\over a}\right)^4
                     \left({r_0l\over2\rmax}\right)^{3+n} \qquad -3<n<0 
.\eqno(5.24)$$
which is equivalent to 2nd expression in (5.18). The reason for the divergence
at $n=0$ is because the correlation length, $r_0$, is ill-defined for $n\ge0$.
Apart from this $n$-dependent divergence of the correlation length the
$l$-dependence is correct for $n<1$.  For power law spectra the shear two-point
correlation is given by
$$\eqalign{
C_{\overline{\kappa}}(\vartheta)
&\approx{81\sqrt{\pi}\over2}\,(11-3n)(6-n)\,{\Gamma(1-n)\over\Gamma(8-n)}\,
                     {\Gamma({2+n\over2})\over\Gamma({3+n\over2})}\,
                     \left({H_0r_0\over a}\right)^{3+n}
                     \left({H_0\rmax\over a}\right)^{1-n}
                     {1\over\vartheta^{2+n}}                                \cr
C_{\overline{\gamma}}(\vartheta,\varphi,\varphi')
&\approx C_{\overline{\kappa}}(\vartheta)\,
\left(\cos2(\varphi-\varphi')
      +{(2+n)(4+n)\over n(n-2)}\,\cos2(\varphi+\varphi')\right)
           }\qquad -2<n<0.\eqno(5.25)$$
The correlation functions diverge for $n<-2$ due to an infrared divergence.
The integrals which give these correlation functions are formally convergent at
large $l$ for any $n$, however for $n>-1$ the contribution to  the integral is
dominated by the behaviour of the integrand at $l\gg1/\vartheta$.  Thus for
large $n$ the results obtained are sensitive to our assumption of exact power
law behaviour over a broad range of scales.

One must be careful about evaluating these expressions for integer values of
$n$.  While it may appear that there is a singularity at $n=0$, note that
$C_{\overline{\kappa}}$ is zero at $n=0$, and $C_{\overline{\gamma}}$ is
smooth at $n=0$.  Note that these power-law spectra cause the correlation
function be be singular at zero separation which leads to certain pathologies.
For example $C_{\overline{\kappa}}$ is negative for all $\vartheta$ if $n>0$
while the coefficient of $\cos2(\varphi+\varphi')$ in $C_{\overline{\gamma}}$
has a positive divergence as $\vartheta\rightarrow0$.  If we were to regulate
this divergence at small scales we would find that $C_{\overline{\kappa}}$
would be positive and finite at $\vartheta=0$ while the coefficient of
$\cos2(\varphi+\varphi')$ in $C_{\overline{\gamma}}$ would go to zero.

	While the power spectrum of mass fluctuations may indeed be a power law
down to very small scales, we can only measure shear averaged over a finite
region of the sky and this averaging decreases our sensitivity to the the
divergent density fluctuations at large scales.  On small scales, the
convolution of the shear pattern with a Gaussian beam corresponds to
multiplying the angular power spectrum by $\exp(-l^2\sigma^2)$ where $\sigma$
is the ``Gaussian width'' of the beam, which is related to the FWHM by
$\sigma_{\rm fwhm}=2\sqrt{2\ln2}\,\sigma$.  For a Gaussian beam we obtain
$$\overline{C_{\overline{\kappa}}}(\vartheta)
\approx81\sqrt{\pi}\,(11-3n)\,(6-n)\,
         {\Gamma({2+n\over2})\Gamma(-{n\over2})\Gamma(1-n)
                       \over\Gamma({3+n\over2})\Gamma(8-n)}\,
         \left({H_0\rmax\over a}\right)^4
         \left({r_0\over2\rmax}\right)^{3+n}
         {}_1F_1({2+n\over2};1;-{\vartheta^2\over4\sigma^2})\,
         {1\over\sigma^{2+n}}
\eqno(5.26)$$
where ${}_1F_1$ is a generalized hypergeometric function.  Here we have used
the notation $\overline{C_{\overline{x}}}(\vartheta)$ to indicate that we are
taking the correlation of the smoothed quantities.  Simpler expressions can be
found for the variance of zero-lag (i.e. $\vartheta=0$) smoothed quantities :
$$\eqalign{
&\overline{C_{\overline{\kappa}}}(0)
\approx81\sqrt{\pi}\,(11-3n)\,(6-n)\,l^{1+n}
       {\Gamma({2+n\over2})\Gamma(-{n\over2})\Gamma(1-n)
                     \over\Gamma({3+n\over2})\Gamma(8-n)}\,
       \left({H_0\rmax\over a}\right)^4
       \left({r_0\over2\rmax}\right)^{3+n}{1\over\sigma^{2+n}}        \cr
&\overline{C_{\overline{\gamma}}}(0,\varphi,\varphi')\approx
 \overline{C_{\overline{\kappa}}}(0)\,\cos2(\varphi-\varphi') 
         } \qquad -2<n<0
.\eqno(5.27)$$

\header{6. Accuracy of Shear Measurements}

	When the shear is weak, i.e. when $\gamma\ll1$, the way one goes about
measuring shear is by looking for alignments in the apparent orientation of
background galaxies.  The implicit assumption is that there is no significant
intrinsic alignment of galaxy ellipticities and any apparent alignment must be
due to a coherent shear in the image. As discussed above this is not liable to
be exactly true but is liable to be a excellent approximation.  One can measure
the apparent shape of galaxy, $g$, using the ellipticity tensor (Kaiser 1992)
which is
$$e^g_{ab}={Q^g\,Q^g_{ab}-Q^g_aQ^g_b\over Q^g\,Q^g_{cc}-Q^g_cQ^g_c}
      -{1\over2}\delta_{ab}
       ={1\over2}e^g\left(\eqalign{\cos2\chi^g\ &\phantom{-}\sin2\chi^g   \cr
                                   \sin2\chi^g\ &         - \cos2\chi^g}\right)
\eqno(6.1)$$
where $Q^g$, $Q^g_b$, $Q^g_{ab}$ give the 0th, 1st, and 2nd moments of the
galaxy brightness distribution on the sky.  Here $\chi^g$ is position angle of
the galaxy and $e^g\in[0,1]$ it's the ellipticity.  If the true galaxy position
angles are randomly oriented then the expected ellipticity is
$$\langle e^g_{ab}\rangle_\rmg=(1-{1\over2}\overline{e^2})\,\gamma_{ab}
\eqno(6.2)$$
in the weak lensing approximation. Here $\langle\cdots\rangle_\rmg$ indicates
an ensemble average over galaxy orientations, and $\overline{e^2}$ gives the
mean square ellipticity. Given the range of possible ellipticities the
proportionality constant between ellipticity and shear varies by only a factor
of 2 between perfectly circular galaxies, $e^g=0$ and highly 
elongated galaxies, $e^g=1$.  The variation from the expected value is
$$\langle(e^g_{ab}   -\langle e^g_{ab}   \rangle_\rmg)
         (e^{g'}_{cd}-\langle e^{g'}_{cd}\rangle_\rmg)\rangle_\rmg
={1\over4}\overline{e^2}\delta_{gg'}(        g_{ac}       g_{bd}
                                     +\epsilon_{ad}\epsilon_{bc})
\eqno(6.3)$$
where again $g_{ac}$ is the metric on the sphere of the sky and $\epsilon_{ab}$
the Levi-Civita symbol.  For perfectly circular galaxies, $e^g=0$, the apparent
position angle and shape gives the shear precisely, while for non-circular
galaxies there is considerable variance.  Typical intrinsic galaxy shapes have
$\overline{e^2}\sim(0.3)^2$. Note that if one makes {\it random} measurement
errors in the shapes of galaxies then one should add this in quadrature to the
deviation from the expected value in eq~(6.3).  Given the fairly large
intrinsic error due to galaxy non-circularness one would have to make rather
large measurements errors to significantly increase ones uncertainty in the
determination of the shear.  Roughly speaking one needs
$N\sim\overline{e^2}/[8(1-{1\over2}\overline{e^2})^2\gamma^2]$ objects to
obtain a determination of the mean shear in a patch of the sky with S/N$\sim1$
if one has perfect measurements. Since typically one is looking for
$\gamma\sim0.01$ one needs significantly more than 100 perfectly measured
galaxies or even more when one includes measurement errors.  Of course the main
problem in practice will be non-random measurement errors, i.e. errors in
galaxy shapes which are correlated between different galaxies.

	Below we will consider estimating the shear from a uniform all-sky
survey of galaxies.  We will allow for an arbitrary weighting of different
galaxy types, given by a parameter $w_g$ for each galaxy.  This weighting might
depend on the apparent shape (not orientation!) of the galaxies, the colors, or
the apparent magnitude.  Given this weighting there are a variety of different
definitions of the number of galaxies in one survey, in particular one may
define it with different powers of $w_g$, i.e.
$$N_{\rmg n}={\left\langle\sum_g w_g^n\right\rangle_\rmg\over4\pi}
   =\left\langle\sum_g w_g^n\,\delta^{(2)}(\hatbfn-\hatbfn_g)\right\rangle_\rmg
\eqno(6.4)$$
where $\delta$-functions are normalized such that 
$$\int d^2\hatbfn\,\delta^{(2)}(\hatbfn-\hatbfn_g)=1
.\eqno(6.5)$$
and $\langle\cdots\rangle_\rmg$ indicates an ensemble average over galaxy
positions and orientations, but not over realizations of the shear
distribution.  This is only possible because we will be assuming that the
galaxy distribution and shear distribution are independent.  We do not expect
this to be exactly correct, however it is probably a fairly good approximation
since the galaxies which sheared are so far from the galaxies associated with
the mass-distribution which is doing the shearing, that the correlations are
fairly weak.

	An estimator of the radially weighted shear is given by
$$\widehat{\overline{\gamma}}_{ab}(\hatbfn)
          ={\sum_g w_g e^g_{ab}\delta^{(2)}(\hatbfn-\hatbfn_g)\over
            N_{\rmg1}\left(1-{1\over2}\overline{e^2}\right)}
.\eqno(6.6)$$
One should, of course, not take the $\delta$-functions as meaning the shear in
concentrated in small patches of the sky, rather one should average this 
estimator over an angular patch large enough to contain many galaxies, and
thereby obtain an estimator of the angle-averaged shear.  Using the weak
lensing formula, (6.2), we find that
$$\langle\widehat{\overline{\gamma}}_{ab}(\hatbfn)\rangle_\rmg
=\left\langle\gamma_{ab}(r_g,\hatbfn)\right\rangle_\rmg\,
  {\left\langle\sum_g w_g\,\delta^{(2)}(\hatbfn-\hatbfn_g)\right\rangle_\rmg
   \over N_{\rmg1}}
=\int_0^\infty dr V(r)\,\gamma_{ab}(r,\hatbfn)
=\overline{\gamma}_{ab}(\hatbfn)
\eqno(6.7)$$
where we have defined the visibility function, $V(r)$:
$$V(r)={\left\langle\sum_g w_g \delta(r-r_g)\right\rangle_\rmg
                                           \over4\pi N_{\rmg1}}
.\eqno(6.8)$$
Here we have assumed isotropy and homogeneity on the sky, so that the radial
and angular distribution of galaxies factorize.  This shows that
$\widehat{\overline{\gamma}}_{ab}(\hatbfn)$ is an unbiased estimator of
$\overline{\gamma}_{ab}(\hatbfn)$.  Of course one needs to know the
distribution of galaxies in ones sample to determine $V(r)$ and thus to know 
what $\widehat{\overline{\gamma}}_{ab}(\hatbfn)$ is an unbiased estimator for.
In the weak lensing limit we may use eq~(6.3) to determine the mean square
error in this estimator, which is
$$\left\langle(\widehat{\overline{\gamma}}_{ab}(r ,\hatbfn )
                       -\overline{\gamma}_{ ab}(r ,\hatbfn ))
            \,(\widehat{\overline{\gamma}}_{cd}(r',\hatbfn')
                       -\overline{\gamma}_{ cd}(r',\hatbfn'))\right\rangle_\rmg
={N_{\rmg2}\over4{N_{\rmg1}}^2}\,{\overline{e^2}
         \over\left(1-{1\over2}\overline{e^2}\right)^2}\,
     (g_{ac}g_{bd}+\epsilon_{ac}\epsilon_{bd})\,\delta^{(2)}(\hatbfn -\hatbfn')
.\eqno(6.9)$$
Comparing this correlation function with $C_{\overline{\gamma}}$ of \S4 we find
the correlation function of the error in the estimator of eq~(6.6):
$$C^{\rm sn}_{\overline{\gamma}}(\vartheta,\varphi,\varphi')
={N_{\rmg2}\over4{N_{\rmg1}}^2}\,{\overline{e^2}
 \over\left(1-{1\over2}\overline{e^2}\right)^2}\,\cos2(\varphi-\varphi')
 \,\lim_{\varepsilon\rightarrow0}\delta(1-\varepsilon-\cos\vartheta)
\eqno(6.10)$$
where the superscript ``sn'' refers to the {\it shot noise} from the finite
number of galaxies.  We may also compute the power spectrum for the shot noise
by substituting $C^{\rm sn}_{\overline{\gamma}}$ into (4.31):
$$C_l^{\oplus \rm sn}=C_l^{\otimes\rm sn}
={\pi\over4}\,{l(l+1)\over(l+2)(l-1)}\,
  {N_{\rmg2}\over{N_{\rmg1}}^2}\,
  {\overline{e^2}\over\left(1-{1\over2}\overline{e^2}\right)^2}\,
.\eqno(6.11)$$
We see that the shot noise contributes equally to the scalar and pseudo-scalar
part of the error.  The shot noise is white noise, i.e. on small scales
($l\gg1$) both $C_l^{\oplus\rm sn}$ and $C_l^{\otimes\rm sn}$ become
$l$-independent. Substituting (6.11) into (4.27) we may compute the mean
square error in the shear for all modes with $l<L$ is
$$\langle\overline{\gamma^{\rm sn}_{l\le L}}^2\rangle
={N_{\rmg2}\over8{N_{\rmg1}}^2}\,
 {\overline{e^2}\over\left(1-{1\over2}\overline{e^2}\right)^2}\,
 \sum_{l=2}^L(2l+1)
={N_{\rmg2}\over8{N_{\rmg1}}^2}\,
 {\overline{e^2}\over\left(1-{1\over2}\overline{e^2}\right)^2}\,
 (L+3)(L-1) \qquad L\ge2
.\eqno(6.12)$$
At small angles the mean square shot-noise in a Gaussian beam is
$$\langle\overline{\gamma^{\rm sn}}^2\rangle\approx
  {N_{\rmg2}\over8{N_{\rmg1}}^2}\,
  {\overline{e^2}\over\left(1-{1\over2}\overline{e^2}\right)^2}\,
  \int_0^\infty dl\,2l\,e^{-l^2\sigma^2}
 =\sqrt{\pi\ln2}\,{N_{\rmg2}\over{N_{\rmg1}}^2}\,
  {\overline{e^2}\over\left(1-{1\over2}\overline{e^2}\right)^2}\,
  {1\over\sigma_{\rm fwhm}^2}
.\eqno(6.13)$$
where we have used $\sigma_{\rm fwhm}=2\sqrt{2\ln2}\,\sigma$.

\subheader{Estimators for $C_l^\oplus$ and $C_l^\oplus$}

	A sum of unbiased estimators of some quantities is itself an unbiased
estimator of the sum.  Thus combining (6.6) with (4.12) we can construct
unbiased estimators of the spherical harmonic amplitudes which describe the
shear field
$$\eqalign{
\widehat{\phi_{\s(l,m)}^\oplus }^\rmg=-&2\,{(l-2)!\over(l+2)!}\,
    {1\over N_{\rmg1}\left(1-{1\over2}\overline{e^2}\right)}
    \sum_g w_g\,Y_{\s(l,m)}^*{}^{:ab}(\hatbfn_g)\,e^g_{ab}                  \cr
\widehat{\phi_{\s(l,m)}^\otimes}^\rmg=-&2\,{(l-2)!\over(l+2)!}\,
    {1\over N_{\rmg1}\left(1-{1\over2}\overline{e^2}\right)}
    \sum_g w_g\,Y_{\s(l,m)}^*{}^{:ac}(\hatbfn_g)\,e^g_{ad}\,\epsilon_c{}^d
           }.\eqno(6.14)$$
The superscript g emphasizes that these are estimators derived from a galaxy
sample. One can construct estimators of the $C_l^\oplus$ and $C_l^\otimes$
using eq~(4.13)
$$\widehat{C_l^\oplus }^\rmg=-C_l^{\oplus \rm sn}+{l^2(l+1)^2\over4(2l+1)}
  \sum_{m=-l}^l\left|\widehat{\phi_{\s(l,m)}^\oplus }^\rmg\right|^2 \qquad
  \widehat{C_l^\otimes}^\rmg=-C_l^{\otimes\rm sn}+{l^2(l+1)^2\over4(2l+1)}
  \sum_{m=-l}^l\left|\widehat{\phi_{\s(l,m)}^\otimes}^\rmg\right|^2
.\eqno(6.15)$$
These are unbiased because the shot noise has been subtracted and differ from
those in eq~(4.13) since those were estimators derived from perfect knowledge
of the shear pattern on the sky, while these include additional uncertainties
due to finite galaxy sampling.  Of course real surveys will probably never
cover the entire sky and other estimators must constructed to deal with finite
sky coverage.

	The mean square deviation of $\widehat{C_l^\oplus }^\rmg$ and
$\widehat{C_l^\otimes}^\rmg$ from $C_l^\oplus$ and $C_l^\otimes$ will depend
on the 4-point correlation function of the shear as well as the clustering
properties of the galaxies.  We will not go into such detail here, instead we
assume Gaussian statistics, i.e. set the reduced 4-point function to zero, and
assume Poisson sampling of the shear field.  These approximations are liable to
be good ones for deep samples on large angular scales.  In any case, better
estimates of the uncertainties should be computed. With these assumptions
we find
$$\eqalign{
\left\langle\left(\widehat{C_{l }^\oplus }^\rmg-C_{l }^\oplus \right)\,
            \left(\widehat{C_{l'}^\oplus }^\rmg-C_{l'}^\oplus \right)
              \right\rangle
&=\delta_{ll'}\,{2\over2l+1}\,({C_{l }^\oplus}^2+{C_l^{\oplus \rm sn}}^2)   \cr
\left\langle\left(\widehat{C_{l }^\otimes}^\rmg-C_{l }^\otimes\right)\,
            \left(\widehat{C_{l'}^\otimes}^\rmg-C_{l'}^\otimes\right)
              \right\rangle
&=\delta_{ll'}\,{2\over2l+1}\,({C_{l }^\otimes}^2+{C_l^{\otimes\rm sn}}^2)  \cr
\left\langle\left(\widehat{C_{l }^\oplus }^\rmg-C_{l }^\oplus \right)\,
            \left(\widehat{C_{l'}^\otimes}^\rmg-C_{l'}^\otimes\right)
              \right\rangle
&=0        }\eqno(6.16)$$
so that the 1-$\sigma$ fractional uncertainty is
$$     {\Delta{C_l}^\oplus \over C_l^\oplus}=\sqrt{{2\over2l+1}\,
               \left(1+{{C_l^{\oplus \rm sn}}^2\over{C_{l }^\oplus }^2}\right)}
\qquad {\Delta{C_l}^\otimes\over C_l^\otimes}=\sqrt{{2\over2l+1}\,
               \left(1+{{C_l^{\otimes\rm sn}}^2\over{C_{l }^\otimes}^2}\right)}
.\eqno(6.17)$$
The 1st term in the radical represents cosmic variance which is more dominant
for small $l$, while the second is uncertainty due to shot noise which is more
dominant at large $l$.  With the above estimate of the error one can define a
total signal-to-noise:
$$\eqalign{
\left({\rmS^\oplus \over\rmN^\oplus }\right)^2
  &=\sum_l\left({C_l^\oplus \over\Delta C_l^\oplus }\right)^2 
   =\sum_l{2l+1\over2}\,{{C_l^\oplus }^2\over
                         {C_l^\oplus }^2+{C_l^{\oplus \rm sn}}^2}           \cr
\left({\rmS^\otimes\over\rmN^\otimes}\right)^2
  &=\sum_l\left({C_l^\otimes\over\Delta C_l^\otimes}\right)^2
   =\sum_l{2l+1\over2}\,{{C_l^\otimes}^2\over
                         {C_l^\otimes}^2+{C_l^{\otimes\rm sn}}^2}
           }\eqno(6.18)$$
which should be greater than unity in the range of $l$ of interest if one is
capable of detecting a significant shear.  Note that these signal-to-noise
ratios include cosmic variance.

\header{7. A Realistic Example}

	Now let us estimate the angular power spectrum of the shear we might
expect to find in our own universe.  Clearly there are large uncertainties in
the distribution in mass in our universe so we cannot predict precisely what
one will see.  In fact the utility of weak lensing is to measure the mass
distribution.  Here we give a rather simple empirically based model, which is
meant to be optimistic about the size of the shear which one might find.  We
are also not very realistic about the data one is liable to collect.  For
example we assume an all-sky volume limited sample of galaxies, while one is
really liable to obtain a partial-sky magnitude-limited sample.  More detailed
studies clearly should be done.

	On small scales the correlation function of optically selected galaxies
appears to be approximately a power law (Peebles 1980) as was assumed for the
mass in eq~(5.22-27).  Taking parameters. $r_0=5\,h^{-1}$Mpc and $n=-1.2$,
which are close to the values obtained observationally (Humit \etal 1996) we
find
$$\eqalign{C_{\overline{\kappa}}(\vartheta)
&\approx(0.0180)^2        \left(1-{1\over\sqrt{1+\zmax}}\right)^{2.2}
                          \left({1^\circ\over\vartheta}\right)^{0.8}
 \approx(0.0084)^2 \zmax^{2.2}\left({1^\circ\over\vartheta}\right)^{0.8}    \cr
C_{\overline{\gamma}}(\vartheta,\varphi,\varphi')
&\approx C_{\overline{\kappa}}(\vartheta)\,\left(\cos2(\varphi-\varphi')
                                        +0.5833\,\cos2(\varphi+\varphi')\right)
                                                                            \cr
\overline{C_{\overline{\kappa}}}(0)
&\approx(0.0235)^2        \left(1-{1\over\sqrt{1+\zmax}}\right)^{2.2}
                          \left({1^\circ\over\sigma_{\rm fwhm}}\right)^{0.8}
 \approx(0.0110)^2\zmax^{2.2}\left({1^\circ\over\sigma_{\rm fwhm}}\right)^{0.8}
                                                                            \cr
\overline{C_{\overline{\gamma}}}(0,\varphi,\varphi')
&\approx\overline{C_{\overline{\kappa}}}(0)\,\cos2(\varphi-\varphi')        \cr
l^2\overline{C_l^\oplus}
&\approx\left({l\over1.8\times10^5}\right)^{0.8}
                                    \left(1-{1\over\sqrt{1+\zmax}}\right)^{2.2}
 \approx\left({l\over1.2\times10^6}\right)^{0.8}\zmax^{2.2}
           }.\eqno(7.1)$$
where expressions proportional to $\zmax^{2.2}$ are appropriate for
$\zmax\ll1$.  This power law behaviour certainly does not persist to very
large scales and it is generally believed that the spectrum turns over to
something close to a Harrison-Zel'dovich spectrum ($n=1$) at large scales.  As
a model of the power spectrum of density inhomogeneities on both small and
large scales we will use
$$P_\delta(k)
   ={A k\over(1+(k\rHZ)^2)^{1.1}}\qquad
A={\rHZ^{2.2}r_0^{1.8}\Gamma(0.6)\over2^{1.8}\pi\sqrt{\pi}\Gamma(0.9)}\,
.\eqno(7.2)$$
where we again use $r_0=5\,h^{-1}$Mpc and $n=-1.2$ and supplement it with
$\rHZ=38.8h^{-1}$Mpc which matches the COBE normalization for adiabatic initial
conditions.


	In fig~4 is plotted $C_l^\oplus$ from this model for various survey
depths, $\zmax$, and compare this with the result we would obtain if we used
the asymptotic approximation of eq~(5.9) combined with pure  power law spectrum
of eqs~(5.22-24) on all scales. This figure illustrates that the approximations
used to derive eq~(7.1) are accurate on all angular scales for shallow surveys,
and on small angular scales power law model for deeper surveys. In fig~5 we
plot the rms shear contributed by angular scales larger than a given scale, as
a function of the minimal scale, again for various values of depth.  On the
largest angular scales one may only find a shear of $\sim10^{-4}$.  We will
argue below that there will be no significant signal for $\zmax\simlt0.1$ so we
see that for interesting surveys one is looking for a shear
$\simgt3\times10^{-4}$. For such a level of shear one needs $\simgt10^5$
galaxies with good shape information to detect it.  One might need many more if
the galaxy shapes have large statistical or systematic errors.  Of course as
one goes deeper one needs fewer galaxies.  One critical issue is whether one is
able to make absolute measurements of the shear.  If one can then one is
sensitive to all angular scale larger than the area one surveys.  Absolute
measurements are possible in principle as one can use stellar images to correct
for instrumentally or atmospherically induced shear (Kaiser, Squires, and
Broadhurst 1995).  If for some reason one is limited to differential
measurements of shear then for small survey areas one loses much of the signal.
Except for small $l$ the $\gamma_{\rm rms}\approx\kappa_{\rm rms}$ so that
fig~5 also gives the image amplification. Broadhurst, Taylor, and Peacock
(1995) have proposed a methodology to statistically measure amplification due
to gravitational lensing and applied it in Broadhurst (1995).

	Some will argue that we the simple mass model used here is probably an
overestimate of the magnitude of the shear. The author considers it an
optimistic estimate, but not wildly so.  For non-adiabatic perturbations the
large scale density inhomogeneities are less than assumed here.  Furthermore it
has long been argued that the mass inhomogeneities on the $8\,h^{-1}$Mpc scale
are smaller than the galaxy inhomogeneities (e.g. Henry \& Arnaud 1991, White,
Efstathiou, and Frenk 1993), so that we may also be overestimating the lensing
on small scales as well.  Furthermore the model power spectrum used here
clearly has some contribution from non-linear clustering, which is not quite
consistent with the linear evolution of density perturbations we have assumed
in \S5.  Corrections for the non-linear evolution of the power spectrum will be
important for deep samples, $z\simgt0.5$, on small angular scales. If galaxies
are biased tracers of the mass distribution or if the universe is not flat
then, with fixed $\delta\rho/\rho$, the lensing shear at small redshifts scales
proportionate to $\Omega_0/b$, where $b$ is the bias factor.  The reader may
wish to apply this correction using his favorite values, which for many
probably means reducing the predicted shear by at most a factor of 2 on small
angular scales.

	Here is considered a volume limited survey, up to redshift $\zmax$, and
assume no evolution in the number density, which is given by $\Phi^\star_\rmg$,
and use uniform weighting of all galaxies within the survey volume, i.e.
$w_g=1$.  Using eq~(6.8) we find
$$N_\rmg\equiv N_{\rmg1}=N_{\rmg2}=\Phi^\star_\rmg\,
             \left[{2c\over H_0}\left(1-{1\over\sqrt{1+\zmax}}\right)\right]^3
         =2.6\times10^9\left(1-{1\over\sqrt{1+z}}\right)^3
\eqno(7.3)$$
where we have taken $\Phi^\star_\rmg=0.012\,h^3/\Mpc^3$ which is the value of
this parameter in the Schechter which describes observed galaxies (Loveday
\etal 1992).  Taking $\overline{e^2}\approx(0.3)^2$ we find
$$C_l^{\oplus \rm sn}=C_l^{\otimes\rm sn}
=3.0\times10^{-11}\,{l(l+1)\over(l+2)(l-1)}\,
                                         \left(1-{1\over\sqrt{1+z}}\right)^{-3}
\eqno(7.4)$$
so that
$$\sqrt{\langle\overline{\gamma^{\rm sn}_{l\le L}}^2\rangle}
=6.9\times10^{-6}\,\sqrt{(L+3)(L-1)}\,
                   \left(1-{1\over\sqrt{1+z}}\right)^{-{3\over2}}
.\eqno(7.5)$$
The minimum shot noise occurs when one includes only the quadrupole and at
maximum depth, i.e. $L=2$ and $z=\infty$, which yields 
$\sqrt{\langle\overline{\gamma^{\rm sn}_{l=2}}^2\rangle}=4.8\times10^{-6}$.
This is extremely small, much smaller than the expected signal on these scales.
While going to very large redshifts may never be feasible, going to $z=1$,
which seems feasible, brings this up to $3.0\times10^{-5}$.  Measurements to
date have detected shear at the level of a few percent.  We see that by doing
very large area and very deep surveys one may in principle measure the shear
down to a level up to 1000 times smaller.  One may run into uncorrectable
systematic errors long before one reaches this level, however it is certainly
worth the effort get as close to the shot noise limit as possible.

\subheader{Comparison of Shot Noise With Signal}

	It is interesting to compare the shot-noise ``signal'' with that which
one might expect to obtain from density fluctuations.  Of course density
fluctuations produce no pseudo-scalar shear, so we need only compare the scalar
component of the shear.  Comparing eq~(5.20) with eq~(6.11) we find
$$\eqalign{{\overline{C_l^\oplus}\over C_l^{\oplus \rm sn}}
&=144\pi\,{(l+2)!\over(l-2)!}\,{{N_{\rmg1}}^2\over N_{\rmg2}}\,
  {\left(1-{1\over2}\overline{e^2}\right)^2\over\overline{e^2}}\,
  \left({H_0\over a}\right)^4
             \int_0^\infty{dk\over k^2}\,|\overline{I_l(k)}|^2\,P_\delta(k) \cr
&{\longrightarrow \atop{\scriptstyle l\gg1}}
  {9\over2}\pi\,l^3\,N_\rmg
  {\left(1-{1\over2}\overline{e^2}\right)^2\over\overline{e^2}}\,
  \left({H_0\over a}\right)^4\int_{l/r_{\rm max}}^\infty{dk\over k^2}\,
  \left({2\over l}+{l^2\over(kr_{\rm max})^3}-{3\over kr_{\rm max}}\right)^2
                                                                 P_\delta(k)
           }\eqno(7.6)$$
where we have used eq~(A7) for the small angle limit and eq~(5.20) for the
volume-limited survey.  This ratio is plotted in fig~6 and we see that for our
model this signal-to-noise exceeds unity in each angular wavenumber, $l$, for
all-sky surveys with $\zmax\simgt0.2$.  Of course, as one decreases the sky
coverage this signal-to-noise goes down.

	If fig~7 is plotted the fractional uncertainty in a measurement of
$C_l^\oplus$ according to eq~(6.18).  This includes cosmic variance, i.e. the
additional uncertainty in one's determination of the global average of
quantities due to the sample variance associated from looking at the universe
from one vantage point. From this fractional uncertainty one can compute the 
total signal-to-noise defined in eq~(6.18).  This signal-to-noise includes
from cosmic variance, unlike the ratio of eq~(7.6). It is perhaps most
interesting to consider the ${\rmS^\oplus\over\rmN^\oplus}$ where it is close
to unity.  This leads us to consider shallow all-sky surveys where the
contribution to ${\rmS^\oplus\over\rmN^\oplus}$ from each individual $l$ is
much less than unity. From fig~7 it is clear that
${\rmS^\oplus\over\rmN^\oplus}$ should grow greater than unity for
$0.1\simlt\zmax<0.2$. For this range of $\zmax$ it is clear from fig~4 that the
power law approximations of eqs~(7.1) are a good approximation for nearly all
$l$.  We see from fig~7 that cosmic variance gives a negligible contribution to
$\Delta C_l^\oplus$ for all $l$ with $\zmax$ this small.  Furthermore it is
clear that one can approximate the sum of eq~(6.18) with an integral and take
the large-$l$ limit of both $C_l^\oplus$ and $C_l^{\oplus\rm sn}$.  Combining
all of these approximations we find that for an all-sky survey 
$$\left({\rmS^\oplus \over\rmN^\oplus}\right)^2
\approx\int_2^\infty dl\,l\,{{C_l^\oplus}^2\over{C_l^{\oplus\rm sn}}^2}
\approx \left({\zmax\over0.11}\right)^{10.4} \qquad \zmax\simlt0.2
.\eqno(7.7)$$
The $l$-integral converges relatively slowly, so that half the contribution
comes from $l\le10$ and half from $l>10$. The depth at which this
signal-to-noise exceeds unity is not strongly dependent on the assumed
amplitude of inhomogeneities, for example if we were decrease the density
inhomogeneities by a factor of 2 this would change value of $\zmax$ at which
${\rmS^\oplus \over\rmN^\oplus}$ equals unity from 0.11 to 0.15.  For
fractional sky coverage, $f_{\rm sky}$, we expect ${\rmS^\oplus
\over\rmN^\oplus}\simpropto f_{\rm sky}^{1/2}$, although this depends in detail
on the geometry of the region surveyed.  If one makes absolute shear
measurements on a small patch of the sky then one has limited ability to
determine on which angular scale the shear is generated. The extremely strong
dependence on $\zmax$ exhibited by eq~(7.7) breaks down for $\zmax\simgt0.2$ as
one starts to sample the turnover in the density power spectrum.

\header{8. Summary and Future Development}

	In this paper was presented a formalism of tensor spherical harmonics
on the sphere which may be used to describe a weak lensing shear pattern over
the entire celestial sphere.  Many useful formula are presented in this paper
but little in the form of derivations.  A more complete exposition can be found
on the world wide web at 
http://www-astro-theory.fnal.gov/Personal/stebbins/WeakLens/ .  It was shown
that in the all-sky shear pattern may be decomposed into geometrically distinct
components: scalar shear and pseudo-scalar shear.  In the weak lensing limit
only scalar shear is produced by density inhomogeneities, and this is liable to
be the only significant component of the shear field.  Pseudo-scalar shear may
be produced by vector and tensor perturbations, although the amplitude is
liable to be negligible.  On small scales one may also get significant
pseudo-scalar shear when the lensing becomes strong due to more than one mass
concentration along the line-of-sight.  Given that one expects negligible
pseudo-scalar shear on the sky, it's primary use is liable to be as a gauge of
measurement errors.

	In \S5 the formalism was applied specifically to density
inhomogeneities in an Einstein-deSitter cosmology, and in \S6 to the study the
shot noise one will obtain when trying to measure the shear with a finite
number of galaxies.  In \S7 we apply the formalism derived to a model of the
density perturbations in our universe.  We illustrate that there is, in
principle, a significant signal for wide area galaxy surveys even if one does
not have a very deep galaxy sample.  For example an all-sky survey of galaxies
with redshifts less than 0.2 should exhibit a very significant signal of shear
from gravitational lensing if one could reduce systematic errors below the
$5\times10^{-4}$ level in shear.  A concerted observational effort should be
made to see just how small a shear can be reliably measured with present
technology from the ground. It should be noted that shallow surveys do benefit
from the fact that the galaxies subtend a significantly larger angle than in
deep surveys and thus there is less of a problem with seeing a finite
pixelization.  Given the large signal-to-noise which is present it is clear
that some day such shallow large area shear maps will be made, but one may have
to wait for large space-based optical surveys of the sky to be made.

	The estimates of the expected shear presented in this paper are rather
crude in that they assumed a volume limited survey and did not properly take
into account the evolution of nonlinear clustering. Furthermore the error
estimates assumed Gaussian statistics.  Shallow surveys even over large areas
sample the nonlinear clustering regime which is non-Gaussian.  Proper estimates
of the accuracy with which one can measure the shear should include the proper
sample variance for the non-Gaussian density field. Again further study is
needed.

	Both shear and the linear polarization of light are described by a
traceless symmetric rank-2 tensor on the celestial sphere and most of the
mathematic formulae presented here can equally well be applied to a description
of the linear polarization light on the celestial sphere.  Kamionkowski,
Kosowsky, and Stebbins (1996) have applied the tensor harmonic decomposition to
a description of the cosmic microwave background radiation (CMBR).  Just as
with shear, density inhomogeneities (i.e. scalar modes) cannot, in linear
theory, produce pseudo-scalar linear polarization of light.  However in
contrast to shear, vector and tensor modes can produce polarization of
amplitude comparable to that produced by density inhomogeneities.  Thus one may
interpret any pseudo-scalar component of linear polarization of the CMBR as a
direct detection of vector and/or tensor modes.  This is in contrast to CMBR
anisotropy where one cannot distinguish anisotropy induced by scalar modes from
that produced by vector or tensor.

\smallskip
\noindent {\bf Acknowledgements: }
The author would like to thank Josh Frieman, Donn MacMinn, Uros Seljak, and
Nick Kaiser for useful conversations. Special thanks to Marc Kamionkowski who
has looked over some of the formalism.  This work was supported by the NASA
grant NAG 5-2788.

\bigskip

\header{Appendix: Calculating $I_l(x)$}

	In this paper we have introduced the function 
$$\eqalign{I_l(x)&=\int_0^1 {dy\over y}\,(1-y)\,j_l(xy)                     \cr
&={\sqrt{\pi}\over2^{l+1}\Gamma({3\over2}+l)}\,x^l
\left({1\over l   }\,{}_1F_2({l  \over2};{l+2\over2},{3\over2}+l;-{x^2\over4})
     -{1\over(l+1)}\,{}_1F_2({l+1\over2};{l+3\over2},{3\over2}+l;-{x^2\over4})
     \right)
           }\eqno(\rmA1)$$
where ${}_1F_2$ is a generalized Hypergeometric function which can be written
in terms of sines, cosines, and Sine integrals.  Specific cases are
$$\eqalign{
I_1(x)=&{-2x     + x\,\cos x+\sin x+x^2\Si x\over2x^2}                      \cr
I_2(x)=&{1\over3}-{x\,\cos x-\sin x+x^2\Si x\over2x^3}                      \cr
I_3(x)=&{-16x^3-3x(10-x^2)\,\cos x+3(10+x^2)\,\sin x +3x^4\Si x\over24x^4}  \cr
I_4(x)=&{2\over15}-{3x(14+x^2)\cos x-(42-11x^2)\sin x+3x^4\Si x\over 8x^5}
           }\eqno(\rmA2)$$
where the sine integral function, $\Si$, is 
$$\Si(x)=\int_0^x {dx\over x}\,\sin x
.\eqno(\rmA3)$$
These expressions become increasingly more complicated for large $l$ and can
involve enormous cancelation between terms.  While symbolic algebra program do
provide a generalized Hypergeometric functions, they are generally noisy for
large arguments.  Luckily one can derive accurate methods of calculating
$I_l(x)$ based on the asymptotic expansions of this function. Using the Taylor
series expansion of the spherical Bessel function in the definition of $I_l$ we
find
$$I_l(x)
=\sum_{n=0}^\infty{(-1)^n\,x^{2n+l}\over2^n(2(l+n)+1)!!\,n!\,(2n+l)\,(2n+l+1)}
.\eqno(\rmA4)$$
For large $x$ it we can expand $I_l(x)$ in powers of $1/x$ yielding
$$I_l(x)={\sqrt{\pi}\over l}
         \left( {1\over 2}{\Gamma({l+2\over2})\over\Gamma({l+3\over2})}
               -{1\over x}{\Gamma({l+1\over2})\over\Gamma({l  \over2})}\right)
         +\calO({1\over x^2})
.\eqno(\rmA5)$$
Between the series of (A4) for $x\simlt l$ and the series (A5) for $x\simgt l$
one can compute $I_l(x)$ accurately for all values of $x$ as long as $l$ is not
too large.  While these series involves significant cancelation for $x\sim l$
one can use symbolic algebra programs to exactly compute the series up to a
given order for rational values of $x$ and interpolate in between.

\subheader{Asymptotic Approximation and Small-Angle Limit}

	For large $l$ one finds that $I_l(x)$ is well approximated by just the
asymptotic form of (A5) where this is positive, and where (A5) is negative
$I_l(x)$ is negligibly small, i.e. 
$$I_l(x)\approx\left(a_l-{b_l\over x}\right)\,
                                  \calH\left(x-{b_l\over a_l}\right) \qquad
a_l={\sqrt{\pi}\over2l}{\Gamma({l+2\over2})\over\Gamma({l+3\over2})} \qquad
b_l={\sqrt{\pi}\over l}{\Gamma({l+1\over2})\over\Gamma({l  \over2})}
\eqno(\rmA6)$$
where $\calH()$ is the Lorentz-Heaviside function which is unity for positive
argument and zero otherwise.  Taking the large $l$ limit of the factorial
ratios this becomes
$$I_l(x)\approx\sqrt{\pi\over2l}\,\left({1\over l}-{1\over x}\right)
                                \,\calH(x-l) \qquad l\gg1
.\eqno(\rmA7)$$
The the fractional error of (A7) goes to zero for large $l$ and $x>l$ while the
absolute error goes to zero for large $l$ and $x<l$.

\subheader{Volume Average: $\overline{I_l}$}

	In this paper we have considered the mean shear from uniform distance
limited in an Einstein-deSitter cosmology where the gravitational potential is
assumed to be constant.  In this case one use $\overline{I_k}$ defined in
eq~(5.12) with the visibility function of eq~(4.42) to compute the mean shear.
The Taylor series of the resultant function is 
$$\overline{I_l}(\xmax)=3\sum_{n=0}^\infty
     {(-1)^n\,\xmax^{2n+l}\over2^n\,(2(l+n)+1)!!\,n!\,(2n+l)(2n+l+1)\,(2n+l+3)}
\qquad \xmax\equiv k\rmax
.\eqno(\rmA8)$$
Note that by making $\overline{I_l}$ a function of $k\rmax$ rather than just
$k$ we are using a slightly different notation than in the body of the paper.
One may defined an asymptotic approximation just as in eq~(A6) which is
$$\overline{I_l}(\xmax)\approx
 \left(a_l-{3\over2}{b_l\over\xmax}+{1\over2}{b_l^3\over a_l^2\xmax^3}\right)\,
 \calH\left(\xmax-{b_l\over a_l}\right)
\eqno(\rmA9)$$
where $a_l$ and $b_l$ are defined just as in eq~(A6).  The large $l$ limit of
this function is
$$\overline{I_l}(\xmax)\approx{1\over l}\,\sqrt{\pi\over2l}\,
    \left(1-{3\over2}{l\over\xmax}+{1\over2}{l^3\over\xmax^3}\right)\,
    \calH(\xmax-l) \qquad l\gg1
.\eqno(\rmA10)$$
Numerically the function $\bar{I_l}$ as a function of $\xmax$ is not very
different than the $I_l$ as a function of $x$.

\vfill\eject

\header{FIGURE CAPTIONS}

\noindent
{\bf Figure 1:} In the top left panel is shown a {\it scalar} pattern of
gravitational deflections.  Scalar deflections are parallel (or anti-parallel)
to the direction of variation of the deflection.  In the top right panel is
shown a {\it pseudo-scalar} pattern of gravitational deflections. Pseudo-scalar
deflections are $\pm90^\circ$ from the direction of variation of the
deflection. For an arbitrary pattern of gravitational deflections one will
transform the scalar components of the deflections into pseudo-scalar
components and vice versa by rotating the deflection direction by $90^\circ$.
Plotted in the bottom left panel is a scalar pattern of ellipticities generated
by applying the deflections in the top left panel to a square array of circles.
Scalar shear produces ellipticity patterns which are parallel and perpendicular
to the (horizontal) direction of variation of the ellipticities.  Scalar shear
produces images which are magnified or demagnified. In the bottom right panel
is shown the pattern of ellipticities induced by a the deflections in the top
left panel. Pseudo-scalar ellipticities oscillate between being $+45^\circ$ and
$-45^\circ$ from the direction of variation.  In the weak lensing limit
pseudo-scalar shear does not amplify the images. One will transform the scalar
components of the shear pattern into pseudo-scalar components of shear and vice
versa by rotating the object ellipticities by $45^\circ$.  Such a
transformation will not change the magnification or the displacement of the
objects on the sky, so rotated scalar shear will still exhibit
(de)magnification while rotated pseudo-scalar shear will not.  While shear
patterns cannot always be treated as the result of a deflection it is usually
a good approximation to do so.

\noindent
{\bf Figure 2:} Here is illustrated the meaning of the variables in the 2-point
correlation function $C_\gamma$, $C_{\gamma\kappa}$, and $C_{\gamma\omega}$.
The thick solid line represents the geodesic connecting the two points on the
celestial sphere in question, $\hatbfn$ and $\hatbfn'$, whose length is
$\vartheta$.  The thin solid line at $\hatbfn$ gives the orientation of the
component of the shear that is being correlated, which is rotated $\varphi$
from the direction of the geodesic.  For the shear-shear correlation function,
$C_\gamma$, we also need the orientation of the component of the shear at
$\hatbfn'$ which is given by $\varphi'$.  The handedness of the coordinate
system is important $C_{\gamma\omega}$ since it changes sign if
$\varphi\rightarrow-\varphi$.  The rotation of $\varphi$ is to the right in a
right-handed coordinate system. Note that the usual spherical polar
coordinates, $(\theta,\phi)$ are usually left-handed on the sky since we view
the celestial sphere from the {\it inside}.

\noindent
{\bf Figure 3:} Plotted as a function of $l$, for a power law spectrum of
density perturbations, is the ratio of the true $C_l^\oplus$ to that computed
with the {\it asymptotic approximation} of eq~(5.9).  The power law index, $n$,
is 0.9 for the crosses, 0 for the circles, -1 for the  triangles, and -2 for
the squares.  The errors continue to become smaller for larger $l$.  The ratio
for $l=1$, which contributes nothing to the shear, is not shown for $n\le-1$
since the true value diverges for these spectral indices.  The asymptotic
approximation is accurate except for very small $l$.

\noindent
{\bf Figure 4:} Plotted as a function of $l$, for a volume limited survey in an
Einstein-deSitter universe is $\overline{C}_l^\oplus$ obtained for an unbiased
COBE-normalized model of adiabatic density fluctuations given by eq~(7.3).  The
yellow, green blue, cyan, magenta, and red points are for a limiting survey
redshift of $\zmax=$0.1, 0.2, 0.5, 1, 2, and 5 respectively.  The corresponding
solid lines give the $\overline{C}_l^\oplus$ for the same limiting redshift but
using the asymptotic approximation and assuming the the small scale power law
index $n=-1.2$ on all scales.  The deviation of the solid lines from the points
is primarily due to the turnover in the spectrum at large scales, which greater
for deeper surveys and smaller $l$ because these probe larger comoving scales.
The large $l$ behaviour is described by eq~(7.2), i.e. $C_l^\oplus\propto
l^{-1.2}\zmax^{2.2}$, while for deep enough surveys, $\zmax\simgt1$, eq~(5.16)
describes the behaviour, i.e. $C_l\simpropto l$ with only a weak depth
dependence.

\noindent
{\bf Figure 5:} Plotted versus $L$  is the rms shear contributed by modes with
angular wavenumber $l<L$.  As in fig~3 these are for the power spectrum model
of eq~(7.3) and the yellow, green blue, cyan, magenta, and red points are for a
limiting survey redshift of $\zmax=$0.1, 0.2, 0.5, 1, 2, and 5 respectively.
The the solid lines are to guide the eye.  For $l\gg1$ this also gives the rms
expansion, $\kappa$, which is gives the source amplification.

\noindent
{\bf Figure 6:} Plotted versus $l$ is the ratio of the shear power
$\overline{C}_l^\oplus$ to that induced by finite galaxy sampling in a volume
limited survey with with rms ellipticity 0.3 and galaxy density $0.012(h/{\rm
Mpc})^3$.  As in figs~3\&4 these are for the power spectrum model of eq~(7.3)
and the yellow, green blue, cyan, magenta, and red points are for a limiting
survey redshift of $\zmax=$0.1, 0.2, 0.5, 1, 2, and 5 respectively.

\noindent
{\bf Figure 7:} Plotted versus $l$ is an estimate of the fractional uncertainty
(1-$\sigma$) in the measurement of $C_l^\oplus$ from a volume limited sample
with rms ellipticity 0.3 and galaxy density $0.012(h/{\rm Mpc})^3$.  This
uncertainty, taken from eq~(6.17), includes the effects of finite galaxy
sampling and cosmic variance. As in figs~$3-5$ these are for the power spectrum
model of eq~(7.?).  The yellow, green blue, and cyan points are for a limiting
survey redshift of $\zmax=$0.1, 0.2, 0.5, and 1.  The smallest fractional
uncertainty, for $\zmax=1$ is dominated by cosmic variance for the $l$'s
plotted.  One cannot significantly reduce this by going to a greater depth.
Less deep surveys are significantly degraded by finite galaxy numbers.  A
sample with $\zmax=0.1$ survey would yield a marginally significant result by
combining all $l$-modes.  One would not obtain a significant signal for $\zmax$
significantly less than 0.1.

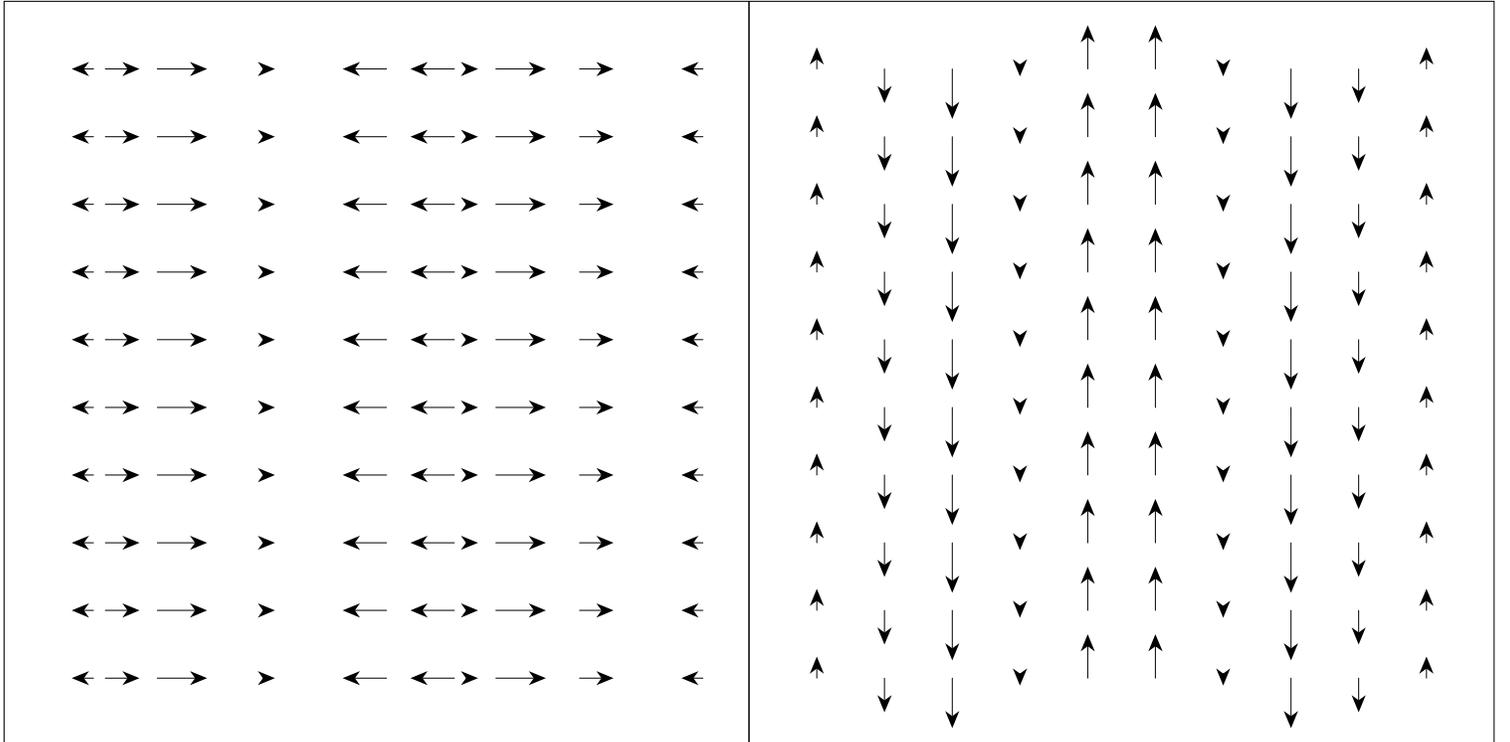
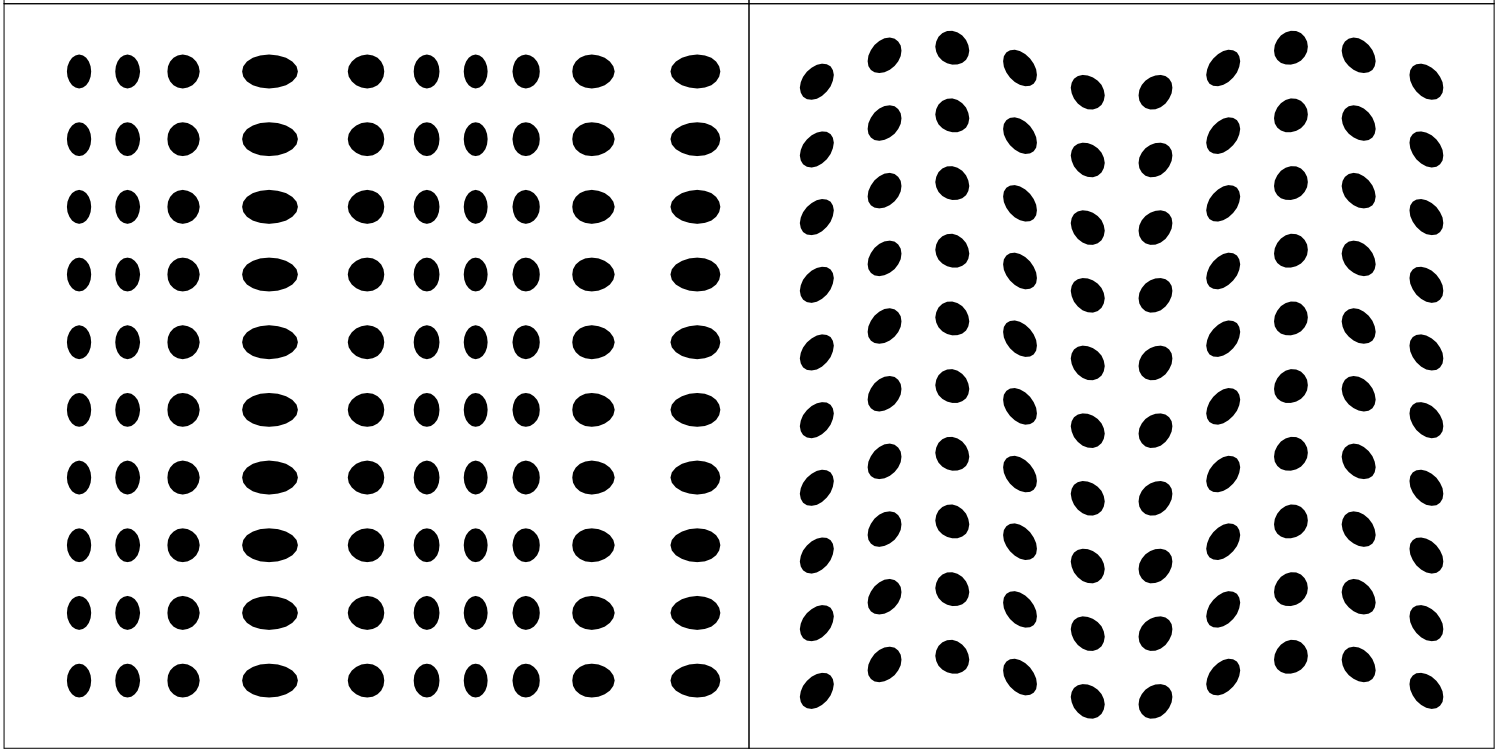
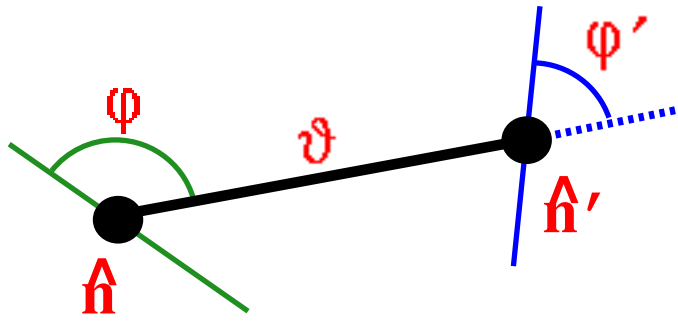
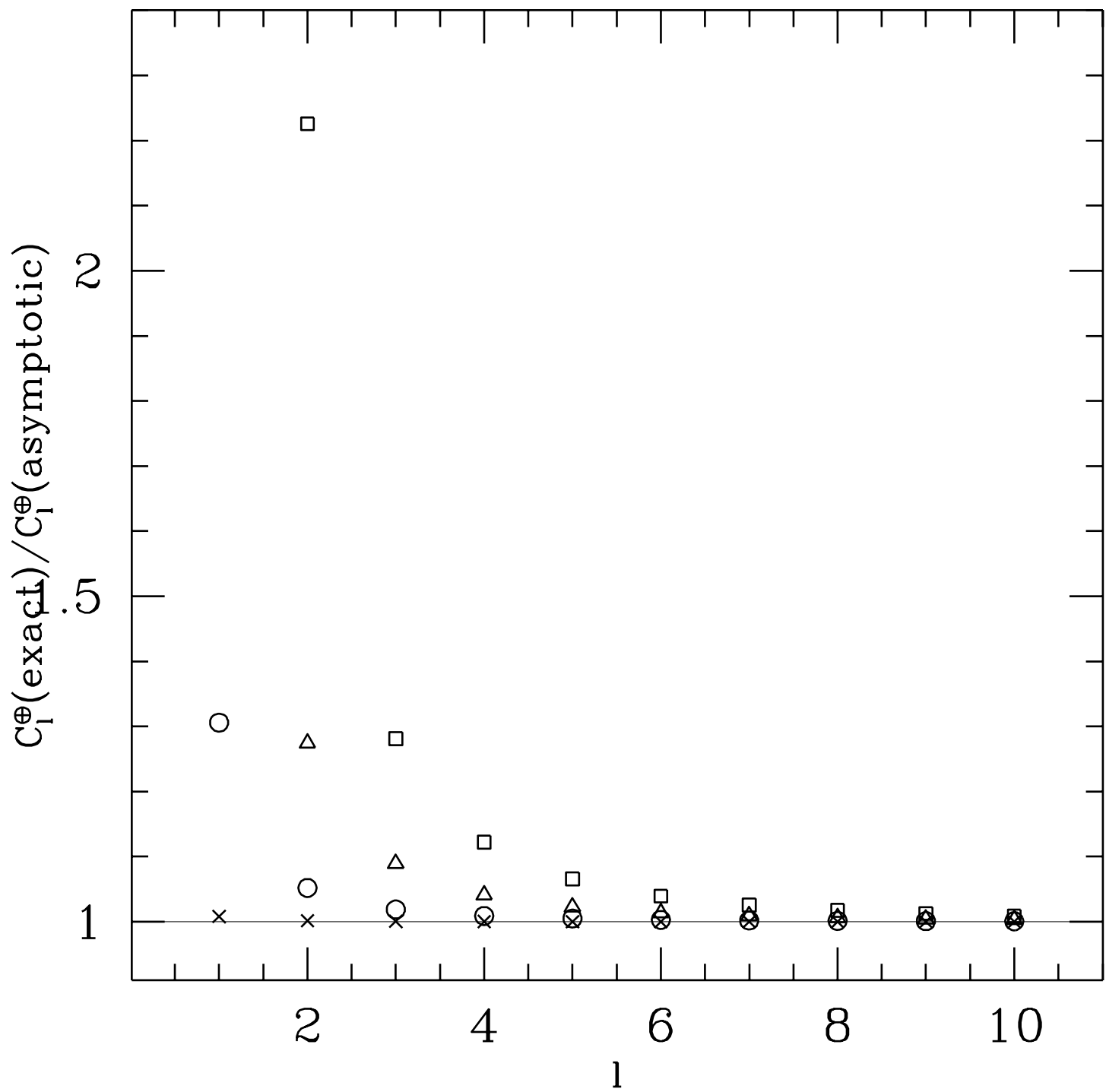
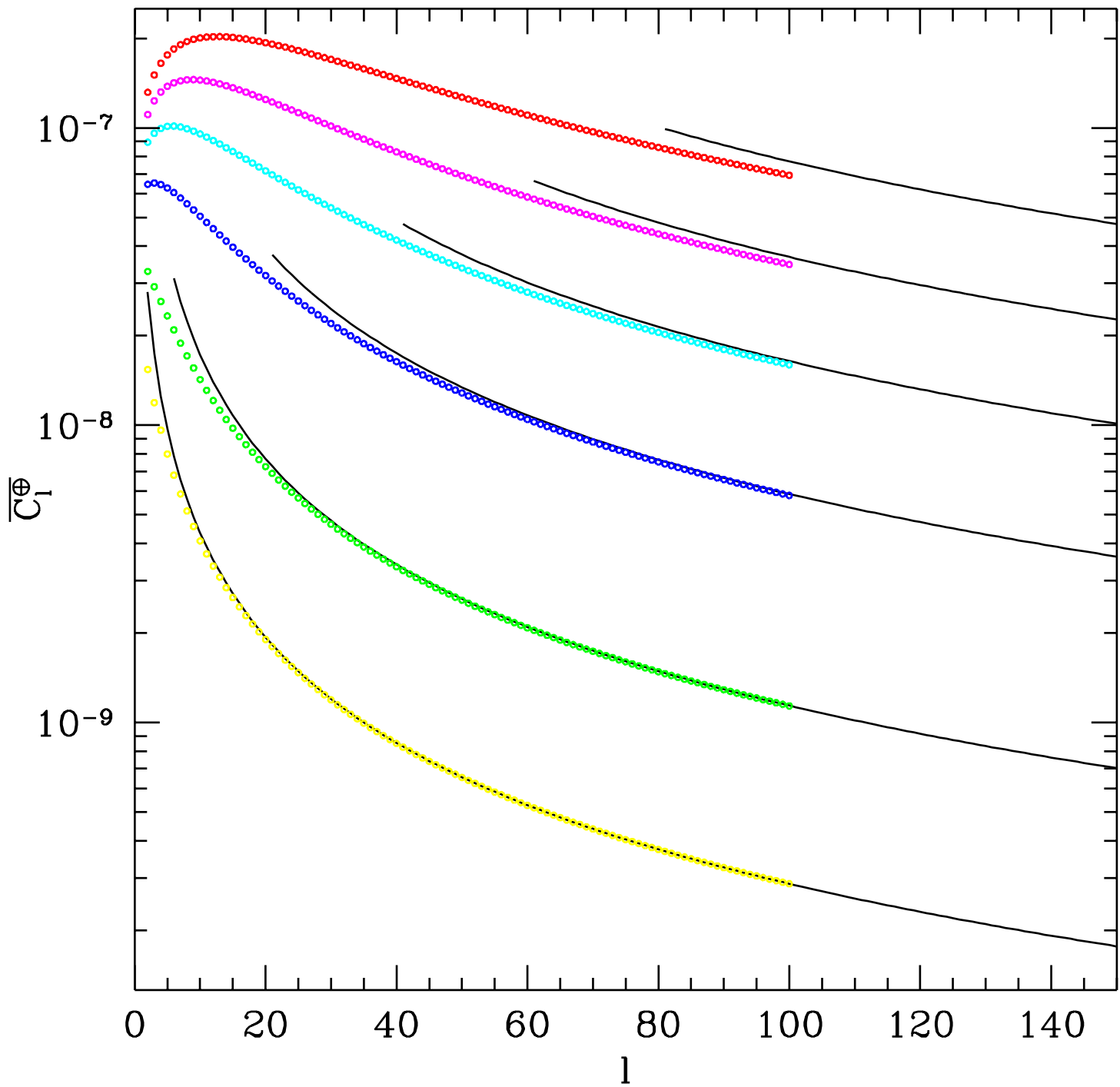
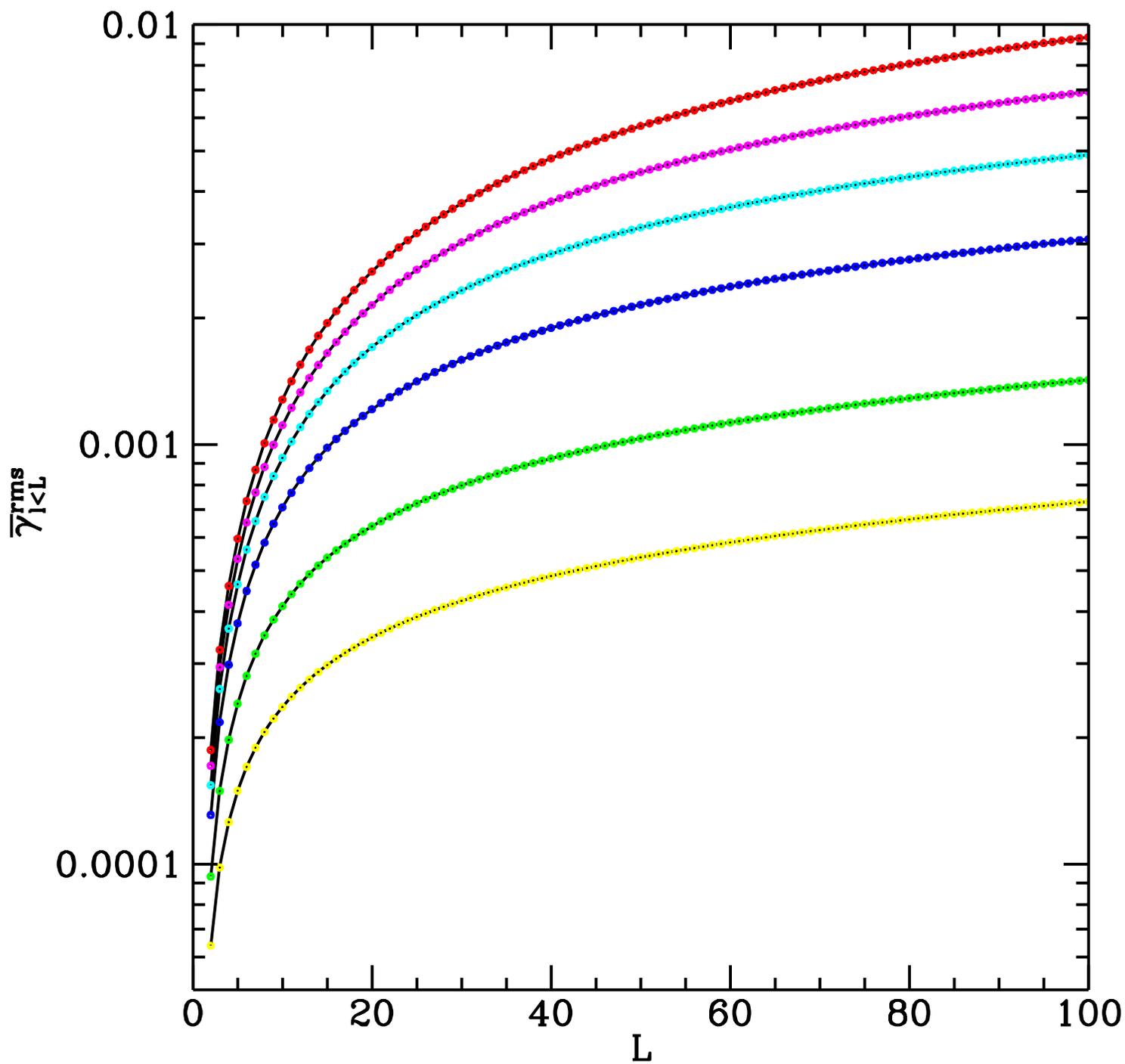
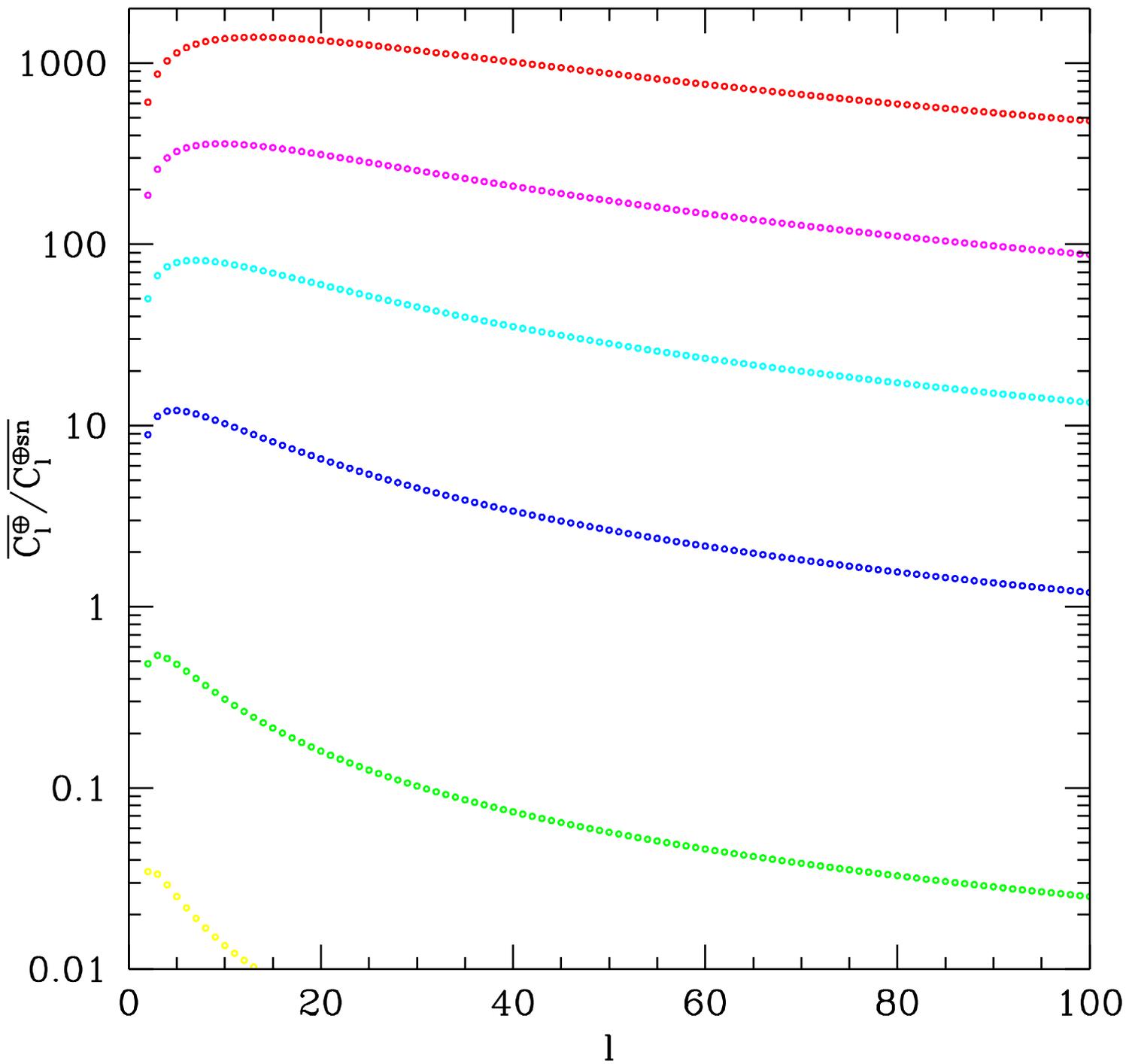
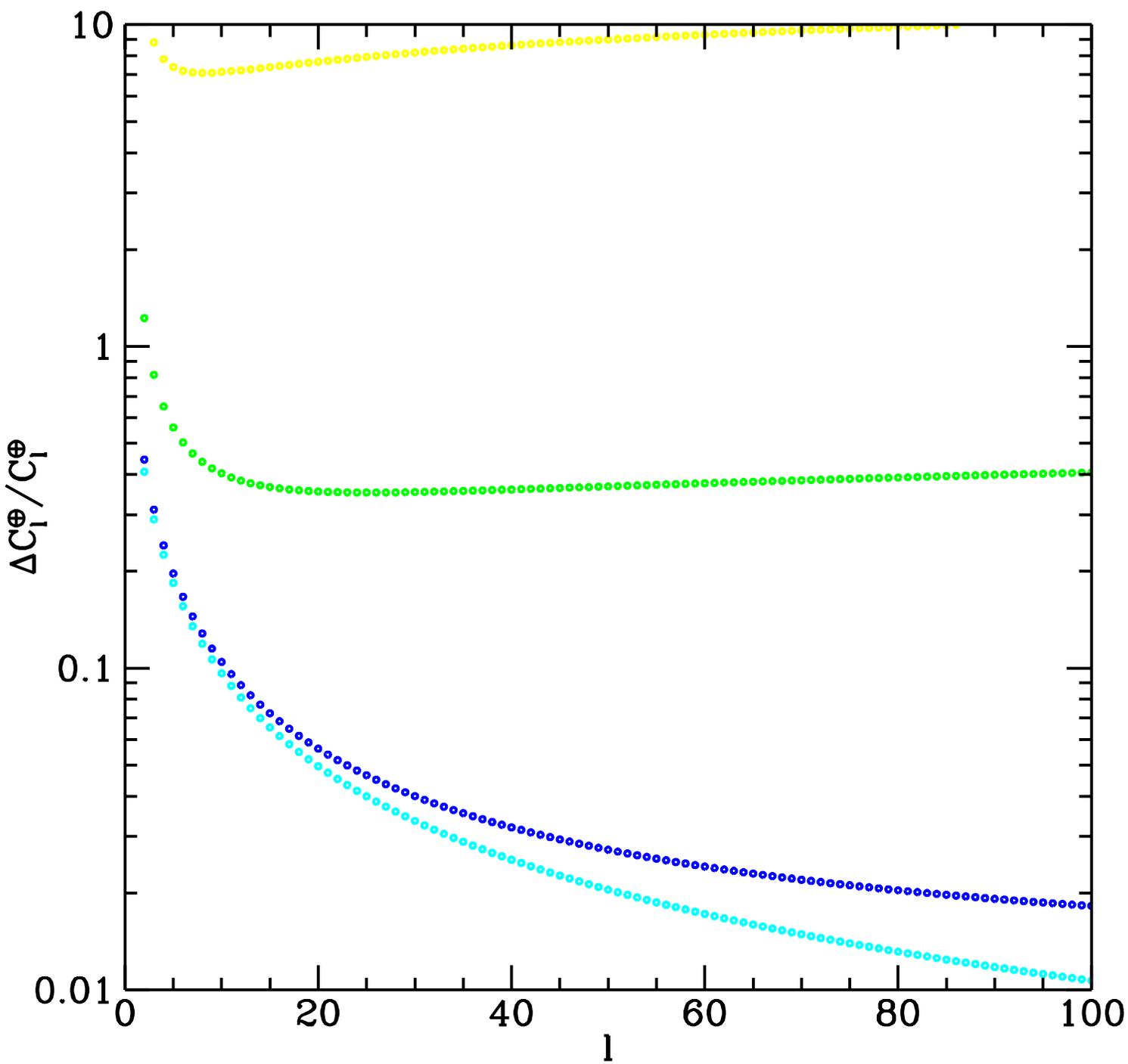
\end